\documentclass[a4paper,11pt]{article}
\usepackage{jheppub} % for details on the use of the package, please see the JINST-author-manual

\usepackage{enumitem} % `nosep' which removes the vertical space between items in an enumerate environment

 \usepackage{adjustbox} 
\usepackage{datetime2}
\usepackage{cleveref}
\crefname{equation}{Eq.}{Eqs.}
\crefname{section}{Section}{Sections}
\crefname{subsection}{Section}{Sections}
\crefname{appendix}{Appendix}{Appendices}
\usepackage{orcidlink}
\usepackage{slashed}
\usepackage{tikz}
\newcommand{\linegraph}[2][1.2]{%
\begin{tikzpicture}[baseline=-0.5ex]
    \foreach \lab [count=\i from 1, remember=\lab as \prev] in {#2} {%
        \ifnum\i>1
            \pgfmathtruncatemacro{\j}{\i-1}
            \draw ({(\j-1)*#1},0) -- ({(\i-1)*#1},0)
                node[midway, above=1pt] {$e_{\prev\lab}$};
        \fi
        \node[circle, fill, inner sep=2pt, label=below:{$k_{\lab}$}]
            at ({(\i-1)*#1},0) {};
    }%
\end{tikzpicture}%
}

\tikzset{
    edgelabel/.style={midway, fill=white, inner sep=1pt}
}

\newcommand{\stargraphfour}[4]{%
\begin{tikzpicture}[baseline=-0.5ex, scale=1]
    \coordinate (v1) at (0,0);
    \coordinate (v2) at (0,1);
    \coordinate (v3) at (-0.9,-0.7);
    \coordinate (v4) at (0.9,-0.7);

    \draw (v1) -- (v2) node[edgelabel, right] {$e_{#1#2}$};
    \draw (v1) -- (v3) node[edgelabel, above left] {$e_{#1#3}$};
    \draw (v1) -- (v4) node[edgelabel, above right] {$e_{#1#4}$};

    \node[momnode] at (v1) {};
    \node[below=6pt] at (v1) {$k_{#1}$};

    \node[momnode, label=above:{$k_{#2}$}] at (v2) {};
    \node[momnode, label=below left:{$k_{#3}$}] at (v3) {};
    \node[momnode, label=below right:{$k_{#4}$}] at (v4) {};
\end{tikzpicture}%
}

\tikzset{
    momnode/.style={circle, fill, inner sep=1.8pt},
    graphline/.style={line width=0.4pt},
}

\newcommand{\forkgraphfive}[5]{%
\begin{tikzpicture}[baseline=-0.5ex, x=0.82cm, y=0.82cm]
    \coordinate (v1) at (0,0);
    \coordinate (v2) at (-1.35,0.95);
    \coordinate (v3) at (-1.35,-0.95);
    \coordinate (v4) at (0.95,0);
    \coordinate (v5) at (1.90,0);

    \draw[graphline] (v1) -- (v2);
    \draw[graphline] (v1) -- (v3);
    \draw[graphline] (v1) -- (v4);
    \draw[graphline] (v4) -- (v5);

    \node at (-1,0.38) {$ e_{#1#2}$};
    \node at (-1,-0.38) {$ e_{#1#3}$};
    \node at (0.47,0.20) {$ e_{#1#4}$};
    \node at (1.42,0.20) {$ e_{#4#5}$};

    \node[momnode] at (v1) {};
    \node[below=2pt] at (v1) {$ k_{#1}$};

    \node[momnode] at (v2) {};
    \node[above left=2pt] at (v2) {$ k_{#2}$};

    \node[momnode] at (v3) {};
    \node[below left=2pt] at (v3) {$ k_{#3}$};

    \node[momnode] at (v4) {};
    \node[below=2pt] at (v4) {$k_{#4}$};

    \node[momnode] at (v5) {};
    \node[right=2pt] at (v5) {$ k_{#5}$};
\end{tikzpicture}%
}
\let\amsmathbb\mathbb
\usepackage[bbgreekl]{mathbbol}
\let\mathbb\amsmathbb

\definecolor{americanrose}{rgb}{1.0, 0.01, 0.24}
\definecolor{cadmiumgreen}{rgb}{0.0, 0.42, 0.24}
\definecolor{darkorange}{rgb}{0.8, 0.549, 0.0}

\newcommand{\dd}{d}
\newcommand{\hd}{\hat{d}}

\newcommand{\hdelta}{\hat{\delta}}

\newcommand{\bu}{\bar{u}}

\newcommand{\bm}{\bar{m}}
\newcommand{\xinf}{x_{\infty}} % conjugate to the momentum transfer, transverse to the incoming velocities
\newcommand{\xinfhat}{\hat{x}_{\infty}} % unit vector along \xinf
\newcommand{\xinfnorm}{|\xinf|} % length of \xinf, \sqrt{-\xinf^2}
\newcommand{\bbX}{\mathcal{R}} % generating vector of rooted trees
\newcommand{\rdelta}{\bar{\delta}} % reduced phase; \bbdelta comes from the mathbbol package
\newcommand{\misangle}{\varpi}

\newcommand{\cO}{\mathcal{O}}

\newcommand{\cA}{\mathcal{A}}

\newcommand{\Disc}{\mathrm{Disc}}
\newcommand{\Res}{\mathrm{Res}}

\newcommand{\sint}{\!\int\!}

\def\probe{h}
\def\heavy{H}

\makeatletter
\newcommand{\smallbullet}{} % for safety
\DeclareRobustCommand\smallbullet{%
  \mathord{\mathpalette\smallbullet@{0.8}}%
}
\newcommand{\smallbullet@}[2]{%
  \vcenter{\hbox{\scalebox{#2}{$\m@th#1\bullet$}}}%
}
\makeatother
\title{\boldmath The Hidden Simplicity of the Classical Limit}

\author[a, b]{Yilber Fabian Bautista \orcidlink{0000-0001-6255-5675},}
\author[b]{Stefano De Angelis \orcidlink{0000-0002-3613-9480},}
\author[b]{David A. Kosower \orcidlink{0000-0002-9087-0071},}
\author[a, b]{Donal O'Connell \orcidlink{0000-0001-8886-6391},}
\author[b, c]{Matteo Sergola \orcidlink{0000-0002-5305-7612}}
\affiliation[a]{Higgs Centre for Theoretical Physics, School of Physics and Astronomy, The University of Edinburgh, Edinburgh EH9 3JZ, Scotland, UK}
\affiliation[b]{Institut de Physique Théorique, CEA, CNRS, Université Paris-Saclay, F-91191 Gif-sur-Yvette cedex, France}
\affiliation[c]{Mani L. Bhaumik Institute for Theoretical Physics, University of California at Los Angeles, Los Angeles, CA 90095, USA}

\emailAdd{yfabian.bautista@ed.ac.uk}
\emailAdd{stefano.de-angelis@ipht.fr}
\emailAdd{david.kosower@ipht.fr}
\emailAdd{donal@ed.ac.uk}
\emailAdd{msergola@physics.ucla.edu}

\abstract{
We study the classical limit of two-to-two massive-scalar amplitudes at all orders of
perturbation theory.  We compute these amplitudes in the probe limit for several
theories: a family of scalar theories, electrodynamics, and 
$\mathcal{N}=8$ supergravity with half-BPS matter.
We show that related multiple-emission classical amplitudes obey a simple factorization derivable
from the usual factorization of quantum amplitudes. 
The classical factorization, which reflects the second-order nature of classical dynamics, holds in more general theories --- including gravity.
We further show that both the multiple-emission and the two-to-two classical amplitudes have a simple
organization in terms of weighted trees.  This allows a recursive computation to all orders in the probe limit.
The probe's trajectory emerges as a generating function of trees.
The required boundary conditions are given by the Feynman $i \varepsilon$ prescription, leading to a novel worldline Green's function.
In our approach, the rich classical dynamics arises from simple combinatorics.
}
\begin{document}
\maketitle
\flushbottom

\section{Introduction}
\label{sec:intro}

The last decade has witnessed a blossoming of new ideas and calculations 
originating in scattering amplitudes applied to gravitational-wave physics.  
These new ideas have dramatically supplemented more traditional approaches based
directly on General Relativity.
 Researchers have applied the new methods to the classical gravitational two-body problem in the weak-field, small-angle (eikonal) regime~\cite{Damour:2017zjx}. These applications
have led to a surge of results in the relativistic perturbative expansion of the two-body problem.   Most notably, these developments have advanced fixed-order calculations
of the potential to the current state of the art 
at $\cO(G^5)$~\cite{Bern:2019nnu,Damour:2020tta,DiVecchia:2021ndb,Bjerrum-Bohr:2021din,Jakobsen:2022fcj,%
Bern:2021dqo,Dlapa:2021npj,Bern:2021yeh,Dlapa:2021vgp,Dlapa:2022lmu,%
Damgaard:2023ttc,Driesse:2024xad,Driesse:2024feo,Bern:2025wyd,Driesse:2026qiz}.

The successes of scattering-amplitude and effective-field-theory methods in obtaining
perturbative results beyond those available from traditional methods nonetheless
leave computational bottlenecks in several regimes of great interest. The simplest example
is the scattering of a light probe off a heavy object. Observables in this regime admit a
well-defined, gauge-invariant expansion in the mass ratio of the two bodies, the so-called
self-force expansion.  At leading order, many examples are integrable, as is also
true in the Schwarzschild and Kerr spacetimes.  The examples can also be treated to all perturbative orders in the coupling --- through either classical 
equations of motion,
or approximations to quantum-mechanical wave equations. The probe limit and the self-force expansion are subjects of active investigation using
scattering-amplitude and effective-field-theory 
methods~\cite{Cheung:2020gbf,Cheung:2023lnj,Kosmopoulos:2023bwc,Cheung:2024byb},
and have been used to reconstruct the metric and probe 
motion~\cite{Mougiakakos:2020laz,Mougiakakos:2024nku,Mougiakakos:2024lif,Damgaard:2024fqj}
in a Schwarzschild background.
The methods that led to the state-of-the-art calculations mentioned above struggle
to reproduce probe-scattering results in a simple way.
Our goal is to bridge this gap and shed light on the emergence of
simple classical results from quantum scattering amplitudes.  Our work will
reveal hidden structure in the classical limit that helps explain the resulting
simplicity.

Precision scattering calculations help us reconstruct the interactions between the 
two bodies; at low orders, this knowledge can be used in bound-system calculations.
This appears to break down starting at $\cO(G^4)$, with
the combination of nonlinearity and the long-range nature of gravity
making knowledge transfer from the scattering to the bound domain
challenging.    From a purely formal point of view,
the dynamics at any point in spacetime depends on that point's entire past light cone~\cite{Foures-Bruhat:1952grw, Poisson:2003nc}. In the two-body problem, this 
appears in interactions that depend on the system's entire past history, 
rather than only on the bodies' instantaneous configuration~\cite{Blanchet:1985sp,Blanchet:1987wq}. 
These interactions, known as \textit{tails} or nonlocal interactions, require careful treatment~\cite{Galley:2009px,Galley:2015kus}.  In a real-world context,
observer limitations and background noise will impose cutoffs on the extent
of these interactions.
Systematic methods for separating local and nonlocal contributions have been developed~\cite{Dlapa:2024cje,Bini:2024tft,Dlapa:2025biy,Bini:2025zvk,Dlapa:2026oyq}. 
Such a separation could make it possible to use relativistic local interactions in 
template-generation pipelines, supplementing them with nonlocal contributions 
computed using other techniques for bound systems.

Other approaches to the gravitational two-body problem avoid the need to map between scattering and bound states because they apply to arbitrary orbits. In the 
so-called multipolar post-Minkowskian (MPM) formalism~\cite{Thorne:1980ru,Blanchet:1985sp,Blanchet:1987wq,Blanchet:1989ki,Blanchet:1992br,Blanchet:2013haa}, the emitted radiation is expressed in terms of the source's multipole moments, which are functionals of the two bodies' worldlines along arbitrary trajectories. The same expressions thereby yield waveforms and other observables for both bound and unbound orbits.  This is demonstrated in recent applications to hyperbolic 
encounters~\cite{Bini:2023fiz,Bini:2024rsy,Georgoudis:2024pdz,Bini:2026dvn}. 
Similarly, post-Newtonian effective field theory describes the interaction between the bodies~\cite{Goldberger:2004jt}, the radiation emitted by a generic multipolar source~\cite{Goldberger:2009qd,Ross:2012fc}, and the radiation's back-reaction on the system~\cite{Galley:2009px,Galley:2012hx}, without assuming a particular orbit. In both approaches, however, the source dynamics is treated in the post-Newtonian expansion, 
that is,
in an expansion in the bodies' relative velocity. One of the authors 
has recently extended this construction to generic orbits at relativistic velocities 
within the weak-field regime, without resorting to an expansion in the 
velocity~\cite{DeAngelis:2026ymn}. The extension
builds on the worldline effective field theory of Ref.~\cite{Kalin:2020mvi}.

In this paper, we take a different approach. We focus on simple systems in which most 
higher-order contributions arise from the details of the bodies' worldline trajectories rather than from complicated bulk interactions. In the probe limit, in particular, there are 
no tails: the light body moves in the fixed field of the heavy one, and the complexity at higher orders comes from its trajectory. Our aim is to identify the systematic corrections associated with particular scattering trajectories and to factor out trajectory-dependent effects.

We pursue this goal using the amplitude approach.
We compute observables associated with relativistic two-body collisions and 
accompanying emitted radiation by taking the classical limit of observables
built out of quantum scattering amplitudes. We use the observables-based formalism (\textit{aka\/} 
KMOC)~\cite{Kosower:2018adc, Cristofoli:2021vyo}, which relates changes in expectation values to $S$-matrix elements.  We take the classical limit using the heavy-mass 
effective field theory method~\cite{Damgaard:2019lfh,Aoude:2020onz,Brandhuber:2021kpo, Brandhuber:2021eyq}, in which the scattering bodies' Compton wavelengths, $\hbar/m$, 
are much smaller than the other relevant length scales. By studying $S$-matrix elements in this limit, we are able to reconstruct the eikonal representation of the amplitude~\cite{Cheng:1969eh,Abarbanel:1969ek,Levy:1969cr,tHooft:1987vrq,Amati:1987wq,Amati:1987uf,Muzinich:1987in,Kabat:1992tb,Saotome:2012vy,Akhoury:2013yua,DiVecchia:2020ymx,DiVecchia:2021bdo,Cristofoli:2021jas,Bellazzini:2022wzv,DiVecchia:2023frv}. This 
representation resums perturbative contributions efficiently,
absorbing classically singular terms that would otherwise appear as divergences at
intermediate stages. Moreover, reorganizing the amplitude at the integrand level helps
avoid the usual bottleneck of perturbative calculations: integration-by-parts (IBP) reduction. Although it plays an essential role in state-of-the-art calculations, 
IBP reduction of integrals can conceal cancellations that are only seen at the end of a long
and large-scale calculation.

The probe limit leads immediately to several simplifications in scattering amplitudes.
When one particle is much heavier than the other, for example, the all-loop integrand 
simplifies: the heavy propagators are always cut. Consequently, the all-loop amplitudes in the probe limit are given by ``fan'' diagrams~\cite{Duff:1973zz}. 
We can therefore construct the $(N-1)$-loop integrand from a tree-level $(N+2)$-point amplitude in which the light particle emits $N$ massless messengers. We will refer to these
as multi-emission amplitudes. Perturbative unitarity~\cite{Bern:1994cg} allows us
to obtain the full amplitude by sewing the emitted messengers to the classical source
associated with the heavy particle. These $N$-emission amplitudes are the central objects of our analysis.

We compute the scattering angle in several theories: a family of scalar theories; 
electrodynamics; and $\mathcal N=8$ supergravity. Although these theories differ in their field content and interactions, their scattering angles are surprisingly similar.
We trace these similarities to a common structure in the classical limits of their
scattering amplitudes.  We reproduce in a novel way results that are simple to obtain classically.  We use our approach to reveal new structure that underlies
the simplicity of the scattering angles.  This structure has implications beyond 
the probe limit.
We prove a factorization theorem for the classical limit of multi-emission amplitudes.
In this limit, pairs of poles of the quantum amplitude collide, giving rise
to double poles in linearized propagators.  
These double poles reflect the fact that the classical equations of motion are second order in time derivatives.
Their coefficients are determined by lower-point classical amplitudes
and their derivatives.  This \textit{classical\/} factorization is independent of the helicities of the emitted messengers. 
It holds for General Relativity in addition to the theories we study explicitly.
He, Parra-Martinez, and Castet~\cite{He:2025how,Castet:2026gao} proved
a very similar factorization theorem in the worldline context.

We then use this theorem to derive a recursion relation that constructs 
the classical limit of multi-emission amplitudes at any multiplicity~\cite{Britto:2005fq}. 
Once the messengers are sewn onto the heavy source, the rational part of the 
probe-limit integrand takes a closed form at any loop order as a sum over Cayley trees, 
also known as labeled trees, in all three theories we consider. 
Each Cayley tree is associated with a simple weight with a manifest double-pole structure.
Unlike Feynman diagrams, the weight cannot be decomposed into vertex and propagator factors.
The combinatorics of the trees and their associated weights capture the classical probe dynamics to all orders.

In a scattering amplitude, the boundary conditions of the trajectory are set by the $i\varepsilon$ prescription.
We show that this translates to a particular worldline Green's function: the max kernel.
In a generating function of trees, the max kernel allows us to resum the trajectory and to compute an eikonal-like phase, defined as the logarithm of the amplitude in coordinate space.
This phase can be interpreted as the on-shell action of the probe.
This formulation makes the deflection angle much easier to compute. 

The Cayley-tree representation reveals cancellations between different trees,
leading to integral identities that are no longer manifest after IBP reduction. 
Our framework thus also clarifies the relationship between scattering amplitudes, the radial action, and the equations of motion. Moreover, the resummed scattering dynamics can connect the classical radial action to bound orbits. Analytically continuing this action below the scattering threshold makes contact with the boundary-to-bound map~\cite{Kalin:2019rwq,Kalin:2019inp}, and yields bound-state periastron advances. 
This provides a complementary perspective on how all-order information contained in amplitudes can encode the dynamics of bound orbits.

The paper is organized as follows. In \cref{sec:scalar_model}, we introduce a family of simple scalar theories as our first example of probe-limit scattering, and compute their scattering angles using both classical, amplitude and eikonal methods. This analysis also highlights the role of multi-emission amplitudes. In \cref{sec:factorization}, we prove a factorization property of heavy-mass amplitudes in the classical limit, derive the recursion relation that follows from it, and use it to construct multi-emission amplitudes in electrodynamics. In \cref{sec:graphs}, we apply this formalism to the scalar theories, electrodynamics and $\mathcal N=8$ supergravity, where a novel use of Cayley-tree diagrammatics allows us to reinterpret the multi-emission amplitudes introduced earlier. In \cref{sec:eikonal}, we combine the Cayley trees with the eikonal integration techniques developed in \cref{sec:scalar_model} to fix the $i\varepsilon$ prescription, express the eikonal-like phase as an on-shell action, connect scattering and bound-state dynamics, and highlight integral identities not captured by IBP reduction.  We summarize
and present an outlook for further investigations in \cref{sec:conclusions}. In \cref{app:worldline}, we revisit the scalar theories from the worldline perspective, rederive their scattering angles classically, and illustrate how Cayley trees emerge in worldline computations.

\paragraph{Conventions}
We work with mostly minus signature $(+,-,-,-)$. Factors of $2\pi$ are absorbed with the help of $\hat{\dd}^n p= \dd^n p/(2\pi)^n$, $\hat{\delta}^n(p) =(2\pi)^n \delta ^n (p)$.
We also set $c = 1$ in expressions for amplitudes. We use the standard conventions for Fourier transforms in $d$ dimensions,
\begin{equation}
    f(x)=\int\! \hat\dd^d k\ e^{-ik{\cdot} x} \tilde{f}(k),\, \qquad \tilde{f}(k)=\int\! \dd^d x\ e^{ik{\cdot} x}f(x),
\end{equation}
and, when unspecified, integrals are over $(-\infty, \infty)$. 

We use calligraphic letters for quantum amplitudes and roman letters for 
their classical limits.  We denote the massive $2\to2$ amplitude by 
$\mathcal{M}_{2\to2}$, and the amplitude for the emission of $N$ massless
messengers from a single massive line by $\mathcal{A}_N$.
We denote their classical limits by $M_{2\to2}$ and $A_N$ respectively. 
When writing explicit forms of $M_{2\to2}$, we will display the kinematic labels $h, H$, but otherwise suppress these labels when no confusion arises.
We also do the same when considering generic two-body quantities such as the scattering angle or the eikonal-like phase. In contrast, we suppress such labels when considering $A_N$, as these amplitudes depend only on one   massive line.  Finally, we
follow the convention that scattering
amplitudes are matrix elements of $T$ (rather than of $i T$), 
where the $S$-matrix is $S=1+iT$.

\section{A Family of Scalar Theories}\label{sec:scalar_model}

We start with a collection of theories with massless scalar
messengers.  These are in some ways
simpler than electrodynamics and
supergravity.  We will be studying the resummation of the impulse in
the scattering of two massive scalars
(or equivalently of the deflection angle).  We
consider the limit in which one of the massive scalars is heavy.

We will consider four different theories, differing
in the interactions between the massive scalars and the massless
messenger. In the first, the interaction is purely cubic; in
the second, purely quartic; the third, square scalar, has both cubic
and quartic interactions analogous to (scalar) electrodynamics with
like charges for the massive scalars; and the fourth, pseudophoton,
has both cubic and quartic interactions analogous to (scalar)
electrodynamics with opposite charges.

The original quantum theory in all cases has an action of 
the general form,
\begin{equation}
    \begin{split}
        \label{eq:qft_action_scalar}
S
= \int \dd^d x\,
&\left[
\frac{1}{2}\partial_\mu\phi\,\partial^\mu\phi
+
\frac{1}{2}\partial_\mu \probe\,\partial^\mu \probe
-\frac{1}{2}
\left(
\frac{m_\probe}{\hbar}
+\frac{1}{2}
\frac{g_\probe}{\hbar^{\frac{5-d}{2}}}
\phi
\right)^2
\probe^2
-\frac{1}{4}
\frac{\lambda_\probe}{\hbar^{5-d}}
\phi^2 \probe^2\right.\\
&\quad\left.
+\frac{1}{2}\partial_\mu \heavy\,\partial^\mu \heavy
-\frac{1}{2}
\left(
\frac{m_\heavy}{\hbar}
+\frac{1}{2}
\frac{g_\heavy}{\hbar^{\frac{5-d}{2}}}
\phi
\right)^2
\heavy^2
-\frac{1}{4}
\frac{\lambda_\heavy}{\hbar^{5-d}}
\phi^2 \heavy^2
\right]\,,
    \end{split}
\end{equation}
where $\heavy$ represents the heavy scalar, and $\probe$ the
probe (light massive) scalar. 
The massless messenger is denoted
$\phi$.  We keep $\hbar$ explicit in the first subsection below.

We always set $\lambda_\heavy=-g_\heavy^2/2$, so that the heavy field $\heavy$
couples to $\phi$ only through the cubic interaction $\phi \heavy^2$. 
The four different theories we consider then correspond to four 
choices of the couplings of the probe $\probe$:
\begin{itemize}
    \item[$\smallbullet$] Purely cubic, with $\lambda_\probe=-g_\probe^2/2$. This theory 
    is simple enough to allow some computations to be carried out
    to all orders, as we will show later in this section.
    \item[$\smallbullet$] Purely quartic, with $g_\probe=0$.  
    \item[$\smallbullet$] square scalar: both cubic and quartic interactions, tuned via
    $\lambda_\probe=0$.  The effective mass squared is a perfect
square. We will explore this theory in 
    \cref{sec:graphs}.
    \item[$\smallbullet$] pseudophoton: both cubic and quartic interactions, tuned via
    $\lambda_\probe=-g_\probe^2$. This is analogous to electromagnetic interactions. We will explore this 
    theory in \cref{sec:graphs} as well.
\end{itemize}

We are interested in the simplifications that arise in the 
classical limit,
rather than in specific classical results. 
The corresponding worldline theories describe the 
time dependence of the mass of the probe as an extended object due 
to its motion  
in the background scalar field produced by the heavy scalar.
The classical worldline action of this theory takes the form,
\begin{equation}\label{eq:wqft_scalar_1}
S
=
\int \dd^d x\,
\frac12 \partial_\mu\phi\,\partial^\mu\phi
-
\int \dd\tau\,
m_\probe\left(
\phi(x_\probe(\tau))
\right)\,,
\end{equation}
where $\phi(x_\probe(\tau))$ is the classical messenger field at
the probe. This action is reparametrization invariant, as
explained in \cref{app:worldline}.
The specific functional form of $m_\probe(\phi)$ 
corresponds to the different theories mentioned above.
We give the respective functional forms in 
\cref{tab:scalar_worldline_models} in \cref{app:worldline}. 

\subsection{Classical Angle}\label{subsec:scalar_angles}

As a warm-up exercise in our quantum-to-classical approach, let us
compute the scattering angle for a test mass $m_\probe$ moving in 
the scalar background
$\phi(x)$ generated by the heavy source $\heavy$. 
We could obtain this angle in
several ways: from a classical worldline computation as
described in \cref{app:worldline}; from the classical limit of QFT
amplitudes, an approach we discuss below 
in~\cref{subsec:classical_limit}; or directly
from the solutions to the equations of motion of $\probe(x)$.
In this subsection, we follow the last approach. More precisely, 
we extract the classical angle from the classical limit of the
partial-wave phase shift, following the construction of 
Ref.~\cite{Kol:2021jjc}.  That construction was itself inspired by 
earlier studies~\cite{Ford:1959, Berry:1972na}. 

\def\classical{\textrm{cl}}
In this subsection we restrict our analysis to $d=4$. 
Accordingly, both $g_i$ and $\lambda_i$ are dimensionless, while
\begin{equation}
    [\phi]=[\probe]
    =\mathsf{M}^{1/2}\mathsf{L}^{-1/2}\,,
\end{equation}
where $\mathsf{M}$ and $\mathsf{L}$ denote the dimensions of mass
and length, respectively.
From \cref{eq:qft_action_scalar} we obtain the  equation of 
motion for $\probe(x)$
\begin{equation}
\left[
\partial_t^2-\nabla^2
+
\left(\frac{m_\probe}{\hbar}+\frac{g_\probe}{2\sqrt{\hbar}}\phi(x)\right)^2
+
\frac{\lambda_\probe}{2\hbar}\phi^2(x)
\right]\probe(x)=0\,.
\end{equation}
Assume that $\heavy$ is fixed at the origin, so that
$\phi(x)$ is a Coulomb-like potential depending only on the 
radial coordinate,
\begin{equation}
   \phi(x)=-\frac{\sqrt{\hbar}g_\heavy}{8\pi r }\,.
\end{equation}
This indicates that the classical coupling corresponding to $g_H$ should absorb a $\sqrt{\hbar}$ factor.
Use the separation-of-variables ansatz,
\begin{equation}
\probe(t,\mathbf{x})
=
e^{- i E t/\hbar}\,
Y_{j 0}(\theta,0)\,
\frac{\sqrt{\hbar}u(r)}{r}\ ,
\end{equation}
where the $Y_{j m}(\theta,\varphi)$ are the scalar spherical harmonics
with angular eigenvalues $j(j+1)$.  We specify the four-momentum
of the probe $p_\probe$, and use the notation $E=p_\probe^0$ and
$k=|\mathbf{p}_\probe| = \sqrt{E^2- m_{\probe}^2}$.
The radial function $u(r)$ satisfies
\begin{equation}\label{eq:radial_equation}
   u''(r)+
\left[
\frac{k^2}{\hbar^2}
-
\frac{2\eta_c k}{\hbar r}
-
\frac{\nu(\nu+1)}{r^2}
\right]u(r)=0\ ,
\end{equation}
where
\begin{equation}
\begin{aligned}
\eta_c &= -\frac{m_\probe g_\probe g_\heavy}{16\pi k}\,,\\
    \nu(\nu+1) &=
j(j+1)+\frac{g_\probe^2 g_\heavy^2}{256\pi^2}+\frac{\lambda_\probe g_\heavy^2}{128\pi^2}\,.
\end{aligned}
\label{eq:parameters}
\end{equation}
The radial equation \eqref{eq:radial_equation} is of Coulomb type. 
We can write the 
solution regular at $r=0$ in terms of the regular Whittaker 
function $M_{a,b}(z)$~\cite{NIST:DLMF},
\begin{equation}
    \frac{u(r)}{r} =   \frac{\Gamma (\nu-i \eta_c +1) }{\Gamma (2 \nu+2) }M_{i \eta_c ,\nu+\frac{1}{2}}\left(\frac{2 i k r}{\hbar}\right)\frac{1}{r}\ .
\end{equation}
As usual, we obtain the partial-wave
$S$-matrix element, $S_j=e^{2 i \delta_j/\hbar}$, from the
large-$r$ behavior of the regular solution
\begin{equation}
    u(r) \sim e^{-i k r/\hbar+i \eta_c  \log \left(2 k r/\hbar\right)}-(-1)^j e^{2 i \delta_j/\hbar} e^{i k r/\hbar-i \eta_c  \log \left(2 k r/\hbar\right)}\,,
\end{equation}
where the partial-wave phase shift is
\begin{equation}
    \label{eq:partial_wave_quantum}
    2\delta_j=\hbar\pi  (j-\nu)+i\hbar \log \left(\frac{\Gamma (\nu-i \eta_c +1)}{\Gamma (\nu+i \eta_c +1)}\right)\ .
\end{equation}
The next step is to take the
classical limit, which amounts to considering large partial-wave angular
momenta and large charges. In practice, we rescale,
\begin{equation}
    j\to \frac{J}{\hbar}\,,\quad g_i\to \frac{g_i}{\sqrt{\hbar}}\,,\quad \lambda_i\to \frac{\lambda_i}{\hbar}\ ,
\end{equation}
and keep the leading contribution as $\hbar\to0$.
It is convenient to introduce the combinations
\begin{equation}
    \alpha_\phi^2=\frac{g_\heavy^2(2\lambda_\probe+g_\probe^2)}{256\pi^2}\,,\qquad
    \beta_\phi=\frac{g_\heavy g_\probe}{16\pi}\ .
\end{equation}
We expand the logarithms of the gamma functions using Stirling's formula,
so that the classical phase shift takes the form,
\def\Jasq{\sqrt{\vphantom{\gamma^2}J^2{+}\smash{\alpha_\phi^2}}}
\def\Jbsqpm{\sqrt{\vphantom{\gamma^2}J^2\pm\smash{\beta_\phi^2}}}
\begin{equation}
\begin{aligned}
2\delta_J^{\classical}
=\pi J
&+\frac{\beta_\phi}{\sqrt{\gamma^2{-}1}}
\left[
2{-}\log\left(\frac{
J^2{+}\alpha_\phi^2{+}\frac{\beta_\phi^2}{\gamma^2{-}1}}{\hbar^2}
\right)
\right]
\\
&{-}\Jasq
\left[
\pi{+}2\arctan\left(
\frac{\beta_\phi}
{\sqrt{\gamma^2{-}1}\Jasq}
\right)
\right]\ .
\end{aligned}
\label{eq:class_eikonal_partial_waves}
\end{equation}

We can then obtain the scattering angle by 
differentiating,
\begin{equation}
    \chi=-\frac{\dd(2\delta_J^{\classical})}{\dd J}\,,
\end{equation}
whereupon,
\begin{equation}
\label{eq:angle_scalar}
\chi=
\frac{J}{\Jasq}
\left[
\pi+2\arctan\!\left(
\frac{\beta_\phi}{\sqrt{\gamma^2-1}\Jasq}
\right)
\right]
-\pi\ .
\end{equation}
We can specialize this general result to the four theories
of interest
discussed above.  We summarize the corresponding scattering angles in
\cref{tab:scalar_worldline_angles}.

\begin{table}[ht]
\centering
\begin{tabular}{c|c}
\hline
\hline
Theory & Scattering angle $\chi$ \\
\hline
Pure cubic $(\lambda_\probe=-g_\probe^2/2)$ &
$\displaystyle
2\arctan\biggl(\frac{ \beta_\phi}{J\sqrt{\gamma^2-1}}\biggr)
$
\\[1.2em]
Pure quartic  ($g_\probe=0$)&
$\displaystyle
\pi\frac{J}{\sqrt{J^2+\frac{\lambda_\probe g_\heavy^2}{128\pi^2}}}-\pi
$
\\[1.2em]
square scalar ($\lambda_\probe=0$)&
$\displaystyle
\frac{J}{\sqrt{\vphantom{\gamma^2}J^2+\smash{\beta_\phi^2}}}
\left[
\pi
+
2\arctan\biggl(
\frac{1}{\sqrt{\gamma^2-1}}
\frac{\beta_\phi}{\sqrt{\vphantom{\gamma^2}J^2+\smash{\beta_\phi^2}}}
\biggr)
\right]
-\pi
$
\\[1.2em]
pseudophoton ($\lambda_\probe=-g_\probe^2$)&
$\displaystyle
\frac{J}{\sqrt{\vphantom{\gamma^2}J^2-\smash{\beta_\phi^2}}}
\left[
\pi
+
2\arctan\biggl(
\frac{1}{\sqrt{\gamma^2-1}}
\frac{\beta_\phi}{\sqrt{\vphantom{\gamma^2}J^2-\smash{\beta_\phi^2}}}
\biggr)
\right]
-\pi
$
\\[2.2em]
\hline
\hline
\end{tabular}
\caption{Classical scattering angles for the four classical scalar theories.}
\label{tab:scalar_worldline_angles}
\end{table}
The angle for the pure-quartic theory resembles an angular-momentum-induced
correction, rather than that of a Coulomb-like long-range
interaction. A more interesting case is the pure-cubic theory. Although the
result is fully relativistic, as reflected in the appearance of $\sqrt{\gamma^2-1}=k/m_\probe$, the
angular-momentum dependence inside the arctangent closely resembles the
structure of non-relativistic Coulomb scattering in quantum
mechanics, or that of Newtonian scattering in gravity.
More strikingly, as we shall see in \cref{subsec:supergravity_n8},
this angle is identical to the angle for test-body motion in the 
background of an extremal
half-BPS black hole in $\mathcal{N}=8$ supergravity, 
as given in Refs.~\cite{Caron-Huot:2018ape,Parra-Martinez:2020dzs}
(up to a redefinition of couplings),
\begin{equation}   \label{eq:angle_sugra}
    \chi^{\mathcal{N}=8}
    =
    2\arctan\left[
        \frac{2 G m_\heavy m_\probe (\gamma-\cos\misangle)^2}
        {  \sqrt{\gamma^2-1} \,J}
    \right]\ ,
\end{equation}
where $\misangle$ parametrizes the relative orientation of the charges of the two black holes. This particular scattering angle in maximal supergravity is
as simple as that in a cubic scalar theory.

Finally, the cubic-plus-quartic theory gives an angle that closely resembles
Coulomb scattering \cite{Landau:1975pou}:
\begin{equation}
    \label{eq:angle_coulomb}
    \chi^{\text{Coulomb}}=\frac{J}{\sqrt{J^2-\alpha^2}}\left[\pi+2 \arctan\left(\frac{\gamma \alpha}{\sqrt{\gamma^2-1} \sqrt{J^2-\alpha^2}}\right)\right]-\pi\ .
\end{equation}
From a Feynman-diagrammatic point of view, this scalar theory resembles
scalar electrodynamics in containing both cubic and quartic vertices.
We would expect electrodynamics to be richer thanks to the helicity structure
of the messenger. The similarity of the two results is nonetheless striking. 
We return to the origin of this similarity and the underlying simplifications
in later sections.
In the remainder of the article, we suppress factors of $\hbar$ which we kept explicit in this subsection.

\subsection{Amplitudes, the Classical Limit and a 
Perturbative Eikonal-like Phase}
\label{subsec:classical_limit}
In this subsection, we discuss how the results obtained in the previous subsection emerge from the classical limit of perturbative amplitude computations. We consider the classical limit of the massive two-to-two amplitude $\mathcal{M}_{2\to2}$
\begin{equation}\label{eq:m_quantum}
    \mathcal{M}_{2\to2} (\gamma, -q^2) =
    \vcenter{\hbox{\includegraphics{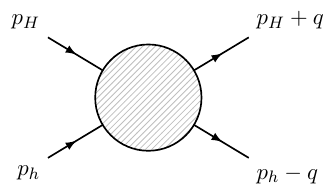}}} \ ,
\end{equation} 
in a self-force type of approximation.
In the classical limit the four-point amplitude admits an organization
in powers of the two masses
(known as \textit{mass polynomiality\/}~\cite{Bini:2020wpo, Bini:2019nra}).  At order $(g_\heavy g_\probe)^N$, the classical
contribution has total mass degree $3-N$, with possible term coefficients
$m_\heavy^{1-n} m_\probe^{n+2-N}$, for all $n=0,\dots,N-1$.  (This statement
excludes the disconnected, classically singular iterations of lower-order 
amplitudes.)  We recover the scattering off a background as discussed above
by retaining only the leading term in the heavy scalar mass $m_\heavy$. The
leading probe contribution then scales as $m_\heavy m_\probe^{2-N}$. We will
show that the logarithm of this amplitude in \textit{coordinate space}
--- an eikonal-like phase ---
reconstructs, order by order, the classical phase shift and the scattering
angle discussed above.

\subsubsection*{Multi-Emission Scalar Amplitudes}
We can reconstruct the long-range contribution from the $(N-1)$-loop
massive amplitude 
$\mathcal{M}_{2\to2}$ from its discontinuity in the total exchanged
momentum. Schematically, the cut construction is,
\begin{equation}
    \label{eq:discontinuity}
    \underset{q^2}{\Disc}\,\mathcal{M}_{2\to2} = \sum_{N=1}^{\infty} \frac{1}{N!}
    \vcenter{\hbox{\includegraphics{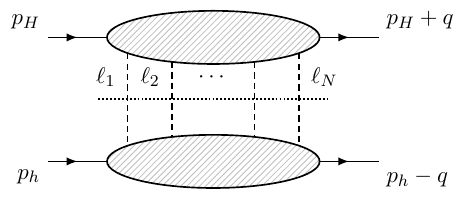}}}\ .
\end{equation}
In the probe limit,\footnote{The order of limits matters: we must first take
the classical limit, and only then the probe limit where the mass ratio 
becomes large.} the heavy-source amplitude factorizes into the product of background source insertions. The leading probe contribution is then
given by the tree-level $N$-emission amplitude,
\begin{equation}
    \label{eq:probe_limit}
    \lim_{\ {m_\probe}/{m_\heavy} \to 0} \underset{q^2}{\Disc}\,\mathcal{M}_{2\to2} \,= \sum_{N=1}^{\infty} \frac{1}{N!}
    \vcenter{\hbox{\includegraphics{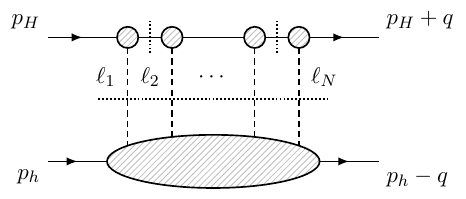}}}\,,
\end{equation}
where the lower amplitude $\cA_N$ is,
\begin{equation}
    \label{eq:a_quantum}
    \cA_{N} (p_h;k_1,\dots ,k_N) =
    \vcenter{\hbox{\includegraphics[scale=1]{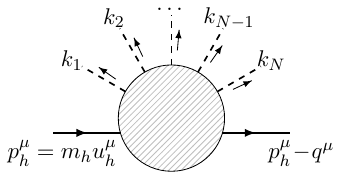}}} \,.
\end{equation} 
Here, all momenta are on shell,
\begin{equation}
        k_i^2 = 0\ , \quad p^2_h = (p_h-q)^2 = m^2_h\ , \quad \sum_{i = 1}^{N} k_i^\mu = q^\mu\ .
\end{equation} 
Contributions arising from off-shell momenta, not
captured by the on-shell unitarity construction above,
are proportional to the messenger virtualities $k_i^2$. 
These factors cancel individual messenger propagators. 
Accordingly, they cannot generate long-range interactions. 
In the classical limit, we can ignore them.

In the classical limit, we have thereby mapped the problem of computing probe contributions from higher-loop amplitudes
(or more precisely, their integrands) to the computation of tree-level amplitudes at arbitrary multiplicity.
We can go beyond the probe limit by including corrections to the factorization of the heavy-mass amplitude.
A class of those corrections is given by a similar tree amplitude for the heavy line.
This class is not sufficient for all classical effects because of radiation reaction --- to include those, we would need to include loop corrections to the tree amplitudes we are discussing.
In this paper, we focus on the radiation-less probe limit, omitting radiation-reaction effects.

How can we take the classical limit of $\cA_N$ efficiently?
To study this question, let us first focus on the simplest case
where $\lambda_i=-g_i^2/2$, so that we have only trivalent
vertices.  All Feynman diagrams then have $N$ massless scalar
messengers emitted from the massive line.   As we have suppressed the
$\hbar$s here, we can instead rephrase the classical limit as
a multisoft limit, taking the momentum of each emitted massless
scalar to be much smaller than the mass of the primary scalar
(the so-called \textit{heavy-mass limit} 
\cite{Damgaard:2019lfh,Aoude:2020onz,Brandhuber:2021kpo,%
Brandhuber:2021eyq}).
For simplicity, in what follows we use the generic
momentum $p^\mu = m u^\mu$ and coupling label $g$ for the scalar multi-emission amplitude, 
and restore labels distinguishing the heavy source from the 
light probe only in the final result.
No confusion should arise as we focus only
on the nontrivial subamplitude  $\mathcal{A}_N$  (\cref{eq:a_quantum}) within this section.
\def\Ksum{k}
We then have:  
\begin{equation}
    \begin{split}
        \cA_N^{\text{scalar-cubic}}&(p; k_1, {\cdots}, k_N) =
        \sum_{\sigma \in S_N} \vcenter{\hbox{\includegraphics{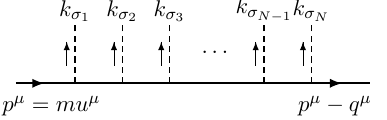}}} \\
        &= - \sum_{\sigma \in S_N}\frac{(g m)^N}{\prod\limits_{j=1}^{N-1} \left[(p-\Ksum_{\sigma_1\dots \sigma_j})^2-m^2 + i\varepsilon\right]} \\
        & = 2m \left(-\frac{g}{2}\right)^N\!\! \left(\frac{m}{\bm}\right)^{N-1} \!\!\!\sum_{\sigma \in S_N} \prod_{i=1}^{N-1} \frac{1}{\left[\bu {\cdot} \Ksum_{\sigma_1\dots \sigma_i} +\frac{\Ksum_{\sigma_1\dots \sigma_i}{\cdot}
        \Ksum_{\sigma_{i+1}\dots \sigma_N}}{2 \bm}-i\varepsilon\right]} \,.
    \end{split}\label{eq:three_point_scalar_quantum}
\end{equation}
We denote the sum of messenger momenta by multiple indices,
\begin{equation}
    \Ksum_{\sigma_1\dots\sigma_i}=\sum_{a=1}^i k_{\sigma_a}\,.
\end{equation}
Furthermore, we introduce shifted variables, denoted with a bar,
\begin{equation}\label{eq:barred_variables}
    p^\mu = \bm \bu^\mu + \frac{q^\mu}{2}\,, \qquad
    \bm^2=m^2-\frac{q^2}{4}\,, \qquad \bu^2=1\,, \qquad \bu {\cdot} q = 0\ .
\end{equation}
Next, the \textit{eikonal identity\/} allows us to obtain a closed form for the leading term in the heavy-mass expansion:
\begin{equation}
    \begin{split}
        \label{eq:source_scalar}
        \cA_{N}^{\text{scalar-cubic}} & =
        2m \left(-\frac{g}{2}\right)^N
        \sum_{\sigma\in S_N}\prod_{j=1}^{N-1}
        \frac{1}{\left[\bu {\cdot} \Ksum_{\sigma_1\dots \sigma_j}-i\varepsilon\right]} + \cO(m^0)\\
        & = -2mi \left(-\frac{i \,g}{2}\right)^N \,
        \prod_{j=1}^{N-1}\hat \delta\!\left(\bu {\cdot} k_{j}\right) +
        \cO(m^0)\,.
    \end{split}
\end{equation}
This form of the eikonal identity can be proved in the standard way, 
using the Schwinger parametrization of the propagators~\cite{Saotome:2012vy} (it holds on the support of the energy-conserving delta function $\hdelta(\bu{\cdot} q)$).
This leading term corresponds to the form of the 
\textit{upper\/} amplitude in
\cref{eq:probe_limit}, where all massive propagators are cut 
to effectively replace the tree-level amplitude by
a product of sources.

Following the strategy of Ref.~\cite{Saotome:2012vy}, we can also obtain the
subleading terms in the classical expansion by postponing the heavy-mass
expansion. We rewrite \cref{eq:three_point_scalar_quantum} in a Schwinger-parameter
representation and then change variables:
\begin{equation}
    \label{eq:schwinger_scalar}
\begin{aligned}
    \hdelta&\biggl(\sum_{i=1}^N \bu {\cdot} k_i \biggr)
    \cA_N^{\text{scalar-cubic}} =
    \\ &\hspace*{4mm}
    2m \left(-\frac{g}{2}\right)^N \left(i \frac{m}{\bm}\right)^{N-1}
    \!\!\!\int_{\mathbb{R}^N} \dd^N t\, \exp\biggl[- i \sum_{j=1}^N t_j \bu
    {\cdot} k_j - \frac{i}{2 \bm} \sum_{\substack{i,j = 1 \\ j<i}}^N |t_i-t_j| k_i {\cdot}
    k_j\biggr] \,.
\end{aligned}
\end{equation}
We include an overall $\hdelta(\bu{\cdot} q)$ on the left-hand side
of \cref{eq:schwinger_scalar} to make manifest the symmetry of the amplitude
under permutations. Alternatively, we can view it as arising from the 
right-hand side: choose any $j$, shift the integration variables 
$t_i \to t_i + t_j$ for all $i\neq j$, and integrate over $t_j$.
This representation was first derived in Ref.~\cite{Ajith:2024fna} and 
is useful for understanding the structure of subleading terms in the 
classical expansion (see also Ref.~\cite{Capatti:2024bid}).  
Because the leading terms in the expansion of
$\cA_N$ are singular, going as $\hbar^{-N}$, subleading terms will
contribute after the cancellation of the singular terms in a physical
observable.

We can now obtain the full expansion in the classical limit
by expanding the exponential in powers of $1/\bm$. The $t_i$ integrals
then give rise to various \textit{local\/} or \textit{rational\/} expressions,
\begin{equation}
\begin{aligned}
    \int_{\mathbb{R}} \dd t |t|^n e^{- i t \omega - \varepsilon |t|} 
    &= -n!\,i^{n+1} \left[
    -\frac{1}{\left(\omega + i \varepsilon\right)^{n+1}}+
    \frac{(-1)^n}{\left(\omega - i \varepsilon\right)^{n+1}}\right] \,.
\end{aligned}
\end{equation}
For $n$ even, this evaluates to
\begin{equation}
\begin{aligned}
        (-1)^{\frac{n}{2}} \hdelta^{(n)}(\omega) \,,
\end{aligned}
\end{equation}
while for $n$ odd we will write this as,
\begin{equation}
\begin{aligned}
\int_{\mathbb{R}} \dd t |t|^n e^{- i t \omega - \varepsilon |t|}  &= 
        (-1)^{\frac{n+1}{2}}\frac{2 n!}{\omega^{n+1}} \,,
\end{aligned}
\end{equation}
where the suppression of $i\varepsilon$
means the use of the principal-value prescription 
for odd $n$,
\begin{equation}
    \frac{1}{\omega^{n+1}} := \frac{1}{2} \left[\frac{1}{(\omega-i\varepsilon)^{n+1}}+\frac{1}{(\omega+i\varepsilon)^{n+1}}\right]\,.
\end{equation}
As we explain below, we will separate terms that are rational in the external invariants from terms that are localized by delta functions or derivatives thereof.
In the classical limit, the massive propagators become eikonal propagators.
At initial stages, they appear to be raised to a variety of powers.
As we now show, integration over the $t_i$ prunes the
set of powers.  Let us expand 
\cref{eq:schwinger_scalar} in powers of ${\bm}^{-1}$ and truncate at 
order $N-1$. At this order, there are $N-1$ nontrivial integrations over
the Schwinger parameters. Hence, for a term to have support everywhere, the 
factors $|t_i - t_j|$ must contain all the $N-1$ independent $t_i$:
if any of the $t_i$ does not appear in these factors, its integral 
yields a delta function in its coefficient.
Thus, none of the $|t_i - t_j|$ can appear raised to a higher power, 
and accordingly the integration over the Schwinger parameters leaves only \textit{squared\/} linearized propagators.
This reflects the fact that the classical equations of motion are second order in time derivatives: a key aspect of our physical understanding of the classical world.
This simplification is already manifest in the classical worldline formulation, as we review in \cref{app:worldline}. 
It is a general feature of causal physical theories, owing to the classical factorization theorem for these multi-emission amplitudes that we present in \cref{sec:factorization}.\footnote{After expanding in the classical limit, keeping track of the $i \varepsilon$, and summing over all permutations, we find that all terms $\propto m^{1-k}$, with $k=0,\dots , N-2$,
have support only on a subspace of external invariants, and so correspond
only to classically singular terms.}

In the classical limit, the tree-level amplitudes of the pure-cubic 
theory at the lowest multiplicities are,
\begin{equation}
    \label{eq:amplitude_scalar}
    \begin{split}
    A_1^{\text{scalar-cubic}} (\bm \bu; k_1)&=-gm\ ,\\
    A_2^{\text{scalar-cubic}} (\bm \bu; k_1, k_2)& =\frac{g^2}{2}\left[
    im\hdelta(\bu{\cdot}k_1)-\frac{k_1{\cdot}k_2}{(\bu{\cdot}k_1)^2}\right]\ ,\\
    A_3^{\text{scalar-cubic}} (\bm \bu; k_1, k_2, k_3)
    &=
    \\ &\hspace*{-15mm}\frac{g^3}{4}\left(m+\frac{q^2}{4m}\right)
    \hdelta(\bu{\cdot}k_1)\hdelta(\bu{\cdot}k_2)
    +\frac{i g^3}{8}\sum_{\sigma\in S_3}
    \frac{k_{\sigma_1}{\cdot}k_{\sigma_2}}{(\bu{\cdot}k_{\sigma_1})^2}\hdelta(\bu{\cdot}k_{\sigma_3})
    \\
    &\hspace*{-15mm}+\frac{g^3}{64m}\sum_{\sigma\in S_3}\left[
    (k_{\sigma_1}{\cdot}k_{\sigma_2})^2\hdelta''(\bu{\cdot}k_{\sigma_1})\hdelta(\bu{\cdot}k_{\sigma_3})
    -\frac{8(k_{\sigma_1}{\cdot}k_{\sigma_2})(k_{\sigma_2}{\cdot}k_{\sigma_3})}{(\bu{\cdot}k_{\sigma_1})^2(\bu{\cdot}k_{\sigma_3})^2}\right]\,.
    \end{split}
\end{equation}
Here, we have included local terms in the amplitudes $A_N$; 
in much of the paper, we will exclude them.
The mass scaling of these amplitudes is not homogeneous, as expected from 
our general discussion. Indeed, the classical expansion includes the terms 
of higher order in the mass, which are classically singular and are
needed for the exponentiation of the classical amplitudes,
but do not contribute to classical observables.

We can generalize \cref{eq:schwinger_scalar} to include contact interactions appearing in \cref{eq:qft_action_scalar}.
We will discuss this further in \cref{sec:graphs,sec:eikonal}.

\subsubsection*{The Four-Point Amplitude and an Eikonal-like Phase}

The multi-emission amplitudes in \cref{eq:schwinger_scalar} determine the probe contribution to the massive $2\to2$ amplitude at order 
$(g_\heavy g_\probe)^N$.  For the heavy source we retain only the leading term in \cref{eq:source_scalar}, while for the probe we keep the barred-mass dependence in \cref{eq:schwinger_scalar}. We are ultimately
interested in the normalized amplitude in coordinate space, given by 
\begin{equation}
    \begin{split}
        \widetilde{\mathcal{M}}_{2\to 2} (\gamma, \xinfnorm) 
           &=
        \int \hd^d q\,
        e^{i q {\cdot} \xinf}\,
        \hdelta\left(2 m_\heavy u_\heavy {\cdot} q + q^2\right)\hdelta\left(2 m_\probe u_\probe {\cdot} q - q^2\right)\,
        \mathcal{M}_{2\to 2} (\gamma, -q^2)\\
        &\simeq \frac{1}{4 m_\heavy m_\probe}\!
        \int \hd^d q\,
        e^{i q {\cdot} \xinf}\,
        \hdelta\left(u_\heavy {\cdot} q\right)\hdelta\left(u_\probe {\cdot} q - \frac{q^2}{2 m_\probe}\right)\,
        \mathcal{M}_{2\to 2} (\gamma, -q^2)\,,
    \end{split}
    \label{eq:impact_parameter_form}
\end{equation}
where we have now reinstated the particle labels $h$ and $H$, with  $\gamma=u_\probe{\cdot}u_\heavy$ the Lorentz factor. It is worth emphasizing that, in this equation, $q$ is the momentum transfer, which is not necessarily small in the classical limit. 
The vector \(x_\infty^\mu\) lies in the plane transverse to the incoming velocities \(u_h\cdot x_\infty=u_H\cdot x_\infty=0\), and has a subtle geometric meaning which will be discussed in \cref{sec:eikonal}. We denote its length by $\xinfnorm\equiv\sqrt{-\xinf^2}$. 
The choice of this parameter is natural if we want the amplitude to be a generating function of observables in the far future. 
The orthogonality choice makes some of the computations easier.
The sign in the phase in \cref{eq:impact_parameter_form} appears to be opposite to the Fourier transform convention, but that is an illusion created by our conventions for $\xinf$.
We will show in \cref{sec:eikonal} that an emergent trajectory will adjust this parameter. 
This definition of $\widetilde{\mathcal{M}}_{2\to 2}$ differs from the one employed in the standard definition of the eikonal~\cite{DiVecchia:2023frv}.
In the second line of \cref{eq:impact_parameter_form}, we have expanded the delta function of the heavier state. The amplitude has perturbative momentum-space expansion
\begin{equation}
    \mathcal{M}_{2\to 2} (\gamma, -q^2) = \sum_{N=1}^\infty \mathcal{M}_{2\to 2}^{(N-1)} (\gamma, -q^2)\ ,
\end{equation}
where as usual the superscript denotes the number of loops. This expansion is usually not well justified for the amplitude, because the coupling ($\beta_\phi$ in this case) is large. On the other hand, the effective coupling (\textit{e.g.}, $\beta_\phi/J$) is smaller and observables have a well-defined perturbative expansion.

\def\cMtt{\mathcal{M}_{2\to2}}
\def\tcMtt{\widetilde{\mathcal{M}}_{2\to2}}
We build the $2\to2$ amplitude (or more precisely the contribution
leading in the heavy mass) at  order $N-1$
by gluing the multi-emission amplitude
$\cA_N$ discussed earlier to a set of heavy sources via massless
propagators.  In performing the $q$ integration 
in \cref{eq:impact_parameter_form}, we must remember that the barred
probe quantities depend on the momentum transfer $q$.\footnote{The analogous dependence of $\bu_\heavy$ and $\bm_\heavy$ is subleading in the probe limit, so we can set $\bu_\heavy=u_\heavy$ and $\bm_\heavy=m_\heavy$. }
Schematically, we have
\begin{equation}\label{eq:apm_source_contracted}
    \cMtt^{(N-1)} \,\,=\,
    \vcenter{\hbox{\includegraphics{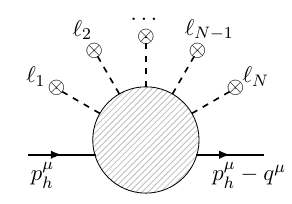}}}\ .
\end{equation}

\def\dips#1{\dd\sigma_{#1}}
Starting with the integrand in
\cref{eq:schwinger_scalar}, we can thus obtain the
integrand for $\cMtt$ by multiplying by the Fourier phase and
the propagators $1/\ell_i^2$.  
We can simplify the presentation by redefining the Schwinger parameters,
\begin{equation}
t_i \to -\frac{\bm_\probe}{m_\probe} t_i\,,    
\end{equation}
in order to remove the factor $\left(m_\probe/\bm_\probe\right)^N$.
Writing out
\begin{equation}
    \bu_\probe^\mu=\frac{m_\probe}{\bm_\probe}u_\probe^\mu 
                - \frac{q^\mu}{2 \bm_\probe}\,,
\end{equation}
rearranges the structure of the exponent in \cref{eq:schwinger_scalar}: 
\begin{equation}
\label{eq:firstMax}
    {|t_i-t_j|} \rightarrow 
    {|t_i-t_j|} + {(t_i+t_j)} = 2\max{(t_i,t_j)}\,.
\end{equation}
To write the result in a compact form,
it is helpful to define the measure,
\begin{equation}
\label{eq:eikonal_measure}
    \dips{i} := \hd^d \ell_i\, \dd t_i\, \hdelta(u_\heavy {\cdot} \ell_i) \frac{e^{i (\xinf + t_i u_\probe) {\cdot} \ell_i
    }}
    {\ell_i^2}\,.
\end{equation} 
We then obtain,
\begin{equation}
    \label{eq:probe_amplitude}
    \begin{split}
        i \tcMtt^{(N-1)}(\gamma,\xinfnorm)
        &=\frac{1}{N!}\left(-i\frac{g_H g_\probe}{4}\right)^N
        \sint\prod_{i=1}^N\dips{i}\;
        \exp\biggl[-\frac{i}{m_\probe}\sum_{\substack{i,j = 1 \\ j<i}}^N
        \max(t_i,t_j)\,\ell_i{\cdot}\ell_j\biggr]\,,
    \end{split}
\end{equation}
where the integration over the momentum transfer $q$ has been performed using the momentum-conservation delta function, $q=\sum_{i=1}^N \ell_i$.
The overall factor of $(-i)^N$ arises from the product of the
factor $i^{2N}$ in the heavy and probe multi-emission amplitudes,
along with the factor of $i^N$ from the messenger propagators.
The overall factor of $1/N!$ is the Bose symmetry factor of the
$N$-messenger exchange. In \cref{eq:probe_amplitude}, we dropped exponent terms proportional to $\ell_i^2$, which would give rise to contact terms.

We can think of the $\max$ function in \cref{eq:probe_amplitude} as a worldline Green's function.
It is not the natural one for classical worldline theories with retarded boundary conditions --- but it is the Green's function picked out by the classical limit of the Feynman $i \varepsilon$ prescription.
We will discuss this in greater detail below.

Let us define an eikonal-like phase $\delta$ as the logarithm of the normalized amplitude in coordinate space:
\begin{equation}
    \label{eq:def_eikonal_phase}
    2 i \delta (\gamma, \xinfnorm) \equiv \log\left(1 + i \widetilde{\mathcal{M}}_{2\to 2} (\gamma, \xinfnorm)\right)\ .
\end{equation}
This definition is analogous to the standard definition of the eikonal phase in the literature.\footnote{Our definition agrees with $\check {\delta}$ in Eq. (6.118) of Ref.~\cite{DiVecchia:2023frv}, up to our use of $\xinf$ instead of the incoming impact parameter $b$. Our $\delta$ thus differs from their eikonal phase. It also differs from the eikonal phase discussed by some of the present authors in Ref.~\cite{Cristofoli:2021jas}. Finally, our definition is distinct from the eikonal phase of Ref.~\cite{Kim:2024svw}, which the authors identify with the classical amplitude with a principal-value prescription.}
Taking the logarithm of the amplitude allows us to avoid subtleties
associated with possible $\hbar/\hbar$ contributions from the 
non-exponentiating (quantum) remainder in the eikonal limit 
(see Ref.~\cite{DiVecchia:2019kta} for a discussion in the context 
of $\mathcal{N}=8$ supergravity).

The phase $\delta$ has a much simpler structure than the amplitude~\eqref{eq:probe_amplitude} in the classical limit, $\ell_i/m_\probe \to 0$. To see this, introduce the shorthand,
\begin{equation}
    w_{i j} = -\frac{i}{m_\probe} \max(t_i,t_j) \ell_i {\cdot} \ell_j\,,
\end{equation}
and define,
\begin{equation}
    e^{w_{i j}} \equiv 1 + f_{i j} \ .
\end{equation}
The pairwise exponential then has a graph expansion
\begin{equation}
    \prod_{j < i} e^{w_{i j}} = 
    \sum_{G \in \mathcal{G}_N}\prod_{e_{ij}\in G}f_{ij}\ ,
\end{equation}
where $\mathcal{G}_N$ is the set of all (possibly disconnected) graphs
with labeled vertices $\{1,\ldots,N\}$, including the empty graph; and
where the edges are $e_{ij}$, each contributing a factor of $f_{ij}$. 
The logarithm of the amplitude then selects the subset $\Gamma_N$
of \textit{connected graphs\/} with $N$ labeled vertices. We
then obtain the expression for our eikonal-like phase,
\begin{equation}
    2i\delta^{(N-1)}(\gamma,\xinfnorm)
    =\frac{1}{N!} \left(-i\frac{g_\heavy g_\probe}{4}\right)^N
    \sint\prod_{i=1}^N\dips{i}
    \sum_{G\in\Gamma_N}\prod_{e_{ij}\in G}f_{ij}\,.
\end{equation}
This result for the logarithm in terms of connected graphs is
known as the 
\textit{Mayer cluster expansion}~\cite{Mayer:1937,Salpeter:1958}. 
Specializing to the classical limit, we have
\begin{equation}
    f_{ij} = w_{i j} + \cO(w_{ij}^2)\ .
\end{equation}
As we are interested in the leading terms in the mass power counting, we 
can restrict the sum over connected graphs to the set $\mathcal{T}_N$ of \textit{tree graphs\/} with $N$ labeled vertices:
\begin{equation}
    \label{eq:eikonal_phase_classical}
    2i\delta_{\mathrm{cl}}^{(N-1)}(\gamma,\xinfnorm)
    =\frac{i(-1)^N}{N!\;m_\probe^{N-1}}
    \left(\frac{g_Hg_\probe}{4}\right)^N
    \sint\prod_{i=1}^N\dips{i}
    \sum_{T\in\mathcal{T}_N}\prod_{e_{ij}\in T}
    \max(t_i,t_j)\ell_i{\cdot}\ell_j\ .
\end{equation}
A similar graph construction has recently appeared~\cite{Brandhuber:2025igz,Guo:2026hyi,Brandhuber:2026dkd} in the context of the Magnus expansion.

In the following sections, we will show how this graph structure emerges naturally from the classical limit of multi-emission amplitudes in other theories, as a consequence of the unitarity of the quantum amplitudes.

\subsubsection*{Perturbative Computation of the Phase} 

We now present a recursive strategy to compute the phase $\delta$ at high
loop orders, and compare its output with the result 
in \cref{eq:class_eikonal_partial_waves} for the purely cubic scalar theory.

The integrals over the $\ell_i$ factorize and can be performed independently. Indeed, all integrals appearing in the computation 
take the form
\begin{equation}
    \label{eq:tensor_integrals}
    V^{\mu_1 \dots \mu_n} (x_i) \equiv \int \hd^d \ell \frac{\hdelta(u_\heavy{\cdot} \ell)}{\ell^2}\, e^{i x_i {\cdot} \ell}\, \ell^{\mu_1} \cdots \ell^{\mu_n} = (-i)^n \nabla_{i}^{\mu_1 \dots \mu_n} \int \hd^d \ell \frac{\hdelta(u_\heavy{\cdot} \ell)}{\ell^2}\, e^{i x_i {\cdot} \ell}\ ,
\end{equation}
where $x_i^\mu = \xinf^\mu+t_i u_\probe^\mu$, and
\begin{equation}
\label{eq:derivative_definition}
\nabla_{i}^{\mu_1 \dots \mu_n} \equiv \frac{\partial^n}{\partial x_{i\, \mu_1} \cdots \partial x_{i\,\mu_n}}\ .
\end{equation}
We can evaluate the scalar integral easily:
\begin{equation}
    \begin{split}
        \label{eq:potential_scalar}
        V (x_i) = \int \hd^d \ell \frac{\hdelta(u_\heavy{\cdot} \ell)}{\ell^2}\, e^{i x_i {\cdot} \ell} =  \frac{\nu_{d}}{\left[(u_\heavy{\cdot} x_i)^2 - x_i^2\right]^{\alpha_d}}
        \,,
    \end{split}
\end{equation}
where, for notational convenience, we have defined
\begin{equation}
    \alpha_d = \frac{d-3}{2}\ , \qquad 
    \nu_{d} = - \frac{\Gamma\left[\alpha_d\right]}{4 \pi^{\alpha_d+1}}\ .
\end{equation}
We also define the variables,
\begin{equation}
    \eta_{\perp}^{\mu\nu} = \eta^{\mu\nu} - u_\heavy^\mu u_\heavy^\nu\ , \qquad y_i^\mu = x_{i \perp}^\mu = \eta_{\perp}^{\mu\nu} x_{i \nu}\ .
\end{equation}
We then have
\begin{equation}
    \begin{split}
        y_{i}^2 = x_i^2 - (u_\heavy{\cdot} x_i)^2\ ,\qquad \nabla_i^{\mu} y_{i}^2 = 2 y_{i}^\mu\ , \qquad \nabla_i^{\mu} y_{i}^\nu = \eta_{\perp}^{\mu\nu}\,.
    \end{split}
\end{equation}

When we perform the $\ell_i$ integrals in \cref{eq:eikonal_phase_classical}
we obtain,
\begin{equation}
\begin{aligned}
    2i\delta_{\mathrm{cl}}^{(N-1)} (\gamma, \xinfnorm)
    =\frac{-i}{N!\,m_\probe^{N-1}}&\left(\frac{g_Hg_\probe}{4}\right)^N
    \!\\&\hspace*{-5mm}\times \int_{\mathbb R^N}\!\dd^Nt\,
    \sum_{T\in\mathcal T_N}\prod_{e_{ij}\in T}
    \max(t_i,t_j)\nabla_i{\cdot}\nabla_j
    \prod_{i=1}^N V(x_i)\,.
\end{aligned}
\end{equation}
 In particular, we have $y_i^\mu = \xinfnorm\, \xinfhat^\mu + t_i \sqrt{\gamma^2-1} \hat{u}_{\probe}^\mu$, where $\xinfhat^\mu$ and $\hat{u}_\probe^\mu$ are the space-like unit vectors orthogonal to $u_\heavy^\mu$:
\begin{equation}
    \xinfhat^\mu = \frac{\xinf^\mu}{\xinfnorm}\ , \qquad \hat{u}_\probe^\mu = \frac{u_\probe^\mu - \gamma u_\heavy^\mu}{\sqrt{\gamma^2-1}}\ , \qquad \hat{u}_\probe^2 = - 1\ .
\end{equation}
We can remove all dimensionful quantities from the integrals by 
rescaling the Schwinger parameters,
\begin{equation}
    t_i = \frac{\xinfnorm}{\sqrt{\gamma^2-1}} \tau_i \ ,
\end{equation}
so that $y_i^\mu = \xinfnorm (\xinfhat^\mu + \tau_i \hat{u}_\probe^\mu)$. The phase then takes the form
\begin{equation}
    \label{eq:dimensionful_eikonal}
    2i\delta_{\mathrm{cl}}^{(N-1)} (\gamma, \xinfnorm)
    = i m_\probe \xinfnorm\sqrt{\gamma^2-1} (-z)^N \mathcal{I}_N\ ,
\end{equation}
where
\begin{equation}
    z = \frac{g_\heavy g_\probe \nu_{d}}{4 m_\probe \xinfnorm^{2\alpha_d} (\gamma^2-1)}\ ,
\end{equation}
and where $\mathcal{I}_N$ is the integral over the dimensionless Schwinger parameters.

\def\xred{\hat x}
\def\xredb{{\hat{\boldsymbol{x}}}}
\def\rredb{{\boldsymbol{r}}}
\def\nablared{\hat \nabla}
\def\nablaredb{{\hat{\boldsymbol{\nabla}}}}
The vectors $y_i^\mu$ lie in the two-dimensional subspace defined
by $\xinfhat^\mu$ and ${\hat u}_\probe$.  It will be convenient to
define a related two-component vector $\xredb$, given by the coefficients
of these two basis vectors.  Its square should be understood to be
taken using the two-dimensional Euclidean metric. The corresponding
derivative $\nablaredb$ follows the same definition as 
in \cref{eq:derivative_definition} but
taking derivatives with respect to $\xredb$; it again carries two-dimensional
indices.  Where needed, we will denote the two-dimensional indices by
Latin-letter superscripts.

We can then write the following expression for $\mathcal{I}_N$,  normalized by $1/N!$:\footnote{Each of the $N-1$ contractions contributes
a minus sign upon switching to the Euclidean metric. 
Together with the original $-i$, this gives an overall $i(-1)^N$.}
\begin{equation}
    \label{eq:fixed_order_eikonal}
    \begin{split}
        \mathcal{I}_N &= \frac{1}{N!} \int_{\mathbb{R}^N} \dd^N \tau \sum_{T\in\mathcal T_N}\biggl(\prod_{e\in T} \max(\tau_i,\tau_j)\nablaredb_i{\cdot}\nablaredb_j \prod_{k=1}^N \left.\frac{1}{(\xredb_k^2)^{\alpha_d}}\biggr)
         \right|_{\xredb_{k} = (1,\tau_k)}\ .
    \end{split}
\end{equation}
The graph-by-graph evaluation of this integral grows very rapidly 
in complexity with the number of vertices. Moreover, individual
graphs have IR divergences which cancel between graphs for $N>3$
(in addition to an overall Coulomb IR divergence).  These individual
divergences manifest in the usual way as poles in the dimensional
regulator $d-4$.  

This combinatorial structure has an interpretation in terms of 
a classical on-shell action~\cite{Nambu:1968rr,Boulware:1968zz}. Indeed, we can define the two-dimensional trajectory
recursively, 
\begin{equation}
    \rredb(\tau) = \rredb_0(\tau) - z \int_{\mathbb R} \dd \tau' \max (\tau,\tau') 
    \left. \nablaredb \frac{1}{(\xredb^2)^{\alpha_d}} 
    \right|_{\xredb = \rredb(\tau')} \,,
    \label{eq:recursive_r}
\end{equation}
where $\rredb_0(\tau) = (1,\tau)$. 
We can think of this as a classical trajectory, in spite of the non-standard $\max$ Green's function.
We can now write a perturbative expansion of this trajectory:
\begin{equation}
    \rredb(\tau) = \rredb_0(\tau) + \sum_{n=1}^\infty z^n \rredb_n(\tau)\,,
\end{equation}
where the $\rredb_n(\tau)$ are determined by the recursive 
relation~\eqref{eq:recursive_r}:
\def\Extract#1{\mathop{\mathcal{C}}\nolimits_{#1}}
\begin{equation}
    \label{eq:recursive_trajectory}
    \begin{split}
        \rredb_{n+1}(\tau) = - \int_{\mathbb R} \dd \tau' \max (\tau,\tau')\, \Extract{n} \biggl( 
        \exp\biggl[\sum_{k=1}^n
           z^k \rredb_k(\tau') \cdot\nablaredb\biggr] 
           \nablaredb \frac{1}{(\xredb^2)^{\alpha_d}} \biggr) 
           \Bigg|_{\xredb = \rredb_0(\tau')}\ ,
    \end{split}
\end{equation}
where $\Extract{n}$ denotes the operation of extracting the 
coefficient of $z^n$ in the expansion. The right-hand side of Eq.~\eqref{eq:recursive_trajectory} involves only the $\rredb_k(\tau)$ with $k\leq n$, so we can solve for the trajectory order by order in $z$.
Given the solution of the recursion, we can compute the set of
integrals defined by \cref{eq:fixed_order_eikonal} together by resumming,
\begin{equation}
    \label{eq:onshell_action_scalar}
    \mathcal{I}(z) = \sum_{N=1}^\infty (-z)^N\, \mathcal{I}_N = - \frac{1}{2} \!\int_{\mathbb{R}} \dd \tau \left[\rredb(\tau)-\rredb_0(\tau)\right]\cdot \ddot{\rredb}(\tau) - z \!\int_{\mathbb{R}} \dd \tau
    \, \frac{1}{\rredb(\tau)^{2\alpha_d}}\ .
\end{equation}
This provides an efficient recursive strategy for computing the 
set of integrals~\eqref{eq:fixed_order_eikonal} for $d>4$.  For $d=4$, 
in contrast, the integrals are IR divergent, and this strategy requires 
a careful treatment of $(d-4)/(d-4)$ terms.

We have thereby recovered the classical dynamics
from the combinatorial structure of the  amplitude, and identified the 
eikonal-like phase defined here with a specific classical on-shell action. 
One expects that any definition of an eikonal phase can be identified with a corresponding classical on-shell action.
The boundary conditions 
are dictated by the integration kernel $\max(\tau,\tau')$. 
Because we have arrived at a classical central-potential problem, we 
can skip the explicit computation of the integrals \eqref{eq:fixed_order_eikonal}; the angle can be computed directly 
using Hamilton--Jacobi theory. To this end, we define the subtracted 
radial action
\begin{equation}
    \Phi (e, j) = \lim_{R \to \infty} \sqrt{2e}\, \biggl(\int_{r_{\text{min}}}^{R} \dd r \sqrt{1 - \frac{j^2}{2 e r^2} - \frac{z}{e} \frac{1}{r^{2\alpha_d}}} - \int_{{j}/{\sqrt{2e}}}^{R} \dd r \sqrt{1 - \frac{j^2}{2 e r^2}} \biggr)\ ,
\end{equation}
where $e$ and $j$ are the normalized energy and angular momentum, respectively. The $\max$ kernel fixes the boundary conditions such that $e = {1}/{2}$ and $j = \cos\chi$, while $r_{\text{min}}$ is the turning point of the radial motion,
given by the solution of,
\begin{equation}
    1 - \frac{j^2}{r_{\text{min}}^2} - \frac{2z}{r_{\text{min}}^{2\alpha_d}} = 0\,.
\end{equation}
In $d=4$, we have $\alpha_4 = \frac{1}{2}$, and so the scattering angle is 
given by
\begin{equation}
    \chi = -2 \frac{\partial \Phi(\frac{1}{2},j)}{\partial j} = 2 \arctan\frac{-z}{j} = 2\arctan\frac{g_\heavy g_\probe}{16\pi J \sqrt{\gamma^2-1}}\ ,
\end{equation}
where $J = j m_\probe \xinfnorm \sqrt{\gamma^2-1}$ is the angular 
momentum. This result matches the one obtained from the partial-wave 
expansion in \cref{tab:scalar_worldline_angles}. As we will show in \cref{sec:eikonal}, the seagull interactions do not change our findings: they simply introduce an extra term in the potential, proportional to $V(x)^2$.

As we will see below, the same combinatorial structure in terms of tree graphs, together with the corresponding on-shell action, also emerges in the classical limit of electrodynamics and $\mathcal{N}=8$ supergravity. We will prove this statement beyond the cubic scalar theory, showing that the same structure emerges in the classical limit of multi-emission amplitudes in all theories without self-interactions. 
We expect similar structures to emerge in theories with self-interactions, such as gravity.

\section{Tree-Level Factorization in the Classical Limit}
\label{sec:factorization}
In the previous section, we discussed how amplitudes for the emission of 
multiple massless scalars from a massive source provide the building blocks 
of the two-to-two amplitude via unitarity methods. 
In more general theories, the analogs of the $N$-scalar-emission tree
amplitudes in \cref{eq:a_quantum} have $N$ external legs of
outgoing radiation of arbitrary helicity, with momenta $k_i$.
In the classical limit, these emission amplitudes obey universal properties inherited from the factorization of tree-level amplitudes in the corresponding quantum theory. In this section, we prove a general factorization statement for their classical limit, 
here understood as a uniformly soft limit for the radiated messengers.
Here we will consider scalar massive legs.
The factorization theorem holds in gravity, in addition to the theories we discuss explicitly.
The discussion also generalizes to spinning matter in the quantum theory,
and we expect the classical factorization to carry over to the spinning case.

\subsection{A Classical Factorization Theorem}
We are interested in the classical limit of multi-emission amplitudes, 
that is, amplitudes in which one massive particle emits multiple 
massless messengers. The classical limit $\hbar\to 0$ is equivalent
to the multiple soft limit of the radiation modes; it can also be viewed
as the \textit{heavy-mass limit\/}~\cite{Damgaard:2019lfh,Brandhuber:2021eyq} for both
massive lines:\footnote{For spinning matter, the classical limit further requires taking the spin to be large while keeping products of the form $S{\cdot} k$ fixed \cite{Maybee:2019jus,Arkani-Hamed:2019ymq, Aoude:2021oqj, Guevara:2018wpp,Chung:2018kqs, Chung:2019duq,Chung:2020rrz,Bern:2020buy,FebresCordero:2022jts, Bautista:2021wfy, Bautista:2022wjf, Guevara:2019fsj,Cangemi:2022bew,Jakobsen:2021zvh,Bautista:2019tdr,Bautista:2019evw}.}
\begin{equation}
\label{eq:classical_limit}
    \frac{k_i^\mu}{m} \sim 0 \,.
\end{equation}
In what follows we shall again suppress the particle labels $h$ and $H$, as we will only consider multi-emission amplitudes in this part.

\def\Ib{\bar{I}}
\def\vIb{\vphantom{\Ib}}
In any local QFT, tree-level amplitudes have simple poles at the kinematic points where an internal leg goes on shell. The residue at each pole is given by a product of lower-point amplitudes.  This property is known as \textit{factorization}. As we show below, in the classical limit of the amplitudes $\cA_N(p;k_1,\dots,k_N)$ pairs of factorization poles collide.
In consequence, the classical amplitudes $A_N$ have \textit{double poles\/}
at kinematic points where the relevant invariant vanishes.
This invariant is the total momentum of a subset $I$ of the 
massless legs contracted with the velocity $\bu^\mu$ of the massive source:\footnote{In classical worldline constructions, double poles arise from the (retarded) worldline 
propagator $\sim (\omega_I+i\varepsilon)^{-2}$, where 
$\omega_I\sim u{\cdot} k_I$ is the energy propagating along the worldline~\cite{Mogull:2020sak}.}
\begin{equation}
    \label{eq:pole_position}
    \bu{\cdot} k_I = 0\,, \quad k_I = \sum_{i \in I} k_i\ .
\end{equation}
These poles have a universal structure, independent of the nature of the massless legs, and their coefficients are determined by lower-point amplitudes.  Next, consider a set $I$ of massless external legs and its complement $\Ib$. In a neighborhood of the kinematic configuration in~\cref{eq:pole_position}, for real or complex kinematics, the classical tree-level amplitude factorizes as,
\begin{equation}
    \label{eq:factorization_theorem}
    \boxed{
        A_{N} = - \frac{k_I {\cdot} k_{\Ib}}{2 (\bm\, \bu {\cdot} k_I)^2} A_{I\vIb} A_{\Ib} - \frac{1}{2 \bm^2\, \bu {\cdot} k_I}
        \left[A_I^{\vphantom{\mu}}  k_I^\mu \frac{\partial A_{\Ib}}{\partial \bu^\mu} - A_{\Ib}^{\vphantom{\mu}} k_{\Ib}^\mu \frac{\partial A_{I}}{\partial \bu^\mu}\right] + \cO ((\bu {\cdot} k_I)^0)
    }\ ,
\end{equation}
where we simplify the subscripts of $A$ from $A_{|I|}$ to $A_{I\vIb}$, and 
we use the barred variables defined in \cref{eq:barred_variables}.\footnote{In general, derivatives with respect to $\bu^\mu$ are not well defined, because of the constraints $\bu^2 = 1$ and $\bu {\cdot} q = 0$, as well as $\bu {\cdot} k_I = 0$ near the pole. Nevertheless, one can show that the resulting ambiguities drop out of the classical factorization theorem; they contribute at most to the finite terms.}

\begin{figure}[t!]
    \centering
    \begin{minipage}{0.48\textwidth}
        \centering
        \includegraphics[width=\textwidth]{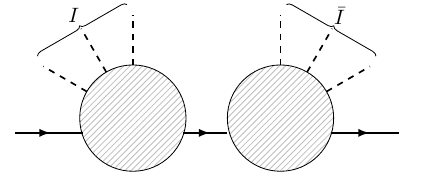}
    \end{minipage}
    \hfill
    \begin{minipage}{0.48\textwidth}
        \centering
        \includegraphics[width=\textwidth]{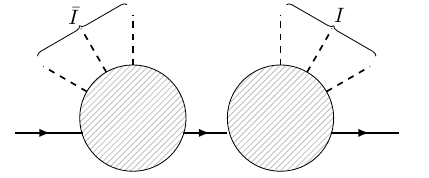}
    \end{minipage}
    \caption{In the classical limit, pairs of factorization channels collide. The two factorization channels differ by the soft momentum exchanged through the intermediate on-shell massive leg.}
    \label{fig:factorization_channel}
\end{figure}

The result in~\cref{eq:factorization_theorem}
follows directly from the simple-pole factorization of 
tree-level amplitudes in the corresponding quantum theory:
\begin{equation}
    \label{eq:tree_factorization}
    \cA_{N} = - \frac{\cA_{I\vIb} \widetilde{\cA}_{\Ib}}{(p - k_I)^2 - m^2} + \cO (((p - k_I)^2 - m^2)^0)\ ,
\end{equation}
where the tilde distinguishes the two lower-point amplitudes: their incoming massive momenta are $p^\mu$ and $p^\mu-k_I^\mu$, respectively. 
At leading order in the classical limit, the two amplitudes do not depend on the soft shift of the massive momentum, which only enters at the next order:
\begin{equation}
    \label{eq:classical_expansion}
    \begin{split}
        \cA_I (p; \{ k_j\}_{j\in I}) &= A_I (\bm \bu; \{ k_j\}_{j\in I}) + \frac{q^\mu}{2\bm}\frac{\partial}{\partial \bu^\mu} A_I (\bm \bu; \{ k_j\}_{j\in I}) + \cO\left(\frac{q^2}{\bm^2}\right)\,,\\
        \widetilde{\cA}_{\Ib} (p; \{ k_j\}_{j\in \Ib}) &= \cA_{\Ib} (p - k_I; \{ k_j\}_{j\in \Ib})\\
        &= A_{\Ib} (\bm \bu; \{ k_j\}_{j\in \Ib}) + \frac{q^\mu-2k_I^\mu}{2\bm}\frac{\partial}{\partial \bu^\mu} A_{\Ib} (\bm \bu; \{ k_j\}_{j\in \Ib}) + \cO\left(\frac{q^2}{\bm^2}\right)\,.
    \end{split}
\end{equation}
In these equations and in following ones, $\bm$ and $q$ are defined by the full $N$-point amplitude, and not by the $|I|$-point subamplitude.
Now consider the two factorization channels that split the radiation modes into $I$ and $\Ib$, emitted in either order from the massive line. Their poles are at
\begin{equation}
    \label{eq:poles}
    0 = (p - k_{I/\Ib})^2 - m^2 = \mp 2 \bm \bu {\cdot} k_{I} - k_{I} {\cdot} k_{\Ib} \simeq \mp 2 \bm \bu {\cdot} k_{I}\,.
\end{equation}
 The two factorization channels  are shown in \cref{fig:factorization_channel}.

In a neighborhood of the two poles, we can write the quantum amplitude as
\begin{equation}
    \begin{split}
        \cA_{N} &\simeq - \frac{\cA_{I} \widetilde {\cA}_{\Ib}}{(p - k_I)^2 - m^2}\ , \\
        \cA_{N} &\simeq - \frac{\cA_{\Ib} \widetilde {\cA}_{I}}{(p - k_{\Ib})^2 - m^2}\ .
    \end{split}
\end{equation}
In the classical limit, there is a neighborhood of $\bu {\cdot} k_{I} = 0$ in which no lone pole dominates the amplitude, but both contribute
comparably, because the term $k_{I} {\cdot} k_{\Ib}$ in \cref{eq:poles} is subleading with respect to $\bm \bu {\cdot} k_{I}$. We thus have,
\begin{equation}
    \label{eq:pole_collision}
    \cA_{N} \simeq - \frac{\cA_{I} \widetilde{\cA}_{\Ib}}{(p - k_I)^2 - m^2} - \frac{\cA_{\Ib} \widetilde{\cA}_{I}}{(p - k_{\Ib})^2 - m^2} \ .
\end{equation}
Expanding the propagators as in~\cref{eq:poles} and the 
lower-point amplitudes as in~\cref{eq:classical_expansion}, we recover 
the factorization theorem in \cref{eq:factorization_theorem}.

The argument above holds for any tree-level amplitude, regardless of the helicities of the massless legs. We also stress that at no point have we used the on-shell conditions of the massless legs. The factorization theorem can therefore be extended to capture the local terms 
(delta functions or derivatives thereof) arising from
the $i\varepsilon$ prescription in the eikonal propagators.\footnote{Indeed, it is not possible to satisfy both $\bu{\cdot} k_i = 0$ and $k_i^2 = 0$ simultaneously within real kinematics; since delta functions are defined only for real arguments, we give up on the on-shell constraint.} For an in--out scattering amplitude, the $i\varepsilon$ prescription is fixed by the Feynman propagators of the quantum theory. Retaining the $i\varepsilon$ in \cref{eq:pole_collision},
\begin{equation}
    \cA_{N} \simeq - \frac{\cA_{I} \widetilde{\cA}_{\Ib}}{(p - k_I)^2 - m^2 +i\varepsilon} - \frac{\cA_{\Ib} \widetilde{\cA}_{I}}{(p - k_{\Ib})^2 - m^2 +i\varepsilon} \ ,
\end{equation} 
and expanding in the classical limit, we obtain
\begin{equation}\label{eq:factorization_theorem_local}
    \begin{split}
        A_{N} & = i \hdelta(\bu {\cdot} k_I) \frac{A_{I\vIb}  A_{\Ib}}{2 \bm} + \frac{i}{4\bm^2} \hdelta(\bu {\cdot} k_I) \left[A_I  k_{\Ib}^\mu \frac{\partial A_{\Ib}}{\partial \bu^\mu} + A_{\Ib} k_I^\mu \frac{\partial A_{I}}{\partial \bu^\mu}\right]\\
        & \hphantom{=} 
        - \frac{k_I{\cdot}  k_{\Ib}}{2 (\bm\, \bu {\cdot}  k_I)^2} A_{I}  A_{\Ib} - \frac{1}{2 \bm^2 \,\bu {\cdot} k_I}\left[A_I  k_I^\mu \frac{\partial A_{\Ib}}{\partial \bu^\mu} - A_{\Ib} k_{\Ib}^\mu \frac{\partial A_{I}}{\partial \bu^\mu}\right] + \cO ((\bu{\cdot}  k_I)^0) \,.
    \end{split}
\end{equation}
Here, too, we have included the local terms in $A_N$.

We see this factorization as an emergent property, in the classical limit, starting with a purely quantum-field theoretical amplitude.
Of course, amplitudes in perturbative quantum field theory involve free propagators which are quadratic in momenta, reflecting the second-derivative nature of the action for quantum fields.
Our factorization theorem demonstrates that the associated classical point-particle effective theories inherit this property, in as much as the eikonal propagators are squared.

In the next subsection, we will use the factorization 
theorem~\eqref{eq:factorization_theorem}
to set up a recursion relation for the classical limit of
tree-level amplitudes at any multiplicity.  This leads to expressions for the integrands of the classical limits of
two-to-two amplitudes at any order in the perturbative expansion. 
More importantly, this recursion allows us to derive explicit all-order
integrands for probe scattering in the scalar theories of the previous section,
in electrodynamics and in $\mathcal{N}=8$ supergravity.

\subsection{A Recursion Relation for Classical Amplitudes}
\label{subsec:recursion}

The classical factorization theorem obtained in the previous subsection
allows us to formulate a recursion relation for tree-level amplitudes with a single classical source and any number of massless messengers. 
Each term that appears will be a mixture of eikonal propagators and delta functions (or derivatives thereof). 
We will describe terms which are built purely from eikonal propagators as ``rational'' terms, while terms involving delta functions will be described as ``local''.
The distinction is important in this classical case, unlike in the usual quantum amplitudes.
In the case of amplitudes, propagators are always defined by the Feynman $+i \varepsilon$ prescription.
In our classical context, this is not so.

In this section, we focus on the rational terms, as the local contributions can be restored straightforwardly in the Schwinger parametrization, as we discuss in \cref{sec:eikonal}.  This is similar to their reconstruction in the
scalar theories as discussed in \cref{subsec:classical_limit}.

We can use Cauchy's theorem in a fashion similar to the 
Britto--Cachazo--Feng--Witten recursion relations~\cite{Britto:2005fq}
to write a classical amplitude with $N$ emissions as a sum over its
massive factorization channels.  To do so, we introduce a 
shift of the massive scalar momenta using a complex shift vector
$\zeta^\mu$ satisfying
\begin{equation}
    \zeta^2=\bu{\cdot} \zeta=\zeta{\cdot} q=0,
    \qquad q=\sum_{i=1}^N k_i\,.
\end{equation}
The shift itself is most easily expressed as a shift of the
barred four-velocity,
\begin{equation}
    \bar{u}^\mu \rightarrow \bar{u}^\mu - w \zeta^\mu\,,
\end{equation}
where $w$ is a complex parameter.  The properties of $\zeta^\mu$ 
ensure that $\bar{u}^\mu$ remains normalized to unity, and that
energy is conserved in the shifted complex kinematics.  We can then
write,
\begin{equation}
    \label{eq:recursion_formula}
    \begin{split}
        A_N (\bm \bu; k_1, \dots, k_N) &= \frac{1}{2\pi i} \oint_{w=0} \frac{\dd w}{w}\, A_N (\bm (\bu - w \zeta); k_1, \dots, k_N) \\
        &= - \sum_{\{I,\Ib\}\in\Pi_2} \, \underset{w=w_I}{\Res}\, \frac{1}{w} A_N (\bm (\bu - w \zeta); k_1, \dots, k_N)\\
        &\hphantom{=} - \underset{w=\infty}{\Res} \, \frac{1}{w} A_N (\bm (\bu - w \zeta); k_1, \dots, k_N)\ ,
    \end{split}
\end{equation}
where $\Pi_2$ denotes the unordered partitions of the massless legs into 
two non-empty sets. As usual, we assume generic kinematics $\{k_i\}$. Each factorization channel is then counted exactly
once, and $w_I = {\bu {\cdot} k_I}/{\zeta {\cdot} k_I}$ is the 
position of the corresponding pole.

In many theories of interest, the classical multi-emission amplitudes have 
vanishing residue at infinity under this shift. They can then be fully 
reconstructed from their factorization channels, once the lower-point amplitudes are known. Examples include the pure-cubic scalar theory of \cref{sec:scalar_model}, the other scalar theories and classical electrodynamics beyond two emissions,\footnote{An alternative way of constructing classical
multi-emission amplitudes in the scalar theory, electrodynamics and gravity
follows from taking the classical limit of the corresponding quantum Berends--Giele
recursion relations~\cite{Berends:1987me} (see also Ref.~\cite{Jakobsen:2023oow}). No closed form for generic-multiplicity classical amplitudes comparable to the results below is known from this recursion.} and certain four-dimensional multi-emission amplitudes in $\mathcal{N}=8$ supergravity. In the rest of this section, we assume that the residue at infinity vanishes.

We denote the shifted amplitude as
\begin{equation}
    A_I(w)=A_{I}\left(\bm (\bu - w \zeta);\{k_j\}_{j\in I}\right)\ ,
\end{equation}
and define the shorthand notation
\begin{equation}
\begin{aligned}
    J_{I \Ib}(w)
    &=
    -\frac{k_I{\cdot} k_{\bar I}}{2\bm^2}
    A_{I\vIb}(w)A_{\bar I}(w),
    \\
    W_{I\bar I}(w)
    &=
    -\frac{1}{2\bm^2}
    \left[
        A_I^{\vphantom{\mu}}(w)\,k_I^\mu
        \frac{\partial A_{\bar I}(w)}
             {\partial\bu^\mu}
        -
        A_{\bar I}^{\vphantom{\mu}}(w)\,k_{\bar I}^\mu
        \frac{\partial A_I(w)}
             {\partial\bu^\mu}
    \right]\ .
\end{aligned}
\label{eq:shorthand}
\end{equation}
Evaluating the residues at the poles $w=w_I$ yields the recursion relation
\begin{equation}
\label{eq:classical_recursion}
\begin{aligned}
    A_N(\bm\bu;k_1,\ldots,k_N)
    = \hspace*{-2mm}
    \sum_{\{I,\bar I\}\in\Pi_2}
    \biggl[
        \frac{J_{I\bar I}(w_I)}
              {(\bu{\cdot} k_I)^2}
        +\frac{1}{\bu{\cdot} k_I}
        \biggl(
            W_{I\bar I}(w_I)
            -
            \frac{1}{\zeta{\cdot} k_I}
            \left.
            \frac{\dd J_{I\bar I}(w)}{\dd w}
            \right|_{w=w_I}
        \biggr)
    \biggr]\,,
\end{aligned}
\end{equation}
where derivatives with respect to four-velocities
in \cref{eq:shorthand} are taken before evaluating the lower-point 
amplitudes at $w=w_I$.

\subsection{Multi-emission Amplitudes in Electrodynamics}
\label{subsec:electrodynamics_amplitudes}

As a first example, we consider
classical amplitudes in electrodynamics. The one-emission seed is
\begin{equation}
    A_1^{\text{Coulomb}} (\bm \bu; k_1) = 2 \, e \,\bm \,\bu{\cdot} \varepsilon_1 \ .
\end{equation}
The recursion relation~\eqref{eq:recursion_formula} will have
a nonvanishing residue at infinity
for the two-emission amplitude
$A_2^{\text{Coulomb}} (\bm \bu; k_1, k_2)$.\footnote{The residue associated with the lone factorization channel depends on the arbitrary shift vector $\zeta^\mu$. Although the four-point amplitude can be reconstructed using other shifts, the particular shift used here is not suitable.} 
The four-point amplitude can nevertheless be computed by other methods, from which we find,
\begin{equation}
    A_2^{\text{Coulomb}} (\bm \bu; k_1, k_2) = 2  e^2\,
\frac{
\bu{\cdot} F_1{\cdot} F_2{\cdot} \bu
}{
(\bu{\cdot} k_1)^2
}\ ,
\end{equation}
where
\begin{equation}
F_i^{\mu\nu} = k_i^\mu \varepsilon_i^\nu - k_i^\nu \varepsilon_i^\mu
\end{equation}
is the linearized field-strength tensor of photon $i$ with polarization $\varepsilon_i^\mu$.
We contract indices in order,
\begin{equation}
    \bu{\cdot} F_1{\cdot} F_2{\cdot} \bu = \bu_\mu F_1^{\mu\nu} F_{2\,\nu\rho} \bu^\rho\ .
\end{equation}
The two-emission amplitude then serves as a second seed for the recursion~\eqref{eq:classical_recursion}; with it, we can
reproduce higher-point amplitudes correctly. In particular, the three-emission amplitude is \cite{Brunello:2024ibk},
\begin{align}
A_3^{\text{Coulomb}} (\bm \bu; k_1, k_2, k_3)&=\frac{  e^3}{\bm}\sum_{\sigma\in S_3}
\frac{
(k_{\sigma_2 \sigma_3}{\cdot} F_{\sigma_1}{\cdot} \bu)\;
(\bu{\cdot} F_{\sigma_2}{\cdot} F_{\sigma_3}{\cdot} \bu)
}
{
(\bu{\cdot} k_{\sigma_1})^2(\bu{\cdot} k_{\sigma_2})(\bu{\cdot} k_{\sigma_3})
}\,.
\label{eq:a3_photons}
\end{align}
This form makes the factorization on double poles manifest. 
We can also write the amplitude with only double poles in the eikonal propagators, in a manifestly gauge-invariant form, as a sum over labeled trees with three vertices. To this end, we introduce the vector
\begin{equation}
\mathcal{Y}_{ab}^\mu = (k_b{\cdot} F_a{\cdot} \bu)\,(F_b{\cdot} \bu)^\mu - (\bu{\cdot} F_a{\cdot} F_b{\cdot} \bu)\,k_b^\mu + (2\,\bu{\cdot} k_a + \bu{\cdot} k_b)\,(\bu{\cdot} F_a{\cdot} F_b)^\mu\,,
\label{eq:y_vector}
\end{equation}
in terms of which
\begin{equation}
A_3^{\text{Coulomb}} (\bm \bu; k_1, k_2, k_3) = \frac{e^3}{2\bm}\sum_{\sigma\in S_3}
\frac{\mathcal{Y}_{\sigma_1\sigma_2}{\cdot} F_{\sigma_3}{\cdot} \bu}{(\bu{\cdot} k_{\sigma_1})^2(\bu{\cdot} k_{\sigma_3})^2}\ .
\label{eq:a3_photons_double}
\end{equation}
These forms assume the standard on-shell condition for the photons, $k_i^2=0$. Although we will sew them off-shell to form two-to-two amplitudes, the discrepancy will give rise only to configuration-space contact terms, which do not contribute to the classical scattering observables.

Iterating the recursion~\eqref{eq:classical_recursion} once more gives the four-emission amplitude. To express it in compact form,
for each label $c$, introduce the scalar
\begin{equation}
\mathcal{Z}_c = -\sum_{i}(\bu{\cdot} F_{i}{\cdot} F_c{\cdot} \bu)\prod_{j \neq i}(k_c{\cdot} F_j{\cdot} \bu) + \sum_{\{i,j\}}(\bu{\cdot} k_i-\bu{\cdot} k_j)\,(\bu{\cdot} F_i{\cdot} F_c{\cdot} F_j{\cdot} \bu)\,(k_c{\cdot} F_d{\cdot} \bu)\ ,
\end{equation}
where $i$ and $j$ run over the three labels different from $c$, the second sum runs over unordered pairs, and $d \neq i, j, c$ is the remaining label. The four-emission amplitude is,
\begin{equation}
\begin{split}
A_4^{\text{Coulomb}} (\bm \bu; k_1, \dots, k_4) = -\frac{e^4}{\bm^2}\sum_{\sigma\in S_4}\Bigg[ & \frac{\mathcal{Y}_{\sigma_1\sigma_2}{\cdot} \mathcal{Y}_{\sigma_4\sigma_3}}{4\,(\bu{\cdot} k_{\sigma_1})^2(\bu{\cdot} \Ksum_{\sigma_1\sigma_2})^2(\bu{\cdot} k_{\sigma_4})^2}\\
& + \frac{\mathcal{Z}_{\sigma_4}}{9\,(\bu{\cdot} k_{\sigma_1})^2(\bu{\cdot} k_{\sigma_2})^2(\bu{\cdot} k_{\sigma_3})^2}\Bigg]\ .
\end{split}
\label{eq:a4_photons}
\end{equation}
The combinatorial factors compensate for the overcounting 
in summing over all permutations.
One can continue to higher multiplicity in a straightforward manner.
It turns out that the integrands arising from contracting
these classical amplitudes with sources can be recast
in a simple form for all multiplicities, as we discuss in the next
section.

\section{Integrands for Massive Scattering from a Cayley-Tree Construction}
\label{sec:graphs}

In \cref{subsec:classical_limit}, we used Schwinger parameters to take the
classical limit of multi-emission amplitudes in the cubic scalar theory.
The resulting classical phase $\delta$, \cref{eq:eikonal_phase_classical},
admits a remarkably simple organization in terms of combinatorial trees.
We also argued that integrating over the Schwinger parameters in this
limit yields both rational and local contributions 
to the amplitudes.  The former contain only squared eikonal propagators,
while the latter contain delta functions as well.

In this section, we work directly in momentum space, where this double-pole structure is manifest, as already illustrated in \cref{eq:amplitude_scalar}. We focus on the rational contributions.\footnote{The local terms can be restored in the eikonal-like phase by going to the Schwinger-parameter representation, as we do in \cref{sec:eikonal}.} This formulation allows us to examine the combinatorial structure of classical amplitudes in the scalar theory in greater detail, and to extend the construction to other theories. In particular, we use the recursion relation derived in \cref{sec:factorization} to
construct amplitudes at arbitrary emission multiplicity in the scalar
theories and in classical electrodynamics.  In the latter
theory, we show that when these amplitudes
are contracted with sources --- that is, when we sum over the intermediate 
polarization states ---
the resulting integrands admit closed representations in terms of
so-called \textit{Cayley trees}, also known as labeled trees~\cite{Cayley:1889,Moon:1970}. We also extend
this organization to probe-limit scattering in half-BPS
black-hole backgrounds in $\mathcal{N}=8$ supergravity
\cite{Caron-Huot:2018ape,Parra-Martinez:2020dzs}.

Amplitudes with photons or $\mathcal{N}=8$ supergravity multiplets generally have a more complicated structure than in the scalar case considered in \cref{subsec:classical_limit}. In particular, their multi-emission amplitudes depend on the polarizations of the emitted massless legs. 
Contracting with classical sources and integrating over the
exchanged momenta (including a factor of $1/N!$ to account for Bose
symmetry) then gives the classical amplitudes
$M_{2\to 2}^{(N-1)}$ and the corresponding phase $\delta$.
This can be carried out at arbitrary multiplicity $N$ and turns out
to give a closed form similar to that obtained in a scalar theory
with an appropriate choice of couplings.\footnote{An inspection of the 
multi-emission amplitudes in electrodynamics even before contraction, \cref{subsec:electrodynamics_amplitudes}, suggests that they too should admit a systematic characterization in terms of Cayley trees even with explicit polarizations. This structure is reminiscent of the results of Ref.~\cite{Brandhuber:2021bsf}, but lies beyond the scope of the 
present paper.}  

We can give different contributions a graphical representation in terms
of trees.  
Each of the multi-emission amplitudes will be represented as a sum over weighted Cayley trees.
Each
massless messenger momentum corresponds to a given vertex of each tree.  
Each edge carries a weight, and the contribution of each graph is
the product of the weights of its edges.  
The weights make the pure double-pole form of the classical limit manifest.

Diagrammatically, each Cayley tree can be mapped to a worldline Feynman diagram, but the associated weight is different.
The worldline Feynman rules, unlike the Cayley trees, lead to contributions from both single and double poles.
Of course, when all diagrams are summed the resulting amplitudes are the same.

We start by showing the Cayley-tree form for the scalar cubic theory.  We then extend it to quartic interactions, classical electrodynamics, and $\mathcal{N}=8$ supergravity.

\subsection{Scalar Theories}
\label{subsec:cayley_scalar}
\subsubsection*{Cubic Theory}
\begin{figure}[t!]
    \centering
    \includegraphics[page=1,width=0.2\textwidth]{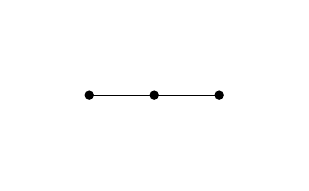}\\[0.5em]
    \includegraphics[page=2,width=0.2\textwidth]{figures/cayley-topologies.pdf}
    \hspace{0.08\textwidth}
    \includegraphics[page=3,width=0.2\textwidth]{figures/cayley-topologies.pdf}\\[0.5em]
    \includegraphics[page=4,width=0.2\textwidth]{figures/cayley-topologies.pdf}
    \hspace{0.04\textwidth}
    \includegraphics[page=5,width=0.2\textwidth]{figures/cayley-topologies.pdf}
    \hspace{0.04\textwidth}
    \includegraphics[page=6,width=0.2\textwidth]{figures/cayley-topologies.pdf}
    \caption{The unlabeled Cayley trees (topologies) contributing to $A_3$, $A_4$, and $A_5$, from top to bottom.}
    \label{fig:cayley_trees}
\end{figure}

\def\connected#1#2{\eb_{#2}(\slashed{#1})}
\def\vb{\bar{v}}
\def\eb{\bar{e}}
\def\AllCayley#1{\mathcal{T}_{#1}}
The rational terms in the amplitudes of the scalar cubic theory,
\cref{eq:amplitude_scalar}, exhibit a clear pattern in both
the contractions of massless momenta in their numerators and the structure
of their matter poles. Explicit calculations at higher multiplicities
reveal the same pattern, suggesting an organization that persists at
arbitrary order.   The set of graphs we will use for the order-$N$
contribution is that of tree graphs of all topologies with $N$
labeled vertices and $N-1$ edges.  We will denote this set
$\mathcal{T}_N$. The first few unlabeled topologies are shown in \cref{fig:cayley_trees}.
We choose one label to be the \textit{root label}.
The rational part of the scalar
multi-emission amplitude with $N$ massless legs can be written
in a compact way as a sum over this set,
\begin{equation}
    \label{eq:n_point_scalar}
   A_N^{\text{scalar-cubic}} (\bm \bu; k_1, \dots, k_N) =
   \sum_{T \in \AllCayley{N}} \omega^{\text{scalar-cubic}}_r(T)\,,
\end{equation}
in terms of a set of rooted weights $\omega_r$.  This is precisely
the momentum space equivalent of the tree sums introduced in \cref{eq:eikonal_phase_classical} in the Schwinger parameter space
representation. 
We have the following expression for the rooted weight of a given
(labeled) tree $T$,
\begin{equation}\label{eq:scalar_cubic_weight}
    \omega^{\text{scalar-cubic}}_r(T):= - \frac{g^N}{2^{N-1}\bm^{N-2} }\prod_{e\in T}
    \frac{k_{i}{\cdot} k_{j}}
    {\bigl(\bu{\cdot} k_{\connected{r}{}}\bigr)^2}\,,
    \qquad
    k^\mu_{\connected{r}{}} \equiv \sum_{a\in \connected{r}{}} k^\mu_a\,,
\end{equation}
with $i$ and $j$ denoting the two vertices of the edge $e$;
and $\connected{r}{}$ denoting the \textit{complementary tree}:
the subtree of $T$ not
containing the root $r$ that is obtained by removing the edge $e$ from $T$.
Any choice of $r$ is suitable, but we must make the same choice in
all terms in \cref{eq:n_point_scalar}.
The Cayley-tree formula in \cref{eq:n_point_scalar}
is remarkably simple. 
It makes manifest the pure double-pole form in the final result.
We give more details 
on the Cayley trees and their connection to the worldline formulation
in \cref{subsec:wlForm}.

The forms~\eqref{eq:amplitude_scalar} for the one-
and two-emission amplitudes follow trivially from
\cref{eq:n_point_scalar} and \cref{eq:scalar_cubic_weight}. Let us therefore consider the three-emission
amplitude, $N=3$. In this case, there is a single tree topology, with
three distinct labelings:
\begin{equation}\label{eq:trees_three}
   \linegraph{1,2,3}\ , \quad \linegraph{1,3,2}\ , \quad \linegraph{2,1,3}\ .
\end{equation}

Consider for instance the first of these, the labeled tree
$T{=}1{-}2{-}3$. According to \cref{eq:scalar_cubic_weight}, for each tree, we must pick a \textit{root}: a privileged vertex. This choice is arbitrary, and the final result does not depend on it. Choose for instance the root vertex to be $r=2$. The tree we are examining
has two edges, $e_{12}$ and $e_{23}$.
Removing the edge $e_{12}$ splits the tree into two subtrees with vertex sets
$\{1\}$ and $\{2,3\}$.
As the root $2$ belongs to the subtree with vertices $\{2,3\}$, 
the subtree not containing the root is $\connected{2}{} = \{1\}$.
The edge $e_{12}$ then contributes with the factor
\begin{equation}
\frac{k_1{\cdot} k_2}{(\bu{\cdot} k_1)^2}\,,
\end{equation}
to the weight of the tree. 
Similarly, removing the edge $e_{23}$ splits the tree into
the subtrees $\{3\}$ and $\{1,2\}$.
The root $2$ belongs to the second subtree, so here the subtree not
containing the root is
$\connected{2}{} =\{3\}$.
The edge $e_{23}$ then contributes the weight,
\begin{equation}
\frac{k_2{\cdot} k_3}{(\bu{\cdot} k_3)^2}\ .
\end{equation}
Putting these two contributions together, 
we obtain the weight~\eqref{eq:scalar_cubic_weight} for the labeled tree $T{=}1{-}2{-}3$:
\begin{equation}
\omega^{\text{scalar-cubic}}_2(1{-}2{-}3)
=- \frac{g^3}{4\bm}
\frac{k_1{\cdot} k_2}{(\bu{\cdot} k_1)^2}
\frac{k_2{\cdot} k_3}{(\bu{\cdot} k_3)^2}\ .
\end{equation}

The remaining two trees yield contributions related by permutations of
the external legs. Summing all three contributions reproduces the
rational part of the three-emission amplitude in
\cref{eq:amplitude_scalar}. We keep the root fixed at $r=2$ throughout.
Different root choices change only the intermediate form of the tree
weights; once energy conservation is imposed, the full amplitude is
independent of the root.

At higher multiplicity, new tree topologies appear. Starting at four emissions, trees contain vertices of higher valence. For example, the four-emission amplitude has two topologies, sequential trees analogous to those in \cref{eq:trees_three}, and a star topology. Following the 
rules~\eqref{eq:scalar_cubic_weight} with $r=4$ as the root, 
a representative star tree yields
\begin{equation}    
\stargraphfour{4}{1}{2}{3}=- \frac{g^4}{8\bm^2}
       \frac{k_1 {\cdot} k_4\, k_2 {\cdot} k_4\, k_3 {\cdot} k_4}{(\bu{\cdot} k_1)^2(\bu{\cdot} k_{2})^2(\bu{\cdot} k_{3})^2}\ .
\end{equation}
At five emissions, we have sequential trees; star trees; and in
addition, fork trees, such as
\begin{equation}\label{eq:fork}
        \forkgraphfive{3}{5}{4}{2}{1} = 
    - \frac{g^5}{16\bm^3}
       \frac{k_1 {\cdot} k_2\, k_2 {\cdot} k_3\, k_3 {\cdot} k_4 \, k_3 {\cdot} k_5}{(\bu{\cdot} k_1)^2(\bu{\cdot} \Ksum_{12})^2(\bu{\cdot} k_{4})^2(\bu{\cdot} \Ksum_{1234})^2}\,,
\end{equation}
where we have chosen $r=5$ as the root.

It is easy to check that the number of labeled trees contributing to amplitudes with up to five emissions is
\begin{equation}
1\,, \qquad1\,, \qquad  3\,, \qquad 12+4=16\,, \qquad 60+5+60=125\,,
\end{equation}
respectively. We listed the three-emission graphs in \cref{eq:trees_three}.
At four emissions, there are $12$ sequential and $4$ star trees; at five emissions, there are $60$ sequential, $5$ star and $60$ fork trees. In general, the number of labeled trees with $N$ vertices is $N^{N-2}$, as given by Cayley's formula. Although this number grows rapidly with the number of emissions, we will show in the next section that the sum over all trees simplifies considerably.  We do not need to enumerate the trees
explicitly, which would be very costly at high multiplicity.
We do this by using the recursion relation~\eqref{eq:recursion_formula}
to prove the expression~\eqref{eq:n_point_scalar} for any $N$.

\subsubsection*{Proving the All-Multiplicity Formula} 
We can prove the all-multiplicity formula~\eqref{eq:n_point_scalar} for 
the classical tree-level amplitudes of the cubic scalar theory 
by induction.  It is clear from the form of \cref{eq:n_point_scalar} and \cref{eq:scalar_cubic_weight}
that when we shift $\bu \to \bu - w \zeta$, the amplitude 
$A_N^{\text{scalar-cubic}}$ scales as $w^{-2(N-1)}$. Analyticity and Cauchy's theorem then determine the amplitude uniquely once we impose the factorization~\eqref{eq:factorization_theorem} at the poles.
Thus, we need only to prove that \cref{eq:n_point_scalar} satisfies the factorization~\eqref{eq:factorization_theorem}, with the three-point
amplitude $A_1^{\text{scalar-cubic}} = - g m$ as a seed. 

\begin{figure}[t!]
    \centering
    \includegraphics{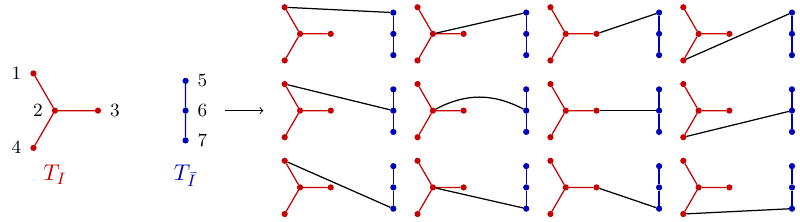}
    \caption{The graphs contributing to the factorization channel at $\bu{\cdot} k_{1 2 3 4} = 0 = \bu \cdot k_{5 6 7}$. The red star on $I=\{1,2,3,4\}$ and the blue line on $\bar I=\{5,6,7\}$
    can be joined by twelve distinguished edges. All twelve bridges carry the same
    double pole $1/(\bu{\cdot}k_{1 2 3 4})^2$, and their numerators $k_a{\cdot}k_b$
    sum to $k_{1 2 3 4}{\cdot}k_{5 6 7}$. We need to sum over all the Cayley trees on the two sides, and for each distinguished pair of subtrees, we have twelve ways of joining them, which all give the same denominator (up to momentum conservation) and the same numerator.}
    \label{fig:bridge}
\end{figure}

Let us study the singular behavior of the amplitude as 
$\bu\cdot k_I\rightarrow 0$ for some subset $I$ of momenta.  
Only a subset of trees can give rise to a singularity:
only trees containing an edge whose removal 
separates the original Cayley tree into two trees with vertex sets
$I$ and $\Ib$ respectively.  Label the vertices of the removed edge by
$a$ and $b$.  The sum over all contributing Cayley trees $\AllCayley{N}$
then factors into separate sums over all trees for $I$, all trees for
$\Ib$, and all choices of $a$ and $b$,
\begin{equation}
    \sum_{T\in\AllCayley{N}} \to \sum_{\substack{a\in I\\ b\in\Ib}}
    \sum_{T_I\in \AllCayley{I}} \sum_{T_{\Ib}\in \AllCayley{\Ib}}\,,
    \label{eq:sum_factorization}
\end{equation}
where the notation $\AllCayley{I}$ denotes all Cayley trees for
the labels in $I$.

Look at one contributing tree $T$, and 
separate its weight~\eqref{eq:scalar_cubic_weight} into three factors,
\def\central{\rho}
\begin{equation}
\begin{aligned}
     \omega_r(T_I\subset T) &= 
     - \frac{g^{|I|}}{2^{|I|-1}\bm^{|I|-2} }\prod_{e\in T_I}
    \frac{k_{i}{\cdot} k_{j}}
    {\bigl(\bu{\cdot} k_{\connected{r}{*}}\bigr)^2}\,,\\
    \central 
    &= -\frac{k_a\cdot k_b}
              {2\bm^2\left(\bu{\cdot}k_{\Ib}\right)^2}
            \,,\\
    \omega_r(T_{\Ib}\subset T) &=
      - \frac{g^{|\Ib|}}{2^{|\Ib|-1}\bm^{|\Ib|-2} }\prod_{e\in T_{\Ib}}
    \frac{k_{i}{\cdot} k_{j}}
    {\bigl(\bu{\cdot} k_{\connected{r}{*}}\bigr)^2}\,,
\end{aligned}
\label{eq:product_factorization}
\end{equation}
where $T_I$ is the subtree of $T$ whose vertices are labeled by
the elements of $I$, and similarly for $T_{\Ib}$.  In these
products, the complementary tree is taken
with respect to the \textit{full\/} Cayley tree $T$, rather than the subtrees
$T_I$ or $T_{\Ib}$.  We denote this by the `$*$' subscript on $\eb$.  

\def\path{P}
Suppose we take the root $r\in I$.
We could then rewrite the complementary tree with respect to the
full tree $T$ in terms of the complementary trees taken with
respect to the subtrees.  For this purpose, it
will be helpful to define the notion of the path $\path(i,j)$
in the tree from vertex $i$ to vertex $j$ (the path is necessarily unique).
For $T_I$, there are two cases
to consider: the first, where removing the edge leaves $r$ and
$a$ in different subtrees; the second, where removing the edge leaves
$r$ and $a$ on the same subtree.  In the former case, the edge lies on
the path from $r$ to $a$, $e\in \path(r,a)$; 
whereas in the
latter case, the path from $r$ to $a$ does not include the edge,
$e\not\in P(r,a)$.  We also define an associated theta symbol,
\begin{equation}
    \theta_{r,a}(e) = \begin{cases} &1,\qquad e\in \path(r,a)\,,\\
    &0,\qquad e\not\in \path(r,a)\,.\end{cases}
\end{equation}
In the first case, we need to add
all the momenta in $\Ib$ to those in the complementary tree with
respect to $I$; in the second case, we add nothing.  We 
can combine the two cases using $\theta_{r,a}(e)$, to obtain,
\begin{equation}
    k_{\connected{r}{*}} = k_{\connected{r}{}} + \theta_{r,a}(e) k_{\Ib}\,.
\end{equation}
In the denominators in \cref{eq:product_factorization}, this sum of momenta is 
contracted with $\bu$,
giving rise to a dependence on $\bu{\cdot}k_{\Ib} = -\bu{\cdot}k_I$.
When we expand in the limit $\bu{\cdot}k_I\to 0$, this dependence will in
turn give rise to subleading poles beyond the double pole introduced by
$\central$.  For this choice of $r$, the complementary tree 
for edges in $T_{\Ib}$ only
contains edges within $T_{\Ib}$, and hence is unchanged.  It is the same complementary tree
that we would obtain were we to have chosen the root to be $b$.

It will be important to maintain the manifest symmetry in
$I\leftrightarrow\Ib$ in order to obtain the symmetric form in
\cref{eq:factorization_theorem}.  This is not straightforward if
we make a specific choice of $r$ in either $I$ or $\Ib$ and then
expand the factors above in powers of $\bu{\cdot}k_I$, so we
proceed a bit differently.

Let us look at the double pole first.  As explained above, the
only difference between $\bu{\cdot}k_{\connected{r}{*}}$ and
$\bu{\cdot}k_{\connected{r}{}}$ consists of terms of $\cO(\bu{\cdot}k_I)$.
In extracting the coefficient of the double pole, we set these terms
to zero, or equivalently simply replace $k_{\connected{r}{*}}$ by
$k_{\connected{r}{}}$.  This in turn replaces,
\begin{equation}
    \begin{aligned}
        \omega_r(T_I\subset T) &\rightarrow \omega_r(T_I)\,,\\
        \omega_r(T_{\Ib}\subset T) = \omega_b(T_{\Ib}\subset T) &\rightarrow \omega_b(T_{\Ib}) =  \omega_{\bar{r}}(T_{\Ib})\,.
    \end{aligned}
\end{equation}
The change of root for $T_{\Ib}$ was explained above.  In the last form,
$\bar{r}\in \Ib$; once we are considering the weight solely with respect
to the subtree, we can choose any fixed root $\bar{r}$.  Doing so decouples the
sum over $b$ from the sum over $\AllCayley{\Ib}$.
Summing the first factor here over all trees $\AllCayley{I}$
gives (using \cref{eq:n_point_scalar}) an amplitude for $I$, while summing the second
factor over $\AllCayley{\Ib}$ gives us an amplitude for $\Ib$.  
We must then sum over all
choices of $a\in I$ and $b\in\Ib$, as any vertex in $I$ could be
connected to any vertex in $\Ib$ to form one of the 
original contributing Cayley trees.
As we approach a pole in $\bu\cdot k_I$, the double pole 
contribution factorizes as follows,
\begin{equation}
    \sum_{\substack{a\in I\\ b\in\Ib}} A_{I\vIb} 
    \frac{(-k_{a}{\cdot} k_{b})}{2 \bm^2 (\bu{\cdot} k_I)^2} A_{\Ib}
    = A_{I\vIb} 
    \frac{(-k_{I}{\cdot} k_{\Ib})}{2 \bm^2 (\bu{\cdot} k_I)^2} A_{\Ib}
\,.
\label{eq:double_pole}
\end{equation}
This matches the first term in \cref{eq:factorization_theorem}.
We give a graphical example of this contribution in \cref{fig:bridge}.

To extract the coefficient of the single pole, we need the coefficient
of the linear term in the expansion of 
the product $\omega_r(T_I\subset T)\,\omega_r(T_{\Ib}\subset T)$.
Let us retain the choice of root $r\in I$.  As discussed above,
we can then replace
\begin{equation}
    \omega_r(T_{\Ib}\subset T) \rightarrow \omega_b(T_{\Ib}\subset T)\,.
    \label{eq:i_substitution}
\end{equation}

\begin{figure}[t]
    \centering
    \includegraphics{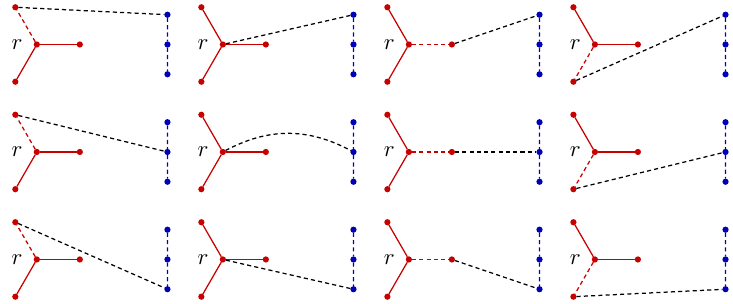}
    \caption{Choose the root to be $r=2$ in the two subtrees in \cref{fig:bridge}. 
    We illustrate the weight computation for each of these: represent the edges which 
    do depend on the momenta of the subtree $5{-}6{-}7$ by dashed lines. The solid edges
    have weights which do not depend on 
    $\bu \cdot k_{5 6 7}$. The red dashed lines represent the edges that contribute to the single-pole factorization upon expansion.}
    \label{fig:bridge_root}
\end{figure}

To understand how we can simplify $\omega_r(T_I\subset T)$, we need to derive
an identity starting with momentum conservation.  Let us consider a tree $T$,
and ask about the relation between the summed momenta of complementary
trees with respect to two different choices of roots $r$ and $v$.
If the edge $e$ is not on the path from
$r$ to $v$, then these momentum sums are identical as are their
contractions with $\bu$.  If the edge is on
the path from $r$ to $v$, then we find the following relation,
\begin{equation}
    \bu{\cdot} k_{\connected{r}{}}+\bu{\cdot} k_{\connected{v}{}}=0\,.
\end{equation}
In our setting, there is an additional term arising from the shift
from the complementary tree with respect to the full tree $T$ to 
the complementary tree with respect to the subtree $T_I$.  We find,
\begin{equation}
    \bu\cdot k_{\connected{r}{}} + \theta_{r,a}(e)\,\bu\cdot k_{\Ib} = \begin{cases}
        \bu\cdot k_{\connected{a}{}} & \theta_{r,a}(e) = 0\\
        -\bu\cdot k_{\connected{a}{}} & \theta_{r,a}(e) = 1
    \end{cases}\,.
    \label{eq:momentum_identity}
\end{equation}
The sign will evaporate when these expressions are squared, so this 
identity allows us to replace roots,
\begin{equation}
    \omega_r(T_I\subset T) \rightarrow \omega_a(T_I\subset T)\,.
    \label{eq:i_bar_substitution}
\end{equation}
(Note that the vertex $a$ plays a special role; we cannot substitute a different
vertex on the right-hand side.)
We give an illustration of the different types of contribution in \cref{fig:bridge_root}.

Using the factorization in \cref{eq:sum_factorization} and applying the substitutions (\ref{eq:i_substitution},\ref{eq:i_bar_substitution}) to  the factors in
\cref{eq:product_factorization} 
we obtain,
\begin{equation}
    \begin{split}
        A_N = - \frac{1}{2 \bm^2 (\bu{\cdot} k_I)^2}
        \sum_{T_I \in \mathcal{T}_I} \sum_{T_{\Ib} \in \mathcal{T}_{\Ib}}
        \sum_{\substack{a\in I\\ b\in\Ib}} k_a {\cdot} k_b\, \omega_a (T_I\subset T)\, \omega_b (T_{\Ib}\subset T) + \cO((\bu {\cdot} k_I)^{0})\,.
    \end{split}
    \label{eq:amplitude_from_subweights}
\end{equation}

We can now expand the denominators in $\omega_a(T_I\subset T)$ using 
the identity~\eqref{eq:momentum_identity}. We replace $k_{\connected{a}{}}$ and
then expand in $\bu{\cdot}k_I$, which gives us a sum that basically runs over
the edges of the path from $r$ to $a$,
\begin{equation}
    \omega_{a} (T_I \subset T) = \left[1 + 2 (\bu {\cdot} k_I) \sum_{e \in T_I} \frac{\theta_{r,a}(e)}{\bu{\cdot} k_{\connected{r}{}}} \right] \omega_{r} (T_I) + \cO ((\bu {\cdot} k_I)^2)\ ,
\label{eq:omega_expansion}
\end{equation}
where $r$ is an arbitrary but fixed root of the tree $T_I$.  While this may seem
to be undoing the change of root we made earlier, that change will allow us to
sum over choices of $a$ below.

We now interchange the sum over $e\in T_I$ inside 
\cref{eq:omega_expansion} with the sum over $a$, and use the result,
\begin{equation}
    \sum_{a\in I} \theta_{r,a}(e) k_a^\mu = k_{\connected{r}{}}^\mu\,,
    \label{eq:summation_identity}
\end{equation}
to obtain,
\begin{equation}
    \label{eq:root_changing}
    \begin{split}
        \sum_{a\in I} k_a^\mu \, \omega_a (T_I \subset T) & = \left[ k_I^\mu + 2 (\bu {\cdot} k_I) \sum_{e \in T_I} \frac{k^\mu_{\connected{r}{}}}{\bu{\cdot} k_{\connected{r}{}}} \right]\! \omega_{r} (T_I) + \cO ((\bu {\cdot} k_I)^2)\\
        & = \left[ k_I^\mu - \bu {\cdot} k_I \frac{\partial}{\partial \bu_\mu} \right]\! \omega_{r} (T_I) + \cO ((\bu {\cdot} k_I)^2)\,.
    \end{split}
\end{equation}
The other weight in \cref{eq:amplitude_from_subweights}, $\omega_b(T_{\Ib}\subset T)$,
satisfies a similar identity with a minus sign in the second term 
because $\bu {\cdot} k_{\Ib} = - \bu {\cdot} k_I$.

This proves the desired factorization~\eqref{eq:factorization_theorem} of the amplitude,
and in turn proves that the amplitude given by
\cref{eq:n_point_scalar,eq:scalar_cubic_weight}
is the \textit{unique} solution to the recursion relation~\eqref{eq:classical_recursion} with a constant single-emission amplitude as the seed.

\subsubsection*{Generalizations to Contact Terms} 
The all-multiplicity formula can be generalized to include four-point contact terms.  Like the two-photon emission amplitude in electrodynamics,
these terms are not BCFW-constructible and must be included separately in a new basic two-emission amplitude $A_2$.  It becomes a new seed for the recursion 
relation~\eqref{eq:classical_recursion}:
\begin{equation}
    A_2 ^{\text{scalar}}(\bm \bu; k_1, k_2) = - g^2 \frac{k_1{\cdot} k_2}{2 (\bu{\cdot} k_1)^2} - \left( \frac{g^2}{2} + \lambda\right)\ .
\end{equation}
The first term would arise from a recursive computation, but we
would be missing the second term. We will not give a detailed
proof, but just state the result.  It generalizes 
\cref{eq:n_point_scalar} in a straightforward manner.
The $N$-emission scalar amplitude is now,
\begin{equation}
    A_N^{\text{scalar}} =  \sum_{T \in \widetilde{\mathcal{T}}_N} \omega_r^{\text{scalar}}(T)\ ,
\end{equation}
where the sum runs over the set $\widetilde{\mathcal{T}}_N$ of \textit{decorated labeled trees} of all topologies with $N$ labeled
vertices and $N-1$ edges, subject to the following rules:
\begin{enumerate}[nosep]
    \item there are two types of edges, ordinary edges $e$ and 
    decorated (contact) edges $\widetilde{e}$;
    \item each vertex has at most one decorated edge attached to it.
\end{enumerate}
The number of decorated edges in different trees runs from zero to the 
maximum allowed by these rules.
The weights of the corresponding labeled trees are
\begin{equation}\label{eq:weight_scalar}
    \omega_r^{\text{scalar}}(T) \equiv- \frac{g^N}{2^{N-1}\bm^{N-2} }
    \prod_{e\in T}
    \frac{ k_i{\cdot} k_j}
    { \left(\bu{\cdot} k_{\connected{r}{}}\right)^2} 
    \prod_{\widetilde{e}\in T} \frac{2\lambda + g^2}{g^2}\,.
\end{equation}
Similarly to \cref{eq:scalar_cubic_weight}, 
$i$ and $j$ denote the two momenta at the ends of edge
$e$ in the first product.
The product over $e$ is over undecorated edges, and the product over $\widetilde{e}$ is over decorated ones.
The second product gives one contact-term factor for each decorated 
edge of the labeled tree.
Overall, each edge contributes one factor.

We expect this structure to generalize to higher-point contact terms, 
as would arise from higher-dimension operators in the effective
theory.  These would correspond to Feynman-diagram vertices of valence higher than four. In this case, we expect each new contact term to
give rise to a new `color' of decoration.  The corresponding decorated trees
may have two or more decorated edges attached to each vertex. Such interactions correspond to potentials $\propto {r^{-2 n \alpha_d}}$ 
with $n\geq 3$, as one can understand from our later discussion in \cref{sec:eikonal}.

\subsection{Electrodynamics}

We now turn to classical photon amplitudes. Like their scalar counterparts,
they exhibit double poles in the eikonal propagators and can be constructed
using the recursion relation~\eqref{eq:classical_recursion}. Their polarization
dependence, however, requires nontrivial gauge-invariant numerators,
as we have already seen in \cref{eq:a3_photons,eq:a4_photons}. 
These numerators obscure the generic
combinatorial structure and make it more difficult to obtain closed
expressions at arbitrary emission multiplicity. On the other hand, this combinatorial structure simplifies considerably upon contraction with sources and summation over the intermediate polarizations.

To illustrate this hidden simplicity, we construct the two-to-two scattering amplitude at $(N-1)$ loops from the classical multi-emission photon amplitudes through \cref{eq:probe_limit}, treating one particle as a classical background source. This corresponds to the probe limit, that is,
zeroth order in the self-force expansion.\footnote{We expect this simplicity to extend beyond the probe limit, where the photons are contracted with a mixture of sources from both external lines, since it stems from the sum over polarizations.} 
Contracting the photon legs with the source replaces their polarization vectors by the source velocity, eliminating explicit polarization labels and simplifying the combinatorial representation.
Here we use the classical photon source 
\begin{equation}
    J_1^\mu(\ell)
=
2ie_\heavy m_\heavy u_\heavy^\mu\,\hat\delta(2m_Hu_\heavy{\cdot} \ell)\ ,
\end{equation}
for a heavy particle with electric charge $e_\heavy$, mass $m_\heavy$ and velocity $u_\heavy$.
Each source attachment thus contributes a factor
$({-i\eta_{\mu\nu}}/{\ell^2})\,J_1^\nu(\ell)
=
{ e_\heavy}
u_{\heavy\mu}\,
\hat\delta(u_\heavy{\cdot} \ell)/{\ell^2}$.
The source-contracted amplitude at leading order is then
\begin{equation}
iM_{2\to2}^{(0)\,\textrm{Coulomb}}
=
4i e_He_\probe m_\heavy\bar m_\probe\,\gamma
\int
\frac{\hat\dd^d\ell_1}{\ell_1^2}
\hat\delta^{(d)}(q-\ell_1)\ ,
\end{equation}
where $\gamma$ is the Lorentz factor and the subscript $\probe$ labels the light particle.
At next-to-leading order, we find
\begin{equation}
i M_{2\to2}^{(1)\,\textrm{Coulomb}}
=-\frac{4i m_\heavy}{2!}(e_He_\probe)^2
\int
\prod_{a=1}^{2}
\frac{\hat \dd^d\ell_a}{\ell_a^2}\,
\hat\delta(u_\heavy{\cdot}\ell_1)\,
\hat\delta^{(d)}(q-\ell_1-\ell_2)
\left[
\gamma^2
\frac{\ell_1{\cdot}\ell_2}{(\bu_\probe{\cdot}\ell_1)^2}
-1
\right]\ .
\end{equation}

At this point, the resemblance to the scalar theory is striking: the tensor structures involving the photon momenta are the same as in the scalar theory. Up to a redefinition of the coupling, which must include a rescaling by appropriate powers of $\gamma$, the integrand is identical to that of the scalar theory. The uniqueness of the all-multiplicity formula in the scalar theory teaches us that the electrodynamic integrand is therefore fixed by the appropriate definitions of the one- and two-emission seed amplitudes. 
At $(N-1)$ loops, we therefore obtain the closed formula,
\begin{equation}
iM_{2\to2}^{(N-1)\,\textrm{Coulomb}}
=\frac{2i m_\heavy}{N!}
\int
\prod_{a=1}^{N}
\frac{\hat{\dd}^{d}\ell_a}{\ell_a^2}
\prod_{a=1}^{N-1}
\hat{\delta}(u_\heavy{\cdot} \ell_a)\,
\hat{\delta}^{(d)}
\left(
q-\sum_{a=1}^{N}\ell_a
\right)
\sum_{T \in \widetilde{\mathcal{T}}_N}
\mathbb{w}_r^{\mathrm{Coulomb}}(T)\ ,\label{eq:all_order_electrodynamics}
\end{equation}
where the  corresponding weight is
\begin{equation}
        \mathbb{w}_r^{\textrm{Coulomb}}(T)
        =\frac{2(e_\heavy e_\probe \gamma)^N}{\bm_\probe^{N-2}} \prod_{e\in T}
        \frac{(-\, \ell_i{\cdot} \ell_j)}
        {\bigl(\bu_\probe{\cdot} \ell_{\connected{r}{}}\bigr)^2}
        \prod_{\widetilde{e}\in T}
        \frac{1}{\gamma^2}\ .
    \label{eq:electrodynamics_tree_weight}
\end{equation}
In \cref{eq:all_order_electrodynamics},
$\widetilde{\mathcal{T}}_N$ denotes all Cayley trees for $N$ momenta,
with all possible decorations subject to the same rules as given above
for the scalar theories with contact terms. As in \cref{eq:weight_scalar}, the first product is over all undecorated edges $e$, and the second one runs over all decorated edges $\widetilde{e}$.
This form exposes the similarity between the photon weight and its scalar-theory
counterpart, \cref{eq:weight_scalar}. As the loop integrals are identical in the two theories, this explains why the corresponding scattering angles, \cref{eq:angle_coulomb,eq:angle_scalar}, 
agree up to a redefinition of the couplings.
In \cref{eq:all_order_electrodynamics}, we use the symbol $\mathbb{w}_r$
for the Cayley weight of trees in the expression for the amplitudes $M_{2\to2}$, as distinguished from the Cayley weights 
$\omega_r$ of the scalar multi-emission amplitudes $A_N$.  The weights
here include the additional coupling factors from the static source. 

At this point, we have determined the integrand only up to the $i\varepsilon$ prescription of the eikonal propagators. We discuss how to reconstruct the full $i\varepsilon$ prescription in \cref{sec:eikonal}.

\subsection{\texorpdfstring{$\mathcal{N}=8$}{N=8} Supergravity}
\label{subsec:supergravity_n8}
As a third example of a theory whose source-contracted amplitudes admit a Cayley-tree construction, we turn to the dynamics of two half-BPS black holes in four-dimensional maximal $\mathcal{N}=8$ supergravity, in the probe limit and with minimal charge misalignment. The scattering angle for this case is given in \cref{eq:angle_sugra}.

Caron-Huot and Zahraee~\cite{Caron-Huot:2018ape} studied
classical black-hole scattering in four-dimensional maximal $\mathcal{N}=8$ supergravity.   They considered the scattering of two extremal half-BPS black holes up to one loop. They computed the scattering angle by modeling the black holes as massive scalars interacting through the exchange of massless bosonic states in the graviton supermultiplet; the appropriate extremal limit suppresses fermionic exchanges. 
Parra-Martinez, Ruf, and Zeng~\cite{Parra-Martinez:2020dzs} extended this
analysis to two loops for black holes with minimal charge misalignment, using the Kaluza--Klein reduction of known massless $D$-dimensional
$\mathcal{N}=8$ supergravity integrands. Upon reduction, the extra-dimensional momenta are identified with the BPS charges.
In the probe limit, the supergravity angle~\eqref{eq:angle_sugra}
exhibits remarkable similarities to the 
scalar result~\eqref{eq:angle_scalar}, to the 
Coulomb result~\eqref{eq:angle_coulomb}, and to 
Newtonian gravity (which is integrable).
The half-BPS condition removes the characteristic features
of gravitational exchange, which require functions such as the 
Appell $F_1$ \cite{Kol:2021jjc,Scharf:2011ii}. 

As we now show, the supergravity problem can be reduced to an analog of
the scalar theory with only a cubic interaction. 
In order to display this connection, let us take the probe and classical
limits of the massive integrands computed in 
Ref.~\cite{Parra-Martinez:2020dzs} up to one loop. We denote by $\misangle$ the single charge-misalignment angle and define
\begin{equation}
\mathcal{C}
  =
  16 \pi G\,
  m_\heavy m_\probe \left(\gamma-\cos\misangle\right)^2\ .
\end{equation}
The one- and two-emission source-contracted integrands are then:
\begin{equation}
    \begin{split}
        i M_{2\to2}^{(0)\,\mathcal{N}=8} & = - 2 i m_\heavy \mathcal{C} \bm_\probe \int \frac{\hd^d \ell_1}{\ell_1^2} \hat\delta^{(d)}(q-\ell_1)\ ,\\
        iM_{2\to2}^{(1)\,\mathcal{N}=8} & = -\frac{1}{2!}\, i m_\heavy \mathcal{C}^2 \int \prod_{i=1}^2 \frac{\hd^d \ell_i}{\ell_i^2}\, \hdelta(u_\heavy{\cdot} \ell_1)\,\hat\delta^{(d)}(q-\ell_1-\ell_2)\left[\frac{\ell_1{\cdot} \ell_2}{(\bu_\probe{\cdot} \ell_1)^2}  \right]\ .
    \end{split}
\end{equation}
As before, we keep only the rational terms. These are the seeds of a Cayley-tree representation of $M_{2\to2}^{\mathcal{N}=8}$: the Cayley formula is analogous to that for 
electrodynamics~\eqref{eq:all_order_electrodynamics}, but the 
labeled-tree weight is now
\begin{equation}\label{eq:weight_n_8}
\mathbb{w}_r^{\,\mathcal N=8}(T)
= -
\frac{\mathcal C^N}{2^{N-1}\bar m_\probe^{N-2}}
\prod_{e\in T}
\,
\frac{\ell_i{\cdot} \ell_j}
{\bigl(\bu_\probe{\cdot} \ell_{\connected{r}{}}\bigr)^2}\,,
\end{equation}
with no contact decorations, just as in the basic pure-cubic scalar theory.

For instance, at three emissions the Cayley-tree formula gives
\begin{equation}
\begin{split}
     iM_{2\to2}^{(2)\,\mathcal{N}=8} = -\frac{1}{3!}\frac{i m_\heavy\mathcal{C}^3}{2\bm_\probe} \int &\prod_{i=1}^3\frac{\hd^d \ell_i}{\ell_i^2} \prod_{i=1}^2\hdelta(u_\heavy{\cdot} \ell_i)\,\hat\delta^{(d)}(q-\ell_1-\ell_2-\ell_3)\\
  & \times \left[   
  \frac{(\ell_1{\cdot} \ell_2)(\ell_1{\cdot} \ell_3)}
    {(\bu_\probe{\cdot} \ell_2)^2 (\bu_\probe{\cdot} \ell_3)^2}
+
  \frac{(\ell_1{\cdot} \ell_2)(\ell_2{\cdot} \ell_3)}
    {(\bu_\probe{\cdot} \ell_1)^2 (\bu_\probe{\cdot} \ell_3)^2}
+
  \frac{(\ell_1{\cdot} \ell_3)(\ell_2{\cdot} \ell_3)}
    {(\bu_\probe{\cdot} \ell_1)^2 (\bu_\probe{\cdot} \ell_2)^2}
  \right]\ ,
  \end{split}
\end{equation}
which agrees with the two-loop
integrand computed in Ref.~\cite{Parra-Martinez:2020dzs} in the probe limit
(up to contact terms). In this result we have chosen a different root for each graph, which is equivalent choosing the same root for all graphs and applying momentum conservation. The Cayley-tree formula~\eqref{eq:weight_n_8} thus provides nontrivial all-order results in supergravity.

\section{The \texorpdfstring{$i\varepsilon$}{i-epsilon} 
Prescription and the On-shell Action}
\label{sec:eikonal}
In the previous sections, we have shown how to compute the classical two-to-two scattering amplitude in the probe limit starting from the multi-emission amplitudes of the light 
particle to all orders in perturbation theory. We also saw that the structure of the simple 
scalar-mediated theory holds more generally and applies to other theories without
nonlinear interactions of the massless messengers.\footnote{Strictly speaking, $\mathcal{N}=8$ supergravity does have nonlinear interactions among the modes of the graviton supermultiplet. However, the half-BPS condition suppresses these interactions. 
This is connected to the no-triangle hypothesis in $\mathcal{N}=8$ supergravity \cite{Bern:1998sv,Bern:2005bb,Bjerrum-Bohr:2005kwk,Bjerrum-Bohr:2006xbk,Bern:2007xj,Bjerrum-Bohr:2008vag,Bjerrum-Bohr:2008qoa,Caron-Huot:2018ape} as applied to diagrams for the probe
limit.} 
In this section, we discuss how to resum the combinatorial objects 
introduced in the previous section in order to recover the classical probe trajectory and on-shell action.
We compute an eikonal-like phase, building on the computation in \cref{subsec:classical_limit}.

\subsection{Restoring Local Terms}
\label{subsec:restoringTerms}

In \cref{sec:graphs}, we computed an integrand for the probe-limit scattering amplitude of these theories up to local terms.  In principle, the local terms can be reconstructed using a generalization of the factorization theorem~\eqref{eq:factorization_theorem_local}. En route to recovering the classical probe trajectory, we now
explain a shortcut to fixing these terms.

The classical factorization theorem~\eqref{eq:factorization_theorem} is formulated in 
terms of the barred variables $\bu$ and $\bm$, and uses a principal-value prescription
for the poles in denominators. In this section, we work with the graph-based integrand of
the $2\to2$ amplitude in the scalar theory.  Its decorated and labeled tree weight is,
\begin{equation}\label{eq:weight_scalar_m2_to_2}
    \mathbb{w}_r^{\text{scalar}}(T) = - \frac{(g_hg_\heavy)^N}{2^{2N-1}\bm_\probe^{N-2}}
    \prod_{e\in T}
    \frac{ \ell_i{\cdot} \ell_j}
    { \bigl(\bu_\probe{\cdot} \ell_{\connected{r}{}}\bigr)^2} 
    \prod_{\widetilde{e}\in T} \frac{2\lambda_\probe + g_\probe^2}{g_\probe^2}\,,
\end{equation}
where as before $i$ and $j$ are the vertices at the ends of the edge $e$.  The
first product is over undecorated edges, and the second over decorated ones, so
that overall each edge contributes one factor.
Generalizations to electrodynamics and $\mathcal{N}=8$ supergravity are straightforward. Because
the local terms are uniquely determined by the rational terms through the factorization theorem, and because these theories have the same tree structure (up to graph decorations), the $i\varepsilon$ prescription in the scalar theory carries
over to the other theories.  We can focus on understanding how to fix the
prescription in the scalar theory.
As a first step, we transform the momentum-space amplitude, which has the Cayley-tree 
form \eqref{eq:all_order_electrodynamics}, to coordinate space with the normalization
of \cref{subsec:classical_limit}. This restores the $N$-th heavy delta function and removes
a factor of $2m_\heavy$:
\begin{equation}
    i \widetilde{M}_{2\to 2}^{(N-1)} = \frac{i}{2\bm_\probe\, N!} \int_{\mathbb{R}} \dd\tau \int \prod_{k=1}^N \left[ \hd^d \ell_k \frac{\hdelta(u_\heavy{\cdot} \ell_k)}{\ell_k^2} e^{i (\xinf + \tau \bu_\probe) {\cdot} \ell_k}\right] \sum_{T \in \widetilde{\mathcal{T}}_N} \mathbb{w}_r(T)\ ,
\label{eq:impact_parameter_2_to_2}
\end{equation}
where the integral over $\tau$ gives a Fourier representation of the energy-conserving delta function $\hat\delta(\bu_\probe{\cdot} q)$.
We now introduce a Schwinger parametrization of the eikonal propagators, similar to
that in \cref{subsec:classical_limit}. Since these propagators are squared and carry a 
principal-value prescription, the appropriate Schwinger parametrization is
\begin{equation}
    \label{eq:schwinger_parametrization}
    \frac{1}{\bigl(\bu_\probe{\cdot} \ell_{\connected{r}{}}\bigr)^2} = 
    - \frac{1}{2}\int_{\mathbb{R}} \dd s_{ij}\, |s_{ij}|\, \exp\bigl[i s_{ij} \bu_\probe{\cdot} \ell_{\connected{r}{}} 
                                 - \varepsilon |s_{ij}|\bigr]\,.
\end{equation}
For reasons that will become clear shortly, the parameters $s_{i j}$ are associated with oriented edges, with the orientation of the edge $e_{ij}$ running from $i$ to $j$. The choice of root $r$ determines the orientation: all edges point toward the root. 
Before performing the loop integrals, we make the following change of variables:
\begin{equation}
    \begin{split}
        s_{i j} = t_i - t_j\ ,\qquad \tau = t_r\ ,
    \end{split}
\end{equation}
where $r$ is the root of the tree. Under this change of variables, each denominator
in the first product in \cref{eq:weight_scalar_m2_to_2} gives rise to a factor of,
\begin{equation}
    -\frac12|t_i-t_j|\,,
\end{equation}
leaving aside a phase to which we return below.  Each edge in the second product
in \cref{eq:weight_scalar_m2_to_2}, over the decorated edges, gives rise to a delta
function,
\begin{equation}
    \delta(t_i-t_j)\,,
\end{equation}
as one of the integrals is then really superfluous.  Finally, we collect
the phases, looking first at a tree with only undecorated edges.  
Here, we find,
\begin{equation}
\begin{aligned}
    \prod_{e\in T} \exp\bigl[i (t_i-t_j) \bu{\cdot}\ell_{\connected{r}{}}\bigr]
    &= \exp\Bigl[i \sum_{e\in T} (t_i-t_j) \bu{\cdot}\ell_{\connected{r}{}}\Bigr]\,;
\end{aligned}
\end{equation}
using \cref{eq:summation_identity}, we can rewrite,
\begin{equation}
\begin{aligned}
\sum_{e\in T} (t_i-t_j) \bu{\cdot}\ell_{\connected{r}{}} &=
\sum_{e\in T}  (t_i-t_j)\sum_{a=1}^N \theta_{r,a}(e)\bu{\cdot}\ell_{a} \\
&= \sum_{a=1}^N (t_a-t_r)\bu{\cdot}\ell_{a}\,.
\end{aligned}
\label{eq:rewritten_phase}
\end{equation}

We can collect the factors other than the phases into 
$\hat{\mathbb{w}}$,\footnote{Each undecorated edge yields a minus sign in the Schwinger 
parametrization~\eqref{eq:schwinger_parametrization}. Extracting a minus sign from each 
decorated edge, we obtain a uniform factor $(-1)^{N-1}$ for all trees with $N$ vertices.}
\begin{equation}
    \label{eq:new_weights_electrodynamics}
    \hat{\mathbb{w}} (T) = \frac{(-g_hg_\heavy)^N}{2^{2N}\bm_\probe^{N-1}} \prod_{e_{ij}\in T} \frac{1}{2}\, |t_i - t_j| \, \ell_i{\cdot} \ell_j \prod_{\widetilde{e}_{ij}\in T} \delta(t_i - t_j) \left(-\frac{2\lambda_\probe + g_\probe^2}{g_\probe^2}\right)\,,
\end{equation}
where we have also absorbed the overall factor of $1/(2 \bm_\probe)$ in
\cref{eq:impact_parameter_2_to_2}.  Combining the original phase in \cref{eq:impact_parameter_2_to_2}
with those rewritten in \cref{eq:rewritten_phase},
we can write the $2\to2$ amplitude in coordinate space in terms of this weight,
as follows,
\begin{equation}
    \label{eq:eikonal_phase_electrodynamics}
    i \widetilde{M}_{2\to 2}^{(N-1)} = \frac{i}{N!} \int_{\mathbb{R}^N} \dd^N t \int \prod_{k=1}^N \left[ \hd^d \ell_k \frac{\hdelta(u_\heavy{\cdot} \ell_k)}{\ell_k^2} e^{i (\xinf + t_k \bu_\probe) {\cdot} \ell_k}\right] \Bigg( \sum_{T \in \widetilde{\mathcal{T}}_N} \hat{\mathbb{w}} (T) +\left\{\substack{\textrm{localized}\\[1mm] \text{contributions}}\right\} \Bigg)\,.
\end{equation}
The result's independence of the choice of root $r$ is now manifest. 
This result agrees with the 
explicit computation for the cubic theory in \cref{eq:probe_amplitude}, up to 
localized terms and the substitution 
$\bu_\probe = \frac{m_\probe}{\bm_\probe} u_\probe - \frac{q}{2 \bm_\probe}$. 
We can match the localized terms as well by changing the $i\varepsilon$ prescription 
of the eikonal propagators in Schwinger parameter space:
\begin{equation}
    \frac{|t_i-t_j|}{2} \to \max (t_i,t_j)\,,
\end{equation}
from which we obtain,
\begin{equation}
    \label{eq:eikonal_phase_max}
    \begin{split}
            2 i \delta_{\mathrm{cl}}^{(N-1)} &= \frac{i}{N!} m_\probe \left(-\frac{g_\probe g_\heavy}{4 m_\probe}\right)^N
\\&\hphantom{=}
\times\!\int\! \prod_{i=1}^N \dd \sigma_{i}\! \sum_{T \in \widetilde{\mathcal{T}}_N} \left[ \prod_{e_{ij}\in T}
    \max (t_i,t_j) \ell_i{\cdot} \ell_j \prod_{\widetilde{e}_{ij}\in T} \delta\left(t_i-t_j\right) \left(-\frac{2\lambda_\probe + g_\probe^2}{g_\probe^2}\right) \right]\,,
    \end{split}
\end{equation}
for the phase (recall that the measure $\dd \sigma_{i}$ was defined in \cref{eq:eikonal_measure}).
In the set of trees $\widetilde{\mathcal{T}}_N$ we could pick any fixed root, or equivalently use the set of unrooted trees.

\subsection{Integration of the Iterative Structures}
\label{subsec:integration}

Now that we have a unified formulation for the phase $\delta$, we want to make contact with the integration strategy used in \cref{subsec:classical_limit} for the scalar theory. 
The standard strategy to perform this integration would be to integrate over the Schwinger parameters $t_i$ first, to recover momentum-space integrals. In turn, these can be reduced using integration-by-parts identities and then computed via differential equations. However, this approach hides several important properties of the integrals.

First of all, in hindsight, we know that there are certain combinations of integrals
that are evanescent in four dimensions, even though they could be divergent when taken in 
isolation. For example, for the cubic scalar theory and $\mathcal{N}=8$ supergravity, we know that the orbits do not precess and hence the contribution to the scattering angle at even orders vanishes \cite{Caron-Huot:2018ape}. This gives us a nontrivial identity between integrals, which is not manifest in the momentum-space representation:
\begin{equation}
    \int\! \prod_{i=1}^N \dd \sigma_{i}\! \sum_{T \in \mathcal{T}_N} \prod_{e\in T}
    \max (t_i,t_j)\, \ell_i{\cdot} \ell_j \propto d - 4\ ,\qquad \text{for even } N\ ,
\end{equation}
which in momentum space would read as
\begin{equation}
    \begin{split}
        \int \hd^d q\, e^{i \xinf {\cdot} q} \hdelta (\bu_\probe {\cdot} q) \int \prod_{k=1}^N &\left[ \hd^d \ell_k \frac{\hdelta(u_\heavy{\cdot} \ell_k)}{\ell_k^2} \right]  \hat\delta^{(d)}\left(q-\sum_{i=1}^N \ell_i\right)  \\&
       \hspace{10mm} \times  \left[ 
        \sum_{T \in \mathcal{T}_N} \prod_{e\in T}
           \frac{ \ell_i{\cdot} \ell_j} { \left(\bu_\probe{\cdot} \ell_{\connected{r}{}}\right)^2} + \left\{\substack{\text{localized}\\[1mm] \text{contributions}}\right\} \right]\! \propto d - 4\ .
    \end{split}
\end{equation}
In the case of electrodynamics and the cubic theory with quartic coupling, contact terms make $N$-even integrals nonzero even in four dimensions. Nonetheless, the identities above
hint at a somewhat different behavior between even and odd loop orders. 
It is interesting that even and odd loop orders yield contributions to scattering angles that respectively lack or have a power of $\pi$. This is obvious from Taylor expanding, for instance, \cref{eq:angle_scalar} or \cref{eq:angle_coulomb}.
Simplifications of the type discussed above are expected to extend beyond the probe limit; some aspects of this simplicity were recently investigated in Ref.~\cite{Bern:2026oqp}.
Moreover, for $N\geq 4$, there are nontrivial cancellations between different graphs. Indeed, a soft messenger, that is, a vertex whose Schwinger time is sent to infinity,
yields at most a logarithmic divergence. Furthermore, there is no divergence for two adjacent vertices which are soft at the same time. The IR poles of a single graph therefore come from sets of non-adjacent vertices, and power counting bounds the divergence of a single graph by $1/(d-4)^{N-2}$ for even $N$ and by $1/(d-4)^{N-3}$ for odd $N$. For instance, at $N=4$ the star and the sequential graphs both diverge as
$1/(d-4)^2$, and these poles cancel only in the sum over all labeled trees. The correct way of performing these integrals is to recover the sum over graphs as a generating function.

In this section, we build on the well-known relation between classical solutions and tree diagrams: classical solutions of the equations of motion generate the tree-level expansion~\cite{Nambu:1968rr,Boulware:1968zz}.

The tree graphs in our computation are \textit{unrooted} graphs: the vertices are all 
treated as equivalent. These graphs can be obtained from \textit{rooted} trees, for which one vertex plays the role of the root. Rooted trees are easier to enumerate due to their recursive structure: all rooted trees with $N$ vertices can be obtained by attaching smaller
rooted trees to a new root vertex, with the smaller trees containing $N-1$ vertices in total.

In our graphical representation, all vertices are \textit{generated} by acting with derivatives (as in \cref{eq:tensor_integrals}) on a single generating 
function:\footnote{The derivatives yield the powers of loop momenta in the numerators, 
as in \cref{subsec:classical_limit}, and as seen in \cref{eq:eikonal_phase_max}.}
\begin{equation}
    U_z(x) = z_1 z V(x) + z_2 \frac{z^2}{2} V(x)^2\ ,
\end{equation}
where $V(x)$ is defined in \cref{eq:potential_scalar}, $z$ is a formal placeholder that counts powers of the coupling, and
\begin{equation}
    z_1 = \frac{g_\probe g_\heavy}{4 m_h}\ ,\qquad z_2 = \frac{g_H^2 (2\lambda_\probe + g_\probe^2)}{16 m_h^2}\ .
\end{equation}
We choose the Schwinger-time Green's function discussed above,
\begin{equation}
\label{eq:maxChoice}
    G(t_i, t_j) = \max(t_i,t_j)\ .
\end{equation}
With these two ingredients, we recursively define a generating vector $\bbX^\mu$:
\begin{equation}
    \label{eq:generating_vector}
    \begin{split}
        \bbX^\mu (z;t_i) & = \sum_{k=0}^\infty \frac{1}{k!} \int \dd t_j\, G(t_i, t_j)\, \nabla_j^{\mu \nu_1 \dots \nu_k} U_z(x_j) \, \bbX_{\nu_1}(z;t_j) \cdots \bbX_{\nu_k}(z;t_j) \\
        & = \int \dd t_j\, G(t_i, t_j)\, \nabla_j^{\mu} U_z(x_j + \bbX(z;t_j))\ ,
    \end{split}
\end{equation}
where the second line is a formal resummation of the first.
Iterating the first line of \cref{eq:generating_vector} gives rise to a rooted tree, the number of vertices corresponding to the order in the expansion, or equivalently, the number of appearances of $U_z$.
For example, trees with at most three vertices arise from iterating up to three times: 
\begin{equation}
    \label{eq:generating_vector_z4}
    \begin{split}
        \bbX^\mu (z;t_i) & = \int \dd t_1\, G(t_i,t_1)\, \nabla_1^{\mu} U_z(x_1) \\
        & + \int \dd^2 t\, G(t_i,t_1)\, G(t_1,t_2)\, \nabla_1^{\mu\nu} U_z(x_1)\, \nabla_{2\,\nu} U_z(x_2) \\
        & + \int \dd^3 t\, G(t_i,t_1)\, G(t_1,t_2) \Big[ G(t_2,t_3)\, \nabla_1^{\mu\nu} U_z(x_1)\, \nabla_{2\,\nu}{}^{\rho} U_z(x_2)\, \nabla_{3\,\rho} U_z(x_3) \\
        & \qquad + \frac{1}{2}\, G(t_1,t_3)\, \nabla_1^{\mu\nu\rho} U_z(x_1)\, \nabla_{2\,\nu} U_z(x_2)\, \nabla_{3\,\rho} U_z(x_3) \Big] 
        + \mathcal{O}(z^4)\ .
    \end{split}
\end{equation}
We thereby see that $\bbX^\mu$ is the generator of rooted trees.
We shall soon see that it can also be understood as a suitable correction to the trajectory of the probe particle.

We have thereby assigned a second graphical meaning to the quantities appearing in the
integrand of the eikonal-like phase~\eqref{eq:eikonal_phase_max}:
\begin{itemize}[nosep]
    \item[$\smallbullet$] each edge $e_{i  j}$ in the graph corresponds to a symmetric Green's function $G(t_i,t_j)$,
    \item[$\smallbullet$] the potential generates the vertices, with the number of derivatives (or equivalently, the number of momenta) associated to each vertex corresponding to the valence of the vertex.
\end{itemize}

We also notice that the generating vector $\bbX^\mu(z;t_i)$ satisfies the equation of motion of a particle in the central potential $U_z(x_i)$. Indeed, we have
\begin{equation}
    \label{eq:greens_function_equation}
    \partial_{t_i}^2 G(t_i, t_j) = \delta(t_i - t_j)\ ,
\end{equation}
which implies
\begin{equation}
    \label{eq:equations_of_motion}
    \partial_{t_i}^2 \bbX^\mu (z;t_i) = \nabla_i^{\mu} U_z(x_i + \bbX(z;t_i))\ ,
\end{equation}
with boundary conditions fixed by the choice~\eqref{eq:maxChoice}:
\begin{equation}
    \label{eq:boundary_conditions}
    \begin{aligned}
        \partial_{t_i} \bbX^\mu(z;t_i) &\to 0\ , &\qquad t_i &\to - \infty\ ,\\
        \bbX^\mu(z;t_i) - t_i\, \partial_{t_i} \bbX^\mu(z;t_i) &\to 0\ , &\qquad t_i &\to + \infty\ .
    \end{aligned}
\end{equation}
We can think of $\bbX^\mu$ as representing the deviation in
the probe's worldline trajectory from the baseline 
\begin{equation}
x^\mu(t) = \xinf^\mu + t u_\probe^\mu \,.
\end{equation}
The asymmetry in the boundary conditions reflects using final-state on-shell conditions in the delta functions appearing in \cref{eq:impact_parameter_form} while constraining $\xinf$ using the initial velocities.\footnote{This differs from the KMOC choice of fixing the impact parameter in the far past.}
We can understand the boundary conditions~\eqref{eq:boundary_conditions} directly
on $\bbX^\mu$ by using the formally resummed form~\eqref{eq:generating_vector},
\begin{equation}
    \partial_{t} \bbX^\mu(z;t) = \int_{-\infty}^{t} \dd \tau\, \nabla^{\mu} U_z(x(\tau) + \bbX(z;\tau))\,,
\end{equation}
from which we find,
\begin{equation}
    \begin{aligned}
        \bbX^\mu(z;t) &\to c^\mu\,, \qquad t \to -\infty\,,\\
        \bbX^\mu(z;t) &\to t (u_{\probe,\textrm{out}}^\mu - u_{\probe}^\mu)\,, \qquad t \to +\infty\,,
    \end{aligned}
\end{equation}
where
\begin{equation}
    \label{eq:incoming_c}
    c^\mu = \int_{-\infty}^{+\infty} \dd t\, t\,\nabla^{\mu} U_z(x(t) + \bbX(z;t))\,.
\end{equation}
The KMOC impact parameter is then $\xinf^\mu +c^\mu$.

We have already noted in \cref{subsec:classical_limit} that the $\max$ choice is not the unique choice of Green's function on the worldline. Even symmetry under interchange of the two Schwinger times on the worldline does not uniquely fix the Green's function. For the family of symmetric kernels,
    $c_1 |t_i-t_j| + c_2 (t_i + t_j)$,
the Green's function equation~\eqref{eq:greens_function_equation} fixes $c_1=1/2$, while $c_2$ parametrizes a shift freedom. In particular, the eikonal-like phase with the $\max(t_i,t_j)$ kernel is IR finite for $N\geq 2$, while the absolute-value kernel does not give a finite result (the two are connected by
an IR-divergent shift of $\xinf$, equivalent to a different choice of boundary conditions).

We would now like to express the phase $\delta$ in terms of $\bbX^\mu$.
As seen in \cref{eq:eikonal_phase_max} we can express the former as a sum over unrooted trees while the latter is expressed as a sum over rooted trees.
To obtain an unrooted tree, we can either link two rooted trees by a single edge (with a symmetry factor of $1/2$) or attach $n$ rooted trees to a new vertex (with a symmetry factor of $1/n!$). Both procedures are valid, but each overcounts the unrooted trees as follows:
\begin{itemize}[nosep]
    \item[$\smallbullet$] Joining trees via a single vertex marks the specific vertex. Thus, in a graph with $N$ vertices, we get an extra multiplicative factor $N$.
    \item[$\smallbullet$] Joining two trees via an edge marks the chosen edge. Thus, in a graph with $N$ vertices, we get an extra multiplicative factor of $N-1$.
\end{itemize}
There is a relative minus sign which compensates for the overcounting:
$N - (N-1) = 1$.
We define the reduced phase
\begin{equation}
    \rdelta(z) = \frac{1}{m_\probe} \sum_{N=1}^\infty z^N\, 2 \delta_{\mathrm{cl}}^{(N-1)} \,,
\end{equation}
which is precisely the sum over these possibilities with the correct symmetry factors:
\begin{equation}
    \label{eq:generating_scalar}
    \rdelta(z) = \frac{1}{2} \int \dd t_i \dd t_j \bbX^\mu(z;t_i) G^{-1}(t_i,t_j) \bbX_\mu(z;t_j) - \int \dd t_i\, U_z(x_i + \bbX(z;t_i)) \ .
\end{equation}
The inverse Green's function in this equation is defined by
\begin{equation}
    \int \dd t_j\, G^{-1}(t_i, t_j)\, G(t_j, t_k) = \delta(t_i - t_k)\ ,
\label{eq:inverse_greens_function}
\end{equation}
so that,
\begin{equation}
    G^{-1}(t_i, t_j) = \partial_{t_j}^2 \delta(t_i - t_j)\ .
\end{equation}
Upon substituting \cref{eq:generating_vector_z4} into \cref{eq:generating_scalar}, every term becomes an unrooted tree. Expanding the potential term around $x_i$, this gives a number of rooted trees attached to the vertex at $t_i$. In the first term, the inverse Green's function cancels one of the two extra edges of the rooted trees. The variation of \cref{eq:generating_scalar} with respect to $\bbX^\mu(z;t_i)$ vanishes by the second line of \cref{eq:generating_vector}.  For example, the error of order $z^4$ in 
\cref{eq:generating_vector_z4} then changes $\rdelta(z)$ only at order $z^8$. 
We then obtain
\begin{equation}
    \label{eq:generating_scalar_z7}
    \rdelta(z) = - \sum_{N=1}^{7} \frac{1}{N!} \int \dd^N t \sum_{T\in\mathcal{T}_N}\, \prod_{e_{ij}\in T} \left[G(t_i,t_j)\, \nabla_i{\cdot}\nabla_j\right] \prod_{k=1}^{N} U_z(x_k) + \mathcal{O}(z^8)\ ,
\end{equation}
which gives the eikonal-like phase exactly through six loops.

We have written the phase $\delta$ as a generating function of unrooted trees. Using the
inverse Green's function~\eqref{eq:inverse_greens_function}, 
we obtain the following expression for the generating function~\eqref{eq:generating_scalar},
\begin{equation}
    \rdelta(z) = \int \dd t \left[\frac{1}{2} \bbX^\mu(z;t) \partial^2_{t}\bbX_\mu(z;t) - U_z(x(t) + \bbX(z;t))\right]\ ,
\end{equation}
which is the action of a particle in the potential $U_z(x_i)$ 
evaluated on shell with the boundary conditions~\eqref{eq:boundary_conditions}.\footnote{As we are working in the mostly-minus convention, the signs
inside the brackets are the correct ones to obtain a Lagrangian.} By going to radial coordinates, one can show using standard arguments from classical mechanics that this action is equivalent to the radial action up to boundary terms
\begin{equation}
    m_h \rdelta(1) = 2\delta = 2\Phi(k,J) + J \chi - \xinfnorm k\sin\chi \ ,
\end{equation}
where $\Phi(k,J)$ is the subtracted radial action,
 \begin{equation}
    \Phi (k, J) = \lim_{R \to \infty}
    \biggl(\int_{r_{\text{min}}}^{R} \dd r
    \sqrt{k^2 - \frac{J^2}{r^2} - 2 m_h^2 \widetilde{U}(r)} - \int_{J/k}^{R} \dd r \sqrt{k^2 - \frac{J^2}{r^2}} \biggr)\ ,
\end{equation}
where $\widetilde{U}(r)$ is the potential $U_1(x)$ evaluated at a
solution of $(u_\heavy{\cdot}x)^2-x^2=r^2$ with $z=1$:
\begin{equation}
    2m_h^2\,\widetilde U(r)=\frac{8\pi\nu_d\,m_h\beta_\phi}{r^{2\alpha_d}}+\frac{16\pi^2\nu_d^2\,\alpha_\phi^2}{r^{4\alpha_d}}\,.
\end{equation} 
Also,
$\chi = - 2 \partial_{J} \Phi(k,J)$ is the scattering angle, 
positive for an attractive potential.

The boundary conditions~\eqref{eq:boundary_conditions} fix 
the momentum and angular momentum
\begin{equation}
    \label{eq:boundary_saddle}
    k = m_\probe \sqrt{\gamma^2-1}\ ,\qquad J = \xinfnorm k\cos\chi\ .
\end{equation}
The computation of the scattering angle from the eikonal-like phase is therefore
\begin{equation}
    \label{eq:eikonal_saddle}
    \sin\chi = -\frac{1}{k} \frac{\partial 2\delta}
        {\partial \xinfnorm}\,.
\end{equation}

Refs.~\cite{Kim:2024svw,Kim:2025gis,Brandhuber:2025igz,Kim:2026wyv,Brandhuber:2026dkd,Guo:2026xaw} study the matrix elements of the logarithm of the S-matrix, following the Magnus expansion~\cite{Magnus:1954zz}, making use of Murua coefficients~\cite{Murua:2006}. 
This expansion involves summing over trees in a manner different from the present paper. 
One might hope to find a single Green's function, along the lines of our max kernel, which yields the Magnus expansion.
However, we cannot find a modification of the max kernel that yields the Magnus expansion. To see this, we write the most general Green's function satisfying \cref{eq:greens_function_equation}, as
\begin{equation}
    \label{eq:general_greens_function}
    \max(t_i,t_j) + c_3 t_i t_j + c_4 t_i + c_5 t_j + c_6\,,
\end{equation}
further generalizing the discussion below \cref{eq:incoming_c}. When substituting the trajectories in \cref{eq:generating_vector} into \cref{eq:generating_scalar} using this general Green's function, writing the potential in terms of momenta as in \cref{eq:potential_scalar}, and performing the integration over Schwinger parameters, we cannot find a choice of $c_3,\dots,c_6$ that matches the Magnus expansion beyond $N=4$. This suggests that any connection between the Magnus expansion and the on-shell action is more subtle for $N>4$, even in integrable theories. 

\subsection{Bound States and Connections to the B2B Map}\label{subsec:b2b}
\begin{figure}[ht!]
    \centering
    \includegraphics{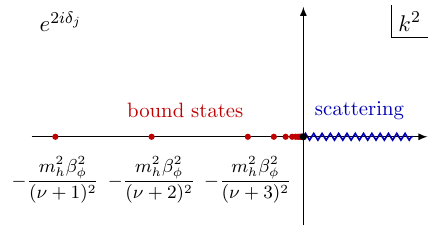}
    \hfill
    \includegraphics{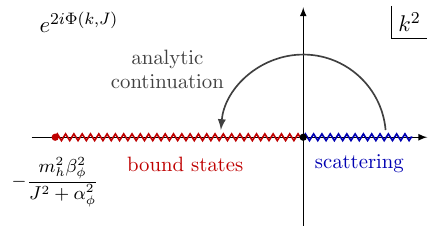}
    \caption{Singularities of the partial-wave $S$-matrix element in the complex $k^2$ plane for an attractive potential, $\beta_\phi>0$ (with $\hbar=1$). Left: the quantum $e^{2i\delta_j}$ of \cref{eq:partial_wave_quantum}, with the unitarity cut along $k^2>0$ and the bound-state poles at $k^2=-m_\probe^2\beta_\phi^2/(\nu+1+n)^2$, $n=0,1,2,\dots$, which accumulate at threshold. Right: the classical $e^{2i\Phi(k,J)}$, in which the poles have merged into a branch cut; the arrow shows the path of analytic continuation from the scattering region to the bound states. These properties have a very long history; for a recent review, see Refs.~\cite{Mizera:2023tfe,Correia:2024yfx} and references therein.}
    \label{fig:k_squared_plane}
\end{figure}

We have computed the phase $\delta$ in the probe limit in coordinate space,
\begin{equation}
    1 + i \widetilde{M}_{2\to 2} (\gamma,\xinfnorm) = \mathcal{N}( \gamma,\xinfnorm)\, e^{2 i \delta(\gamma, \xinfnorm)}\,.
\end{equation}
We have not computed the modulus $\mathcal{N}$: when we take the logarithm of the amplitude, its contribution mixes with the quantum corrections to the eikonal-like phase, which we have neglected. We now move from coordinate space to the partial-wave basis,
that is, to angular-momentum space. The partial waves are defined by the expansion of the momentum-space amplitude in Legendre polynomials,
\begin{equation}
    \label{eq:partial_wave_expansion}
    \frac{i k}{8\pi m_\heavy}\, \mathcal{M}_{2\to 2} \Bigl(\gamma, 4k^2\sin^2\frac{\theta}{2}\Bigr) = \sum_{j=0}^{\infty} J \left(e^{2 i \delta_j}-1\right) P_j(\cos\theta)\ ,
\end{equation}
where $J=j+\frac{1}{2}$ and $\theta$ is the scattering angle in the rest frame of the heavy particle, so that $-q^2=4k^2\sin^2(\theta/2)$. In the large-angular-momentum limit, the sum over $j$ becomes an integral over $J$, and the Legendre polynomials are approximated by Bessel functions uniformly in the scattering angle (see Szeg\H{o}~\cite[Sec.~8.21]{Szego:1975}, and Ref.~\cite{Berry:1969} for an application to semiclassical scattering),
\begin{equation}
    \label{eq:legendre_bessel}
    P_j(\cos\theta) = \left(\frac{\theta}{\sin\theta}\right)^{1/2} \mathrm{J}_0(J\theta) + \cO\bigl(J^{-3/2}\bigr)\,.
\end{equation}
Since $\xinf^\mu$ is transverse to the incoming momentum, the Fourier transform in \cref{eq:impact_parameter_form} involves only the transverse momentum $k\sin\theta$, and the amplitude in coordinate space becomes
\begin{equation}
    \label{eq:partial_waves_to_xinf}
    \mathcal{N}( \gamma,\xinfnorm)\, e^{2 i \delta(\gamma, \xinfnorm)} \simeq 1 + \int_0^{\pi} \dd \theta\, \sqrt{\theta\sin\theta}\;\mathrm{J}_0(k \xinfnorm \sin\theta) \int_0^\infty \dd J\, J\, \mathrm{J}_0(J\theta) \left(e^{2 i \delta_J}-1\right)\ .
\end{equation}
For large $J$ and $k\xinfnorm$, the Bessel functions are
\begin{equation}
    \begin{split}
        \mathrm{J}_0(J\theta)&\simeq\sqrt{\frac{2}{\pi J\theta}}\,\cos\!\left(J\theta-\frac{\pi}{4}\right)\,,\\
        \mathrm{J}_0(k\xinfnorm\sin\theta)&\simeq\sqrt{\frac{2}{\pi k\xinfnorm\sin\theta}}\,\cos\!\left(k\xinfnorm\sin\theta-\frac{\pi}{4}\right)\,.
    \end{split}
\end{equation}
Writing the cosines as sums of exponentials, we can perform the integrals over $\theta$ and $J$ in the stationary-phase approximation, as in the semiclassical analysis of potential scattering~\cite{Ford:1959a} and in Refs.~\cite{Kol:2021jjc,Bautista:2021wfy}. The stationary point is at $\theta=|\chi |$, with $\chi=-\partial(2\delta_J)/\partial J$ and $J=\xinfnorm k\cos\chi$, where the exponent equals $2\delta_J+J\chi-\xinfnorm k\sin\chi$. Comparing with $2\delta=2\Phi+J\chi-\xinfnorm k\sin\chi$ and with \cref{eq:boundary_saddle,eq:eikonal_saddle}, we find
\begin{equation}
    \delta_J = \Phi(k,J)\,, \qquad \mathcal{N}( \gamma,\xinfnorm) = \sqrt{\frac{1}{k}\frac{\partial J}{\partial \xinfnorm}}\,.
\end{equation}
Near four dimensions, the radial action can be computed in closed form, up to terms that vanish as $d\to 4$:
\begin{equation}
    \label{eq:radial_action_d4}
    \begin{split}
        2\Phi(k,J)&= \, \beta_k \left[\frac{2}{d-4}-\gamma_E-\log\pi+2+i\pi+\log(-k^2-i\varepsilon)\right]\\
        &\hphantom{=}+\, i(\alpha_J+i \beta_k)\log(\alpha_J+i \beta_k)-i(\alpha_J-i \beta_k)\log(\alpha_J-i \beta_k)-\pi(\alpha_J-J)\,,
    \end{split}
\end{equation}
where
\begin{equation}
    \alpha_J = \sqrt{\alpha_\phi^2 + J^2}\,, \qquad \beta_k = \frac{m_h\beta_\phi}{i \sqrt{-k^2-i \varepsilon}}\,.
\end{equation}
In the partial-wave basis, we can continue analytically below threshold, $k^2<0$, where the 
bound states are. In the quantum theory, the bound states are the poles of the partial-wave 
$S$-matrix element below threshold (see \cref{eq:partial_wave_quantum}). In the classical limit, these poles become dense and merge into a branch cut, as shown in \cref{fig:k_squared_plane}.
Following the path of analytic continuation shown in \cref{fig:k_squared_plane}, which corresponds to $k^2\to k^2+i\varepsilon$, we obtain the radial action below threshold,
\begin{equation}
    \label{eq:radial_action_bound}
    \begin{split}
        2\Phi(k,J)&=\, 2\pi\left(\frac{m_\probe\beta_\phi}{\sqrt{-k^2}}-\alpha_J\right)+\pi J
        +2i\,\alpha_J\operatorname{arctanh}\frac{\alpha_J\sqrt{-k^2}}{m_\probe\beta_\phi}\\
        &\hphantom{=}-i\,\frac{m_\probe\beta_\phi}{\sqrt{-k^2}}\left[\frac{2}{d-4}-\gamma_E-\log\pi+2-\log\left(\frac{m_\probe^2\beta_\phi^2+\alpha_J^2k^2}{k^4}\right)\right]\,,
    \end{split}
\end{equation}
for $-m_\probe^2\beta_\phi^2/\alpha_J^2<k^2<0$, which is the range of energies in which bound orbits exist.
Its real part is determined by the radial integral over the bound orbit:
\begin{equation}
    \label{eq:b2b_radial_action}
    \begin{split}
        \mathrm{Re}\,2\Phi(k,J) &= 2\int_{r_{\text{min}}}^{r_{\text{max}}}\dd r\,\sqrt{k^2-\frac{J^2}{r^2}-2m_\probe^2\widetilde U(r)}+\pi J\\
        &= 2\Phi(k,J)-2\Phi(k,-J)-\pi J\,,
    \end{split}
\end{equation}
where $r_{\text{min}}<r_{\text{max}}$ are the two turning points of the bound orbit, and the second line is the boundary-to-bound (B2B) map of Refs.~\cite{Kalin:2019rwq,Kalin:2019inp}. 
It gives the radial action of a bound orbit as the analytic continuation of the scattering one in the energy and in the angular momentum.

The observables of the bound orbit follow from $\mathrm{Re}\,\Phi$. For instance, the periastron advance per radial period is
\begin{equation}
    \label{eq:periastron_advance}
    \Delta\psi = - \partial_J \mathrm{Re}\,2\Phi(k,J) - \pi = 2\pi\left(\frac{J}{\alpha_J}-1\right)\,.
\end{equation}

The same analysis applies to electrodynamics and to $\mathcal{N}=8$ supergravity, since their phases $\delta$ differ from that of the scalar theories only by a redefinition of the couplings. Comparing the tree weights~\eqref{eq:electrodynamics_tree_weight} and~\eqref{eq:weight_n_8} with \cref{eq:weight_scalar_m2_to_2}, we find
\begin{equation}
    \label{eq:coupling_map}
    \begin{aligned}
        &\text{electrodynamics:} &\qquad \beta_\phi &\to -\gamma\,\frac{e_\heavy e_\probe}{4\pi}\,, &\qquad \alpha_\phi^2 &\to -\left(\frac{e_\heavy e_\probe}{4\pi}\right)^2\,,\\
        &\mathcal{N}=8\ \text{supergravity:} &\qquad \beta_\phi &\to 2 G m_\heavy m_\probe \left(\gamma-\cos\misangle\right)^2\,, &\qquad \alpha_\phi^2 &\to 0\,.
    \end{aligned}
\end{equation}
With these replacements, the scalar angle~\eqref{eq:angle_scalar} becomes either 
the supergravity angle~\eqref{eq:angle_sugra} or the Coulomb angle~\eqref{eq:angle_coulomb}, 
with $\alpha=-e_\heavy e_\probe/(4\pi)$. In both theories, the potential $\widetilde{U}(r)$ receives no higher-order corrections in the probe limit. In electrodynamics, it is one-loop exact; in $\mathcal{N}=8$ supergravity, the potential is tree-level exact. This difference is visible in the bound orbits. In electrodynamics, \cref{eq:periastron_advance} gives
\begin{equation}\label{eq:periastron_advance_coulomb}
    \Delta\psi = 2\pi\left[\frac{J}{\sqrt{J^2-\left(e_\heavy e_\probe/4\pi\right)^2}}-1\right]\,,
\end{equation}
which is the precession of the relativistic Coulomb orbit~\cite{Landau:1975pou}. In $\mathcal{N}=8$ supergravity, instead, $\alpha_J=J$ and the periastron advance vanishes~\cite{Caron-Huot:2018ape}. Equation \eqref{eq:periastron_advance_coulomb} also reveals the onset of the well-known  Coulomb instability, or \emph{fall to the center} \cite{Landau:1975pou, Boyer:2004}. As the angular momentum approaches the critical value 
\begin{equation}
    J\to J_{\text{crit}} = - \frac{e_H e_h}{4\pi},
\end{equation}
 the periastron advance diverges.
At this value, the centrifugal barrier disappears and the inner turning point reaches the 
origin. The same singularity is also present in the scattering case, as can be seen by 
inspecting the  angle \eqref{eq:angle_coulomb}. Crucially, this behavior emerges only after resumming the amplitude expansion to all loops. The same phenomenon is also present in gravity \cite{Darwin:1959,Bardeen:1972fi,Chandrasekhar:1985kt}, and it is known to worsen when the probe particle has its own intrinsic spin \cite{Flambaum:2025bpx}. 

Finally, the imaginary part of the radial action below threshold has a simple classical meaning. For $k^2<0$, the integrand of the radial action is real only between the two turning points. For $r>r_{\text{max}}$ the radial momentum is imaginary, and $\mathrm{Im}\,\Phi(k,J)$ is the action accumulated in this classically forbidden region. This term gives an exponential suppression of the amplitude for the bound states: the discontinuity of the amplitude below threshold is associated with the overlap squared (tunneling probability) of the bound state with the scattering state.
This is exponentially suppressed in the classical limit.

\section{Outlook and Discussion}\label{sec:conclusions}

In this paper, we have investigated the simplifications that emerge in the classical limit of scattering amplitudes, focusing on the probe limit. 
We derived a classical factorization theorem for multi-emission amplitudes, showing how the collision of quantum factorization poles produces 
\textit{double\/} poles
in the classical limit. An associated recursion relation allows us to organize classical amplitudes in terms of Cayley trees, providing new and compact all-order expressions for the probe-limit integrands.
The classical factorization theorem itself applies immediately to gravity, but the extension of the recursion relation requires more work.
We showed how the
logarithm of the amplitude in coordinate space, an eikonal-like phase, reconstructs the
classical scattering angle, and we uncovered a universal combinatorial structure underlying this correspondence. 

We further showed how this combinatorial organization leads naturally to a resummation of the perturbative series for the phase $\delta$ into the classical on-shell action. A crucial ingredient is the correct treatment of the propagator's $i\varepsilon$ prescription, which fixes the boundary conditions of the classical trajectory and clarifies the relation between scattering amplitudes and worldline dynamics. 
The Feynman $i \varepsilon$ prescription directly yields the max kernel as the Green's function on the worldline.
Our construction also reveals highly 
nontrivial cancellations among different tree graphs, yielding integral identities not manifest 
when employing conventional integration-by-parts reductions.

Our approach renders the underlying factorization properties and combinatorial origin of the classical dynamics explicit. We believe this provides a complementary yet different perspective to recent methods based on the logarithm of the $S$-matrix. It would be interesting to study further the connection between the max-kernel prescription (\ref{eq:firstMax}, \ref{eq:maxChoice}) used here and the
$i \varepsilon$ flow in the Magnus expansion (captured by the Murua coefficients \cite{Kim:2024svw,Brandhuber:2025igz,Kim:2026wyv,Brandhuber:2026dkd,Guo:2026xaw}).
There does not appear to be any modification of the max kernel which matches the Magnus expansion beyond four emissions.
This is true even in integrable models such as half-BPS scattering in $\mathcal{N}=8$ supergravity.
We leave a systematic understanding of the logarithm of the $S$-matrix within our formalism
to future work.

A natural extension is to investigate whether the same structures persist for spinning particles.  Including the spin of the background can be accomplished for instance following Refs.~\cite{Vines:2017hyw,%
Guevara:2018wpp,Chung:2018kqs,%
Bautista:2019tdr,Maybee:2019jus,Guevara:2019fsj,Arkani-Hamed:2019ymq,%
Aoude:2020onz,Bern:2020buy,%
Bautista:2021wfy,Haddad:2021znf,Jakobsen:2021zvh,%
FebresCordero:2022jts,Menezes:2022tcs,Cangemi:2022bew,Bautista:2022wjf,%
Haddad:2023ylx,Chen:2024mlx,Akpinar:2025huz}.
We expect taking the light probe's spin into account will be more challenging.  It would introduce additional tensor structures and couplings in 
multi-emission amplitudes, potentially leading to richer combinatoric
structures and factorization properties.  We expect that our method could also clarify the relationship between the simplicity of classical amplitudes and the integrability of 
spinning dynamics, in light of recent connections between spin effects and the integrability of Kerr orbits \cite{Akpinar:2025bkt}. 
A simple laboratory to test these ideas may be offered by the ``root-Kerr'' system, which is defined in terms of  the single copy of Kerr geometry \cite{Monteiro:2014cda, Arkani-Hamed:2019ymq, Guevara:2020xjx, Monteiro:2021ztt, Menezes:2022tcs, deFirmian:2026mln}.

In this paper, we have focused on global observables.  What about radiation?
Recent years have witnessed much progress in direct fixed-order calculations
of scattering waveforms with amplitude  methods~\cite{Brandhuber:2023hhy,Herderschee:2023fxh,Elkhidir:2023dco,Georgoudis:2023lgf,Caron-Huot:2023vxl,Bohnenblust:2023qmy,Brunello:2024ibk,Bohnenblust:2025gir,Georgoudis:2025vkk,Brunello:2025eso,Kavanagh:2026ivj,DeAngelis:2026kew}.  It would
be interesting to extend the all-orders methods of the present paper to waveforms.
Our calculations here required sewing all $N$ messenger legs to heavy sources,
in order to construct the $2\to 2$ amplitude. 
For radiative observables, we need instead to construct the $2\to3$ amplitude.
One could do this by keeping one of the $N$ messengers as outgoing,
representing real on-shell radiation. The remaining $(N-1)$ messengers would remain
attached to the background. The resulting amplitudes would then provide the building blocks
for perturbative classical radiation. One can ask whether the Cayley-tree structure
uncovered by our analysis would persist in the radiation setting.  If it does indeed
survive, such a construction could establish a direct link between the resummation of scattering waveforms and the classical trajectories reconstructed starting from an eikonal-like phase and coherent states. 

It would be valuable to extend our construction beyond the probe limit, allowing both scattering bodies to contribute dynamically. In this regime, the simple fan-diagram structure of the probe approximation will be replaced by a richer set of diagrams. 
Beyond the probe limit, both massive lines develop nontrivial propagator structures, which account for the recoil of each particle under the influence of the other as well as
radiation reaction~\cite{Peters:1964zz,Damour:1981bh,Iyer:1993xi,Iyer:1995rn,Pati:2002ux,Galley:2009px,Barack:2018yvs,Damour:2020tta,DiVecchia:2021ndb,Herrmann:2021tct,Jakobsen:2022psy,Kalin:2022hph,Dlapa:2022lmu,Bini:2022enm,DiVecchia:2022piu,Elkhidir:2023dco,Porto:2024cwd,Elkhidir:2026njr}.
Two building blocks go beyond the probe limit: the classical factorization theorem of \cref{sec:factorization} and the Cayley-tree organization of multi-emission amplitudes.
The former makes possible a generic recursion relation for the multi-emission amplitudes associated with either massive body. We would then construct the resulting two-body integrands by sewing together these amplitudes, rather than contracting with $N$ static sources.  The open question is whether the Cayley-tree representations of the
individual amplitudes can be combined into a simple graphical representation of the
sewn $2\to2$ amplitude.

The most immediate direction for future work is extending our results to 
gravity.  Here the self-interactions of the gravitational field introduce additional exchange and factorization channels. New vertices and channels will 
lead to a more complicated combinatorial structure of the basic $N$-emission amplitudes.
The fact that the universal factorization holds in gravity as well suggests that the rest of our analysis  will carry over to general relativity.  This would require a generalization of the Cayley-tree
construction to incorporate graviton interactions. It would be interesting to understand
how the gravity eikonal admits an interpretation in terms of 
geodesics alongside a resummation into the classical radial action. 
This would offer an explicit link between the relativistic perturbative series and 
the nonlinear dynamics of gravity.

\acknowledgments

S.D.A. would like to thank Brando Bellazzini, Carlo Heissenberg and Raj Patil for useful discussions on the eikonal, the classical limit of scattering amplitudes and their logarithm(s). M.S., 
Y.F.B.{} and D.O.C.{} would like to thank Luke Lippstreu, Nathan Moynihan and Mao Zeng for helpful discussions. We also thank the organizers of the workshop \href{https://www.ias.edu/sns/workshop-black-holes-theory-observations}{\textit{Black Holes from Theory to Observations}}, held at the Institute for Advanced Study, for inspiration and instruction in the use of AI in research.
D.O.C{} thanks the IPhT for hospitality in the early stages of this work.
We particularly thank Carlo Heissenberg for his careful reading of the manuscript and 
for providing useful comments.

This research was supported by the European Research Council (ERC) under the European Union's research and innovation program grant agreement ERC--AdG--885414 (`Ampl2Einstein').
D.O.C.{} is supported in part by the UK STFC consolidated grant ST/X000494/1.
M.S.{} has been supported in part by the U.S. Department of Energy (DOE) under award number 
DE--SC0009937; by the European Research Council (ERC) Horizon Synergy 
Grant ``Making Sense of the Unexpected in the Gravitational Wave Sky'' grant agreement no. GWSky-101167314; and by the Mani
L. Bhaumik Institute for Theoretical Physics at UCLA.
Funded by the European Union (ERC, Classical$+$, 101200505). 
Views and opinions expressed are however those of the author(s) only and do not necessarily reflect those of the European Union or the European Research Council. Neither the European Union nor the granting authority can be held responsible for them.
For the purpose of open access, the authors grant a Creative Commons Attribution (CC BY) license to any Author Accepted Manuscript version arising from this submission.

\paragraph{Use of AI} We acknowledge the use of AI tools in various stages of the research, literature review, and writing process. The Cayley-tree representation of the tree-level amplitudes was first recognized by ChatGPT 5.5, given the simplified result of the recursion relation for tree amplitudes written in terms of squared eikonal propagators through $N=6$.
The calculations in \cref{subsec:classical_limit} were performed with the 
assistance of various
Anthropic, OpenAI, and Mistral models, after the models were provided with the basic ideas.
The aim of that subsection was to provide an explicit derivation of many of the results
presented or assumed in the subsequent sections. Further, AI models assisted
in the development of the  mathematical proof of the factorization of the tree-level classical amplitude 
using the Cayley-tree representation. 
Finally, we used Gemini 4 Carbon, GPT 6 Astra, and Claude Opus 5.5 to audit the final manuscipt.

\appendix

\section{Worldline Construction of Scalar Theories}\label{app:worldline}
In this appendix, we revisit the scalar theories studied in the main text
from a worldline perspective. Our aims are twofold: 
\begin{itemize}
    \item[$\smallbullet$] to give a classical derivation of the scattering angles 
    presented in
\cref{subsec:scalar_angles}, where they were obtained from the classical
limit of the QFT partial-wave phase shift;
\item[$\smallbullet$] to
illustrate, through examples of worldline amplitudes, the emergence of
Cayley-tree structures in classical scattering.
\end{itemize}
We begin with the worldline action introduced in
\cref{eq:wqft_scalar_1}, specialized to four spacetime dimensions. Here we work
in natural units, $c=\hbar=1$. A convenient Polyakov form of the
reparametrization-invariant point-particle action, under
$s\to s'=f(s)$, is
\begin{align}
S
&=\,
\underbrace{\int \dd^4 x\, \frac12 \partial_\mu\phi\, \partial^\mu\phi}_{\text{bulk}}
-
\underbrace{
\int \dd s\,
\frac{M(\phi(x(s)))}{2}
\left[
e^{-1}(s)\eta_{\mu\nu}\dot{x}^\mu(s)\dot{x}^\nu(s)
+e(s)
\right]
}_{\text{point-particle}}
\,.
\label{eq:pp_worldline}
\end{align}
Here $M(\phi)$ is a generic scalar-dependent mass function. The
einbein transforms as
\begin{equation}
    e'(s')=\frac{ds}{ds'}e(s)\,,
\end{equation}
so that $e(s)ds$ is invariant and the point-particle action is
reparametrization invariant.

Varying the action with respect to the einbein gives the constraint
\begin{equation}
    e^2(s)
    =
    \eta_{\mu\nu}\dot{x}^\mu(s)\dot{x}^\nu(s)\ .
\end{equation}
We may then fix the proper-time gauge $e=1$. In this gauge the
constraint becomes
\begin{equation}\label{eq:geodesic}
    \eta_{\mu\nu}\dot{x}^\mu(s)\dot{x}^\nu(s)=1\ ,
\end{equation}
and the worldline parameter is identified with the proper time,
$s=\tau$. The action becomes
\begin{align}
S
&=\,
\int \dd^4 x\,
\frac12 \partial_\mu\phi\,\partial^\mu\phi
-
\int \dd\tau\,
\frac{M(\phi(x(\tau)))}{2}
\left[
\eta_{\mu\nu}\dot{x}^\mu(\tau)\dot{x}^\nu(\tau)
+1
\right].
\label{eq:pp_worldline_h}
\end{align}

\subsection{Scattering Angle}

We now specialize to radial backgrounds,
$M(\phi(x(\tau)))=M(\phi(r(\tau)))$, with
\begin{equation}
    \phi(r)=-\frac{g_\heavy}{8\pi r}\ .
\end{equation}
We use spherical coordinates $(t,r,\theta,\varphi)$ for worldline
motion.  We take $\varphi$ to be the polar angle, and restrict
attention to the $z=0$ scattering plane ($\varphi=\pi/2$).
The $\theta$ coordinate is the angle within this plane.
For a generic radial mass
function $M(\phi(r))$, the system has two conserved quantities: the
worldline energy $E$, and the angular momentum $j$, which is
orthogonal to the plane of motion. They are
\begin{equation}\label{eq:conserved_quantities}
\begin{split}
    M(\phi(r))\frac{\dd t}{\dd\tau} &= E,\\
    M(\phi(r))r^2\frac{\dd\theta}{\dd\tau} &= j\ .
\end{split}
\end{equation}

The normalization constraint~\eqref{eq:geodesic},
\begin{equation}
    1=
    \left(\frac{\dd x}{\dd\tau}\right)^2
    =
    \left(\frac{\dd t}{\dd\tau}\right)^2
    -
    \left(\frac{\dd r}{\dd\tau}\right)^2
    -
    r^2\left(\frac{\dd\theta}{\dd\tau}\right)^2\ ,
\end{equation}
determines the radial velocity $\dd r/\dd\tau$ up to a sign, corresponding
to the incoming and outgoing branches of the trajectory. Using
\cref{eq:conserved_quantities}, we trade derivatives with respect to
proper time for derivatives with respect to the angular coordinate
$\theta$. Inverting the resulting expression for $\dd r/\dd\theta$ gives
\begin{equation}
    \frac{\dd\theta}{\dd r}
    =
    \pm\frac{j}{r}
    \frac{1}{\sqrt{r^2\left(E^2-M^2(\phi(r))\right)-j^2}}\ .
\end{equation}
Choosing the angular orientation such that $j>0$, the minus sign
describes the incoming branch, along which $r$ decreases as $\theta$
increases, while the plus sign describes the outgoing branch. We use
the negative root when integrating from spatial infinity to the
radial turning point.

We will also use the momentum of the worldline $k=|\boldsymbol{p}_\probe|$ as 
in \cref{sec:scalar_model}. 
The four scalar theories considered in \cref{subsec:scalar_angles} are
summarized in \cref{tab:scalar_worldline_models}.
\begin{table}[ht]
\centering
\begin{tabular}{c|c}
\hline
\hline
Theory & $M^2(\phi)$ \\
\hline
Pure cubic &
$m_\probe^2+m_\probe g_\probe\phi =
m^2_\probe-\dfrac{m_\probe g_\probe g_\heavy}{8\pi r}$
\\[0.8em]
Pure quartic &
$m^2_\probe+\dfrac{\lambda_\probe}{2}\phi^2=
m^2_\probe+\dfrac{\lambda_\probe g_\heavy^2}{128\pi^2 r^2}$\\[0.8em]
square scalar&
$\left(m_\probe+\dfrac{g_\probe}{2}\phi\right)^2 =
m^2_\probe-\dfrac{m_\probe g_\probe g_\heavy}{8\pi r}+ \dfrac{g^2_\probe g_\heavy^2}{256 \pi^2 r^2}$\\[0.8em]
pseudophoton&
$m_\probe^2+m_\probe g_\probe \phi-\left(\dfrac{g_\probe}{2}\phi\right)^2 =m^2_\probe-\dfrac{m_\probe g_\probe g_\heavy}{8\pi r}- \dfrac{g^2_\probe g_\heavy^2}{256 \pi^2 r^2}$
\\[0.8em]
\hline
\hline
\end{tabular}
\caption{Worldline mass functions for the four scalar theories. 
Ref.~\cite{Brandhuber:2026dkd} gives a different but 
equivalent parametrization of the cubic scalar theory.}
\label{tab:scalar_worldline_models}
\end{table}

In all four cases, the angular integral can be written as
\begin{align}
    \Delta\theta
    &=
    \frac{j}{k}
    \int_{r_t}^{\infty}
    \frac{\dd r}{r}
    \frac{1}{\sqrt{(r-r_t)(r+r_1)}}
    \\
    &=
    \frac{\pi j}{k\sqrt{r_t r_1}}
    -
    \frac{2j}{k\sqrt{r_t r_1}}
    \arctan\left(\sqrt{\frac{r_t}{r_1}}\right)\ .
\end{align}
Here $r_t$ is the turning point, while $-r_1$ is the second root of
the radial polynomial. The relevant roots are
\begin{align}
    \text{pure cubic:}\qquad
    r_{t/1} &=
    \mp\frac{m_\probe g_\probe}{2k^2}\frac{g_\heavy }{8\pi}
    +
    \frac{1}{k^2}
    \sqrt{k^2j^2+\frac{m_\probe^2 g_\probe^2}{4}\left(\frac{g_\heavy}{8\pi}\right)^2},
    \\
    \text{pure quartic:}\qquad
    r_t &= r_1 =
    \frac{1}{k}\sqrt{j^2+\frac{\lambda_\probe}{2}\left(\frac{g_\heavy}{8\pi}\right)^2},
    \\
    \text{square scalar:}\qquad
    r_{t/1} &=
    \mp\frac{m_hg_\probe}{2k^2}\frac{g_\heavy}{8\pi}
    +
    \frac{1}{k^2}
    \sqrt{k^2j^2+\frac{g_\probe^2}{4}\left(\frac{g_\heavy}{8\pi}\right)^2(k^2+m_\probe^2)}\,,
    \\
    \text{pseudophoton:}\qquad
   r_{t/1} &=
    \mp\frac{m_hg_\probe}{2k^2}\frac{g_\heavy}{8\pi}
    +
    \frac{1}{k^2}
    \sqrt{k^2j^2+\frac{g_\probe^2}{4}\left(\frac{g_\heavy}{8\pi}\right)^2(-k^2+m_\probe^2)}\,.
\end{align}
The physical scattering angle is,
\begin{equation}
    \chi=2\Delta\theta-\pi\ .
\end{equation}
For the four theories in \cref{tab:scalar_worldline_models} one finds perfect agreement with the results reported in \cref{tab:scalar_worldline_angles} from the partial-wave analysis.

\subsection{Worldline Formulation and Classical Amplitudes}
\label{subsec:wlForm}
In order to connect to the perturbative computations presented in the main text, 
we seek to compute multi-emission amplitudes directly in the classical scalar theory. We thus follow the worldline  effective  field theory construction  \cite{Goldberger:2004jt,Porto:2016pyg,Mogull:2020sak}. As is customary, one allows for  both  the bulk field and the worldline positions to  fluctuate around a chosen background. For the latter, the expansion around the straight-line path is
\begin{equation}
    x^\mu(\tau)  = \xinf^\mu+ u^\mu\tau +z^\mu(\tau)\ .
\end{equation}

Here we specialize to the square scalar theory for simplicity, but analogous
rules can be obtained for the other three theories.
The point-particle action can be divided  into a free contribution 
\begin{equation}\label{eq:freepp}
    S_{\text{pp}} ^0 = -  \int \dd\tau \Big(\underbrace{m}_{\text{const.}}  +\underbrace{m \eta_{\mu\nu} u^{\mu}\dot{z}^{\nu}(\tau)}_{\text{tot. deriv.}} +\frac{m}{2} \eta_{\mu\nu} \dot{z}^{\mu}(\tau)\dot{z}^{\nu}(\tau)\Big)\ ,
\end{equation}
and an interaction part  
\begin{equation}\label{eq:int}
   S_{\text{int}} =  - \frac{g}{2} \int \dd\tau \phi(x(\tau)) \Big(1 + \eta_{\mu\nu} u^{(\mu}\dot{z}^{\nu)}(\tau) +\frac{1}{2} \eta_{\mu\nu} \dot{z}^{(\mu}(\tau)\dot{z}^{\nu)}(\tau)\Big)\ .
\end{equation}
\subsubsection*{Feynman Rules}
Deriving the  Feynman rules is straightforward. 
For the bulk propagator we have the usual expression in position space,
\begin{equation}
    D_{\phi}(x,y) = i\int_\ell \frac{e^{-i \ell {\cdot}(x-y)}}{\ell^2}\,,
\end{equation}
where the $\ell$ subscript denotes integration with respect to $\hd^d\ell$. 
The worldline propagator is obtained from the quadratic form in \cref{eq:freepp}
\begin{equation}
   D_{z}(z^\mu(\tau_1),z^\nu(\tau_2))  = 
   -\frac{i\eta^{\mu\nu}}{m} \int_\omega  \frac{e^{-i\omega (\tau_1-\tau_2)}}
   {(\omega\pm i \varepsilon)^2} 
   = \frac{i\eta^{\mu\nu}}{2m} \Big[|\tau_1-\tau_2| \pm (\tau_1-\tau_2)\Big]\,,
\end{equation}
where the $\omega$ subscript denotes integration with respect to $\hd\omega$.
The interaction vertices can be obtained by exposing the 
$z(\omega)$ dependence via the Fourier transform of the bulk and the worldline fields
\begin{align}
    \phi(x(\tau)) &= \int_k e^{ik{\cdot}(b+u\tau+z(\tau))}\phi(-k) = \sum_{n=0}^\infty\frac{i^n}{n!}\int_k e^{ik{\cdot}(b+u\tau)} (k{\cdot} z(\tau))^n\phi(-k)\,,\\
    \phi(x(\tau)) &= \sum_{n=0}^{\infty} \frac{i^n}{n!}\int_{k, \omega_1, \omega_2, \cdots, \omega_n} e^{ik{\cdot} b} e^{i\big(k{\cdot} u+\sum_{j=1}^n\omega_j\big)\tau} \Big[ \prod_{j=1}^n k{\cdot} z(\omega_j)\Big] \phi(-k)\,,
    \label{eq:bulkfield}
\end{align}
so that  the interaction in \cref{eq:int} becomes 
\begin{equation}
\begin{split}
   S_{\text{int}} &= -\frac{g}{2} \sum_{n=0}^{\infty} \frac{i^n}{n!}\int_{k,\omega_1,\omega_2,\cdots,\omega_n}
   e^{ik{\cdot} b} 
     \hat{\delta}\big(k{\cdot} u +\sum_{i=1}^n\omega_i\big)\phi(-k) \Big[ \prod_{i=1}^n  z^{\rho_i}(\omega_i)\Big]  \\
     &\hphantom{=}
     \times \Big[ \Big( \prod_{i=1}^n k_{\rho_i} \Big) +   \sum_{i=1}^n\omega_i   \Big( \prod_{j\ne i}^n k_{\rho_j} \Big)  u_{\rho_i} + \sum_{i<j}^n \omega_i\omega_j \Big( \prod_{\ell\ne i,j}^n k_{\rho_\ell} \Big) \eta_{\rho_i\rho_j} \Big]\,.
\end{split}
\end{equation}

From this, we can immediately write the $n$-point interaction vertex with $n$ insertions of worldline fluctuations and one bulk field leg
\begin{equation}
\begin{aligned}
\vcenter{\hbox{\includegraphics[page=1]{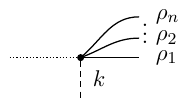}}}
&= i^{n-1}\frac{g}{2} e^{i k{\cdot} b}
\hat{\delta}\big(k{\cdot} u +\sum_{i=1}^n\omega_i\big)
\\
\times &
\bigg[
\Big( \prod_{i=1}^n k_{\rho_i} \Big)
+ \sum_{i=1}^n\omega_i
\Big( \prod_{j\ne i}^n k_{\rho_j} \Big) u_{\rho_i}
+ \sum_{i<j}^n \omega_i\omega_j
\Big( \prod_{\ell\ne i,j}^n k_{\rho_\ell} \Big)
\eta_{\rho_i\rho_j}
\bigg]\,.
\end{aligned}\label{eq:wqft_vertex}
\end{equation}
\subsubsection*{Scalar Multi-Emission Amplitudes}
Worldline Feynman rules can be used directly to compute classical scalar multi-emission
amplitudes. Although the procedure is straightforward, it quickly becomes
cumbersome at higher orders. 
We present the first few orders in the amplitudes anyway. At one emission we have,
\begin{equation}
\begin{aligned}
  \vcenter{\hbox{\includegraphics[page=2]{figures/wqft-diagrams.pdf}}}
  &=-gm \,\hat \delta \bigl(u{\cdot}k_1\bigr)\ ,
\end{aligned}
\end{equation}
where the delta function expresses the conservation of energy along the worldline. In the main text, we do not include the energy-conserving delta function in the definition of the amplitudes $A_N$. At two emissions 
 \begin{equation}
\vcenter{\hbox{\includegraphics[page=3]{figures/wqft-diagrams.pdf}}}
+\, \text{perm.}\,=
-\frac{ g ^2  }{2}\hat\delta \left(u{\cdot}k_1+u{\cdot}k_2\right)
\left(\frac{k_1{\cdot}k_2}{\bigl(u{\cdot}k_1\bigr)^2}+1\right)\ ,
\end{equation}
where the two contributions come from the cubic and quartic coupling from the QFT perspective. At three emissions 
\begin{equation}
\begin{split}
\vcenter{\hbox{\includegraphics[page=4]{figures/wqft-diagrams.pdf}}}
+\, \text{perm.}\,=
   & -\frac{  g^3 }{4m}\left(\frac{k_1{\cdot}k_2 k_2{\cdot}k_3}
   {\bigl(u{\cdot}k_1\bigr)^2 \bigl(u{\cdot}k_3\bigr)^2}+\frac{k_2{\cdot}k_3 k_1{\cdot}k_3}{\bigl(u{\cdot}k_1\bigr)^2 \bigl(u{\cdot}k_2\bigr)^2}\right.\\
   &\qquad
   +\frac{k_1{\cdot}k_2 k_1{\cdot}k_3}{\bigl(u{\cdot}k_2\bigr)^2 \bigl(u{\cdot}k_3\bigr)^2}
+\frac{k_1{\cdot}k_2+k_1{\cdot}k_3}{\bigl(u{\cdot}k_1\bigr)^2}
\left.+\frac{k_1{\cdot}k_2+k_2{\cdot}k_3}{\bigl(u{\cdot}k_2\bigr)^2}\right.\\
&\qquad\left.+\frac{k_1{\cdot}k_3+k_2{\cdot}k_3}{\bigl(u{\cdot}k_3\bigr)^2}\right) \times\hat \delta \left(u{\cdot}k_1+u{\cdot}k_2+u{\cdot}k_3\right)\ .
\end{split}
\end{equation}

While the double-pole structure of the amplitude is expected from the
worldline propagators, arriving at this form for the amplitudes is
nontrivial. The reason is that the powers of energy appearing in the
worldline vertices~\eqref{eq:wqft_vertex} can partially, or even
completely, cancel the worldline propagators. The double-pole form
emerges only after summing over all permutations of the massless legs.

At four and five emissions the following worldline diagrams contribute
\begin{equation}
\vcenter{\hbox{\includegraphics[page=5]{figures/wqft-diagrams.pdf}}}+\, \text{perm.}\,,\qquad\vcenter{\hbox{\includegraphics[page=6]{figures/wqft-diagrams.pdf}}}
+\, \text{perm.}\ .
\end{equation}

The worldline amplitudes presented here are in complete agreement with the Cayley-tree
construction of \cref{eq:weight_scalar} when $\lambda=0$. The
unlabeled trees contributing to the Cayley formula for three, four, and five
emissions were displayed in \cref{fig:cayley_trees}. This collection of trees can
be divided into two classes: those associated with sequential contractions,
which we refer to as \textit{diagrammatic} terms, and those associated with
non-diagrammatic contractions. The former have a direct graphical
interpretation from the quantum field theory perspective, since they arise from
the corresponding Feynman diagrams. The latter instead emerge from
nontrivial algebraic manipulations in the classical limit.

Classical worldline diagrams nevertheless make the origin of the different tree
topologies in the Cayley construction transparent: whether sequential or
non-sequential, their topologies coincide. The comparison is not strictly
one-to-one, however, because a single classical worldline diagram contains more
terms than the corresponding Cayley tree. These additional contributions are
those in which one or more worldline propagators are partially or completely
canceled by the powers of energy appearing in the Feynman 
rule~\eqref{eq:wqft_vertex}. After summing over the relevant permutations, these
extra terms cancel among different diagrams, leaving the amplitude written in the Cayley form. 

We note that the factorization theorem in~\cref{eq:factorization_theorem} reveals a property of 
these amplitudes otherwise hidden in the worldline formulation: 
sequential singularities are nonzero if and only if they arise from diagrammatic 
contributions.

\bibliographystyle{JHEP}
\bibliography{papers}

\end{document}